\documentclass[aps,pra,reprint,amsmath,amssymb,superscriptaddress,onecolumn,longbibliography,nofootinbib,notitlepage]{revtex4-2}
\usepackage{mathtools}
\usepackage{siunitx}
\usepackage[colorlinks,
            linkcolor=black,
            citecolor=cyan,
            urlcolor=magenta]{hyperref}
\usepackage{svg}
\usepackage{float}
\usepackage[braket, qm]{qcircuit}
\usepackage{bbm}

\usepackage{bm}
\usepackage{tocloft}
\usepackage{titletoc}
\usepackage{graphicx}
\usepackage{rotating}
\usepackage[shortlabels]{enumitem}
\usepackage{xcolor}

\begin{document}
\title{Machine vision with small numbers of detected photons per inference}

\author{Shi-Yuan~Ma}
\email{sm2725@cornell.edu}
\email{mashiyua@mit.edu}
\thanks{Present address: Research Laboratory of Electronics, Massachusetts Institute of Technology, Cambridge, MA 02139, USA.}
\affiliation{School of Applied and Engineering Physics, Cornell University, Ithaca, NY 14853, USA}

\author{Jérémie~Laydevant}
\affiliation{School of Applied and Engineering Physics, Cornell University, Ithaca, NY 14853, USA}

\author{Mandar~M.~Sohoni}
\affiliation{School of Applied and Engineering Physics, Cornell University, Ithaca, NY 14853, USA}

\author{Logan~G.~Wright} 
\thanks{Present address: Department of Applied Physics, Yale University, New Haven, CT 06511, USA}
\affiliation{School of Applied and Engineering Physics, Cornell University, Ithaca, NY 14853, USA}
\affiliation{NTT Physics and Informatics Laboratories, NTT Research, Inc., Sunnyvale, CA 94085, USA}

\author{Tianyu~Wang}
\thanks{Present address: Department of Electrical and Computer Engineering, Boston University, Boston, MA 02215, USA.}
\affiliation{School of Applied and Engineering Physics, Cornell University, Ithaca, NY 14853, USA}

\author{Peter~L.~McMahon}
\email{pmcmahon@cornell.edu}
\affiliation{School of Applied and Engineering Physics, Cornell University, Ithaca, NY 14853, USA}

\begin{abstract}
Machine vision, including object recognition and image reconstruction, is a central technology in many consumer devices and scientific instruments. The design of machine-vision systems has been revolutionized by the adoption of end-to-end optimization, in which the optical front end and the post-processing back end are jointly optimized. However, while machine vision currently works extremely well in moderate-light or bright-light situations---where a camera may detect thousands of photons per pixel and billions of photons per frame---it is far more challenging in very low-light situations. We introduce photon-aware neuromorphic sensing (PANS), an approach for end-to-end optimization in highly photon-starved scenarios. The training incorporates knowledge of the low photon budget and the stochastic nature of light detection when the average number of photons per pixel is near or less than 1. We report a proof-of-principle experimental demonstration in which we performed low-light image classification using PANS, achieving 73\% (82\%) accuracy on FashionMNIST with an average of only 4.9 (17) detected photons in total per inference, and 86\% (97\%) on MNIST with 8.6 (29) detected photons---orders of magnitude more photon-efficient than conventional approaches. We also report simulation studies showing how PANS could be applied to other classification, event-detection, and image-reconstruction tasks. By taking into account the statistics of measurement results for non-classical states or alternative sensing hardware, PANS could in principle be adapted to enable high-accuracy results in quantum and other photon-starved setups.
\end{abstract}

\maketitle

\section{Introduction}
\label{sec:intro}

Deep learning has achieved remarkable successes in computer vision \cite{voulodimos2018deep} in scenarios where reliable, well-engineered optical detectors provide high-quality digital data that represents with high fidelity the optical scenes to be processed. 
However, some sensing regimes are fundamentally different: when detection is strongly photon-limited and stochastic, performance depends critically on how information is encoded before it reaches the detector. In
such scenarios, closer integration of physical and digital processes is essential to achieve good task performance.

\begin{figure}[htp]
\centering
\includegraphics [width=.999\textwidth]{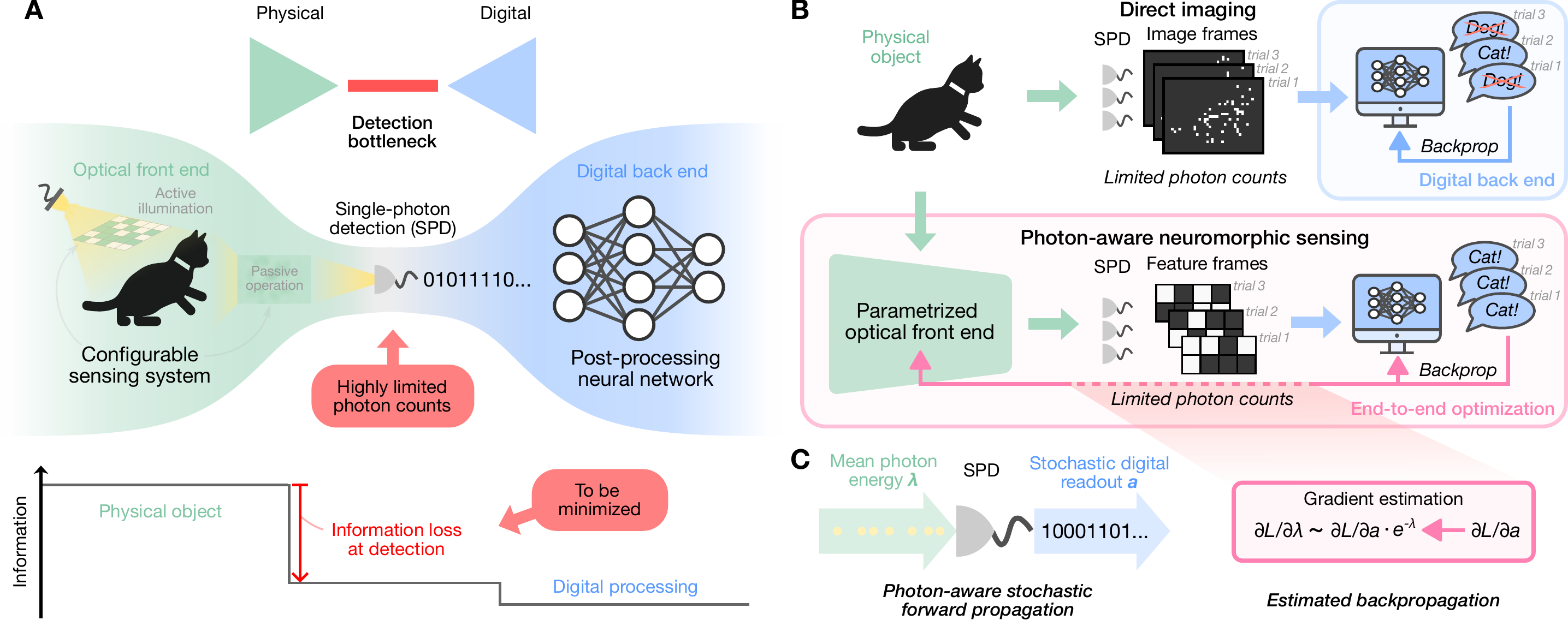}
\caption{\textbf{Detection bottleneck in optical sensing and photon-aware neuromorphic sensing (PANS) under limited photon counts}.
\textbf{A}, Conceptual optical sensing pipeline. An object (illustrated by a cat) interacts with a probe signal (e.g., light) shaped by a configurable optical \textit{front end} (green), which may operate in active (controlled illumination) or passive (incoming-signal modulation) modes.
The detector then converts the incident optical signal into digital data via single-photon detection (SPD).
When photon budgets are highly limited, this conversion presents a \textit{detection bottleneck} with significant information loss (lower schematic), which cannot be recovered by subsequent digital processing.
A digital \textit{back end} (e.g., a post-processing neural network) extracts task-relevant information from the detected data.
\textbf{B}, Direct imaging vs.\ PANS under limited photons. Top: in a conventional direct-imaging pipeline, photon-limited image frames (shown as repeated realizations across independent trials 1, 2, 3) exhibit strong shot noise, making downstream inference challenging (e.g., cat vs.\ dog). Bottom: PANS introduces a parameterized optical front end that transforms the optical field \emph{before} detection, producing photon-efficient feature measurements; the front end and digital back end are jointly optimized end-to-end \emph{through the stochastic detection bottleneck}.
\textbf{C}, Photon detection bottleneck. With mean photon energy $\lambda$, SPD produces a discrete stochastic digital readout $a$. PANS faithfully models this stochastic forward propagation and applies gradient estimation to enable estimated backpropagation (backprop) through the detection bottleneck, allowing end-to-end optimization under photon-budget constraints.
}
\label{fig:intro}
\end{figure}

A general sensing pipeline can be viewed as follows. An object interacts with a physical carrier (e.g., light), producing an analog signal that is measured by a detector and converted into digital data, which are then processed on a digital computer to infer task-relevant information. In resource-scarce settings---such as low signal power or short exposure time (both of which may correspond to detecting only a small number of photons)---the measurement becomes highly non-deterministic and behaves like a lossy channel: only a small fraction of the object’s information survives the detection stage. By the data processing inequality \cite{beaudry2011intuitive}, once information is lost, post-processing in the digital back end cannot recover it. Thus, when this \textit{detection bottleneck} dominates, the only way to preserve more information is to act before detection---i.e., in the physical front end---where we control how object information is presented to the detector (e.g., illumination conditions, light propagation). Conceptually, this can be viewed as an \textit{encoder--decoder} architecture \cite{deb2022fouriernets,yuan2023geometric,wang2025computational,pinkard2024information} (Fig.~\ref{fig:intro}A): a physical encoder determines how information is transformed into measurable signals at detection, followed by a digital decoder that interprets the detected data for the sensing task.

Motivated by this viewpoint, end-to-end optimization (E2E) \cite{sitzmann2018end,metzler2020deep,tseng2021neural,li2021spectrally,zheng2022meta,deb2022fouriernets,pinkard2024information} has become a widely used approach in computational optics, jointly optimizing the physical front end and the digital back end. In particular, neuromorphic sensing \cite{martel2020neural,mennel2020ultrafast,pad2020efficient,wang2023image,zhang2024high,wang2024non,choi2025free} integrates neural-network architectures into physical encoding. 
However, the effectiveness of E2E methods depends on whether the physical process is modeled with sufficient accuracy. When detection resources (e.g., photon counts) are abundant and measurements are reliable, simplified models may suffice; this is the common case where high signal-to-noise ratio (SNR) enables strong digital performance even if the physics is only approximately captured. In contrast, a range of real-world sensing scenarios operate without such redundancy: severely limited optical power (bio-imaging \cite{dixit2003cell,bernas2004minimizing,lichtman2005fluorescence}, minimal-interception setups \cite{sulimany2025quantum}), stringent time constraints (high-throughput or transient sensing \cite{heide2013low}), or both \cite{adan2017flow,schraivogel2022high}. Under such photon-starved conditions, the stochastic nature of detection is not a minor perturbation but the central constraint. The resulting detection bottleneck severely limits information throughput from the physical scene to the digital model and makes end-to-end optimization substantially harder under stringent resource budgets.

Prior work has addressed photon-starved sensing from two complementary directions. On the digital side, many approaches seek to mitigate noise or uncertainty under restricted optical energy \cite{wernick1986image,morris1989pattern,saaf1995photon,chen2017seeing,li2021photon,goyal2021photon}, but cannot affect physical information encoding (e.g., how we illuminate the object or modulate the light field).
On the physical side, decades of work have explored efficient information extraction in low-photon regimes, for example
using nonclassical light states and carefully chosen detection bases \cite{morris2015imaging,zhuang2018distributed,shapiro2020quantum,erhard2020advances,qin2023unconditional,xia2023entanglement,defienne2024advances}, as well as engineered light-matter interactions \cite{yu2014flat,long2019progress,mennel2020ultrafast,wu2022integrated,ma2022intelligent,roques2023biasing}.
These advancements could further benefit if combined with powerful digital back ends in a unified task-specific optimization.

Here, we propose \textit{photon-aware neuromorphic sensing} (PANS), targeting photon-starved scenarios \cite{wernick1986image,morris1989pattern,saaf1995photon,chen2017seeing,li2021photon,morris2015imaging,goyal2021photon} in which detected optical energy (photon counts) is extremely limited, often at the level of a handful to a few tens of detected photons per inference. In this work, we operate in the few-photon-per-inference setting where inference tasks must be performed using only \textit{single-shot} single-photon detection (SPD) measurements for each single-pixel detector in our experimental apparatus, eliminating temporal integration. Our approach consists of two key elements. First, we model the stochastic SPD process as it physically occurs, avoiding approximations that don't faithfully capture the measurement statistics in the photon-starved regime. This enables optimization under true physical constraints, with explicit resource budgets encoded in the loss function, and preserves direct physical meaning of model parameters (e.g., in units of photons rather than arbitrary numerical scales). Second, because standard backpropagation cannot propagate gradients through discrete stochastic measurements, we employ effective gradient estimation techniques that enable end-to-end training despite detection stochasticity.

We validate this framework through experiments and simulations across multiple sensing modalities. Using object classification as a systematic benchmark, we demonstrated experimentally that photon-aware optimization can be used to achieve high accuracy with only a handful of detected photons (2--20 total photons per inference), yielding orders-of-magnitude improvements in photon efficiency over conventional approaches. We then use simulations to explore a broader set of sensing scenarios. For \textit{active} sensing with controlled illumination, we simulated real-time cell classification for flow cytometry and pattern recognition in barcode identification. For \textit{passive} sensing that processes incoming optical signals, we simulated image classification and reconstruction through scattering multimode fibers, transient event detection, tissue perfusion monitoring, and astronomical source classification. Across tasks, PANS enables high performance at photon levels previously considered impractical. 
Together, these results suggest that PANS can accommodate different forms of programmable optical front ends while enforcing the same photon-budgeted optimization principle.
While our work applies the PANS approach to settings with classical light and conventional photon detection hardware, the framework may be compatible with emerging physical approaches such as quantum states of light \cite{shapiro2020quantum,shi2023entanglement,defienne2024advances} and advanced sensing materials \cite{mennel2020ultrafast,ma2022intelligent}. 

\section{Photon-aware neuromorphic sensing (PANS) with highly restricted photon counts}
\label{sec:pans}

The PANS framework is illustrated in Fig.~\ref{fig:intro}B, alongside the conventional ``direct imaging'' approach in which digital processing is applied directly to photon-limited image frames. In PANS, a parameterized optical front end \cite{zheng2022meta,chang2018hybrid,bernstein2023single,chen2023all} transforms the optical field into a task-specific feature space \emph{before} detection, and the resulting detected feature measurements are then processed by a digital back end.

End-to-end optimization of optical--digital pipelines is often effective when photon counts are sufficient and measurements are reliable. Under highly restricted photon budgets---approaching the regime of $\sim$1 detected photon on average per detector readout---measurements become intrinsically stochastic and the variability across independent trials can dominate (Fig.~\ref{fig:intro}B; Fig.~A11). In this regime, training with simplified or deterministic forward models can misrepresent the detection statistics, making optimization substantially more challenging and motivating photon-aware modeling and learning strategies.

\subsection*{Photon-aware modeling of the single-photon detection process}

PANS addresses the detection bottleneck by explicitly modeling photon counting as a stochastic physical process in the forward pass. For classical light, photon arrivals follow Poisson statistics. Given an expected photon number $\lambda$ incident on an ideal single-photon detector (SPD) within a measurement window, the probability of the binary readout $a$ to have a click is $P_{\mathrm{SPD}}(\lambda)=\mathbb{P}(a=1\mid \lambda)=1-e^{-\lambda}$.
We treat this binary click as the activation of a probabilistic neuron \cite{specht1990probabilistic,tang2013learning,peters2018probabilistic,ma2025quantum} and model detection as
$a(\lambda) = \mathbf{1}_{t < P_{\text{SPD}}(\lambda)}$,
where $t \sim \text{Uniform}[0,1]$ and $\mathbf{1}_{\{\cdot\}}$ is the indicator function.

This photon-aware formulation differs from training pipelines that ignore detection noise or approximate it using simplified additive perturbations. Such surrogates can be adequate when photon counts are high ($\lambda\gg 1$), but they become inaccurate in the few-photon regime ($\lambda\sim 1$), where the measurement is intrinsically discrete and strongly non-deterministic. By sampling from the physically correct distribution during every forward pass, the model is trained under the same detection stochasticity it will face at inference, rather than a surrogate noise model.

A key consequence is that $\lambda$ retains direct physical meaning: it represents optical energy in units of photons rather than an arbitrary numerical scale. This enables optimization under true physical constraints, with explicit photon-budget terms encoded in the loss function, and encourages the learned optical encoding to preserve task-relevant information through the lossy detection channel.

\subsection*{Effective gradient estimation for stochastic forward propagation}

Photon-aware modeling introduces a computational challenge: the sampled binary click $a\in\{0,1\}$ is discrete and non-differentiable, so standard backpropagation cannot propagate gradients through the SPD sampling operation. To enable end-to-end optimization through the stochastic detection bottleneck, we employ straight-through estimators (STEs) \cite{bengio2013estimating,hubara2016binarized,ma2025quantum}, which replace the undefined derivative $\partial a/\partial \lambda$ with a surrogate during backpropagation.

In our setting, we find that the naive identity STE ($\partial a/\partial \lambda \approx 1$) is not well matched to the SPD nonlinearity in our regime of interest (when the photon budget is low). Instead, we use a \emph{damped} STE that scales gradients according to photon flux:
\begin{equation}
\frac{\partial \mathcal{L}}{\partial \lambda}
= \frac{\partial \mathcal{L}}{\partial a}\cdot \frac{\partial a}{\partial \lambda}
\approx \frac{\partial \mathcal{L}}{\partial a}\cdot e^{-\lambda}.
\label{eq:ste_gradient_main}
\end{equation}
This implements adaptive gradient scaling: gradients flow in the informative low-flux regime and are naturally suppressed when photon counts increase. Other damping functions with similar qualitative behavior are possible; we adopt $e^{-\lambda}$ for its direct connection to $P_{\mathrm{SPD}}(\lambda)$ (Appendix~1B).

Together with exact stochastic sampling in the forward pass, this estimator enables joint end-to-end training of the optical front end and digital back end through the stochastic detection bottleneck using standard deep-learning frameworks.

\begin{figure}[hbpt]
\centering
\includegraphics [width=.9\textwidth]{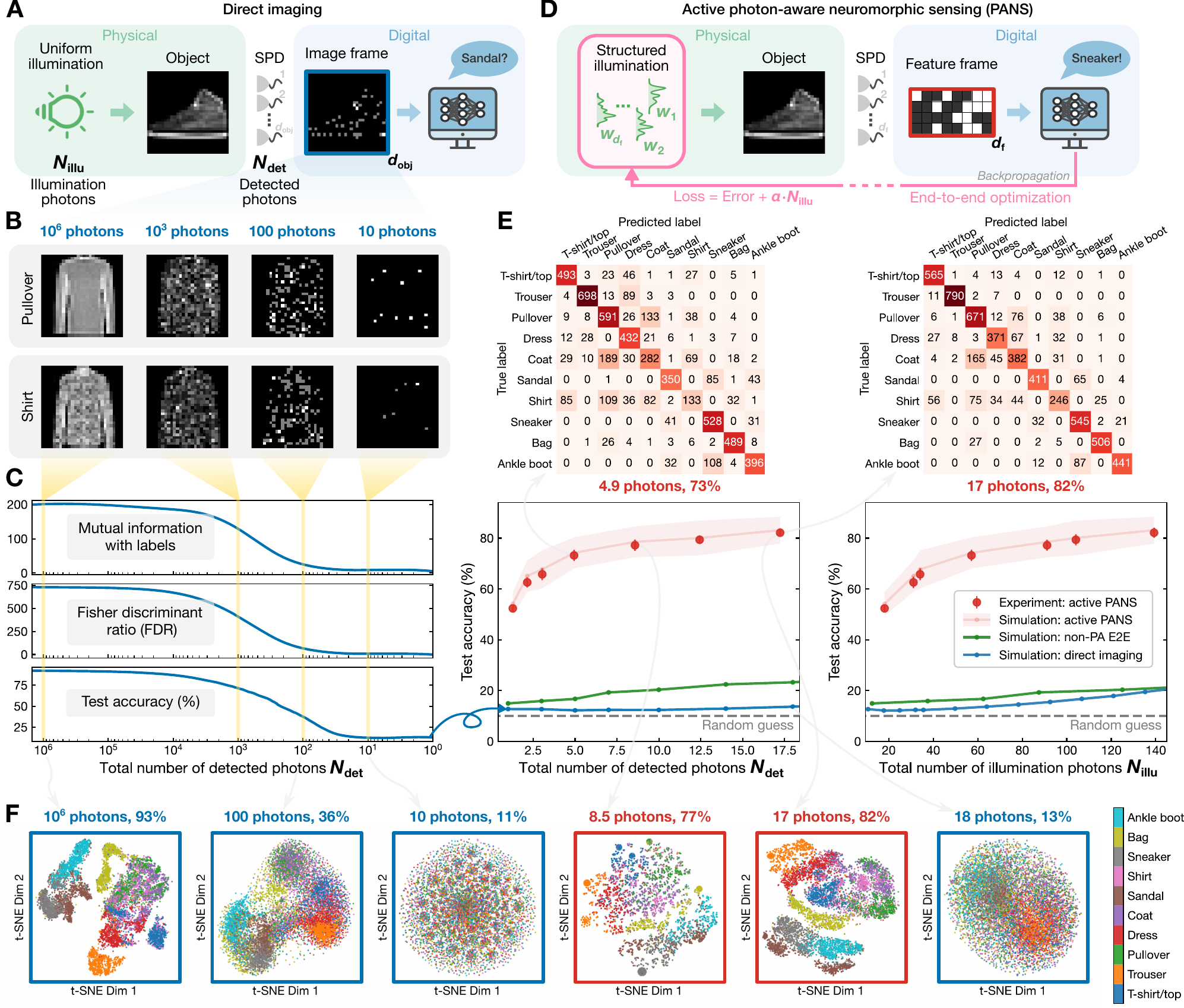}
\caption{\textbf{Active photon-aware neuromorphic sensing (PANS) demonstrated on FashionMNIST object classification.}
\textbf{A}, Direct imaging (conventional approach). Uniform illumination probes the object, and single-photon detectors directly capture an image frame with $d_{\text{obj}}$ pixels. 
$N_\text{illu}$ and $N_\text{det}$ denote average total illumination and detection photon budgets, respectively.
The example object image (a sneaker) is taken from the FashionMNIST dataset.
\textbf{B},  Image frames degrade with decreasing $N_\text{det}$ (denoted above each column) for a pullover (top) and a shirt (bottom).
Frames become increasingly noisy as the photon budget decreases.
\textbf{C}, Quantifying information loss at the detection bottleneck. As $N_\text{det}$ decreases (left to right), three metrics decline: mutual information with labels (top), Fisher discriminant ratio (FDR; middle), and the test accuracy using a convolutional neural network (bottom).
\textbf{D}, Active PANS (our approach). $d_\mathrm{f}$ illumination patterns are projected onto the object, producing a $d_\mathrm{f}$-dimensional feature vector through single-photon detection
(see Fig.~\ref{fig:expresults} and Appendix~12 for details of the experimental protocol). 
\textbf{E}, Experimental results on FashionMNIST. Top: confusion matrices at two different photon budgets. Bottom: Test accuracy vs. $N_{\mathrm{det}}$ (left) and $N_{\mathrm{illu}}$ (right). $N_{\mathrm{illu}}$ is the total illumination incident on the object (uniform for direct imaging; sum of pattern intensities for structured illumination; Appendix~5). Red markers: active PANS experiment (mean $\pm$ std over 30 trials per image) for $d_\text{f} = 3, 4, 6, 10, 16, 24, 32$; light red shade: corresponding simulation (mean $\pm$ 3 std).  Blue curve: direct imaging baseline (from \textbf{C}). Green curve: conventional E2E without photon-aware modeling (non-PA E2E; Appendix~8C).
\textbf{F}, 2D t-SNE \cite{van2008visualizing} visualization comparing feature distributions. Active PANS (red boxes) versus direct imaging (blue boxes) at different $N_\text{det}$ values, with test accuracies shown.}
\label{fig:fashion}
\end{figure}

\section{Quantifying information loss at the detection bottleneck}
\label{sec:info_loss}

When photon budgets are highly limited, the conversion from optical signals to digital readouts becomes a severe information bottleneck: task-relevant information can be irreversibly lost at detection and cannot be recovered by subsequent digital processing. To build intuition, we first quantify how photon shot noise alone degrades the information content of direct-imaging frames as photon counts decrease (see Appendix~8A for more details).

Consider the direct-imaging frame in Fig.~\ref{fig:fashion}A: despite knowing the candidate classes, a photon-limited realization can be visually ambiguous (e.g., resembling a ``sandal'' rather than the true label ``sneaker''). Fig.~\ref{fig:fashion}B illustrates this effect systematically using FashionMNIST examples of a ``pullover'' (top) and a ``shirt'' (bottom). With $N_\mathrm{det}=10^6$ photons per frame, frames closely match the ground truth. As $N_\mathrm{det}$ decreases, frames become increasingly dominated by shot noise; by $N_\mathrm{det}=10$ photons, they contain little visually discernible structure. Even at $N_\mathrm{det}=10^3$ photons, distinguishing ``pullover'' from ``shirt'' is difficult by inspection.

To quantify this information loss, we compute the mutual information (MI) \cite{cover2012elements} between photon-limited frames and class labels, measuring how much label-relevant information remains available for classification (Fig.~\ref{fig:fashion}C, top). We also compute the Fisher discriminant ratio (FDR) \cite{mika1999fisher}, which compares inter-class separation to intra-class variability (middle). Both metrics decrease rapidly once $N_\mathrm{det}$ drops below $\sim10^{4}$--$10^{5}$ photons per frame. An AlexNet-style convolutional network \cite{krizhevsky2012imagenet} trained on these frames shows a corresponding accuracy degradation (bottom), consistent with the optical-energy dependence observed in optical neural network implementations \cite{hamerly2019large,sludds2022delocalized,wang2022optical}.

Crucially, this analysis includes only ideal photon shot noise and excludes additional detector imperfections (e.g., dark counts or read out noise). It therefore represents a \emph{best-case} lower bound on information loss for a given photon budget. Any information not retained through this physical bottleneck is no longer available to subsequent digital processing.

\section{Active PANS using structured illumination}
\label{sec:active}
Active optical sensing designs illumination patterns to estimate object properties efficiently and has a long history in computational optics (e.g., compressive sensing \cite{takhar2006new,duarte2008single}) with applications in LiDAR \cite{gao2018object,lim2003lidar}, transient sensing \cite{heide2013low,winkelbach2002shape}, and biomedical imaging \cite{shi2014fluorescent,liba2017speckle,jin2014light,tian2015computational}. Recent work has explored end-to-end optimization of illumination for specific tasks \cite{horstmeyer2017convolutional,kellman2019physics,zhang2020image,bacca2021deep,hohenester2025optimizing}. We instantiate PANS in this setting by training structured illumination patterns end-to-end under photon-aware modeling of the detection bottleneck.

\subsection*{Learned structured illumination in the optical front end}

In active PANS, $d_\mathrm{f}$ learned illumination patterns are projected onto the object, yielding a $d_\mathrm{f}$-dimensional feature measurement before detection (Fig.~\ref{fig:fashion}D). Each pattern specifies a nonnegative spatial intensity distribution $\vec{w}\in\mathbb{R}_{\ge 0}^{d_\mathrm{obj}}$ applied across the $d_\mathrm{obj}$ object pixels. Under one pattern, the transmitted signal is integrated by a photon counter, producing an expected detected photon number $\lambda = \vec{w}\cdot \vec{x}$ (Fig.~A2), where $\vec{x}\in\mathbb{R}_{\ge 0}^{d_\mathrm{obj}}$ denotes the object transmission (or reflectance). Collecting $d_\mathrm{f}$ patterns forms a matrix $W\in\mathbb{R}_{\ge 0}^{d_\mathrm{f}\times d_\mathrm{obj}}$ that maps object space to a $d_\mathrm{f}$-dimensional feature space, the detected and followed by a digital back end (Appendix~3).

A central goal in photon-starved sensing is to minimize optical energy while maintaining task performance. Because PANS models single-photon detection in physically meaningful units, photon budgets can be imposed directly during training. In particular, we use an objective of the form
\begin{equation}
\mathrm{Loss} = \mathrm{Error} + \alpha\, N_\mathrm{illu},
\end{equation}
where $N_\mathrm{illu}$ is the total illumination photon budget and $\alpha$ controls the accuracy--energy trade-off (Appendix~6D). 
This allows the illumination patterns to be optimized end-to-end under explicit photon constraints.

\begin{figure}[htpb]
\centering
\includegraphics [width=.78\textwidth]{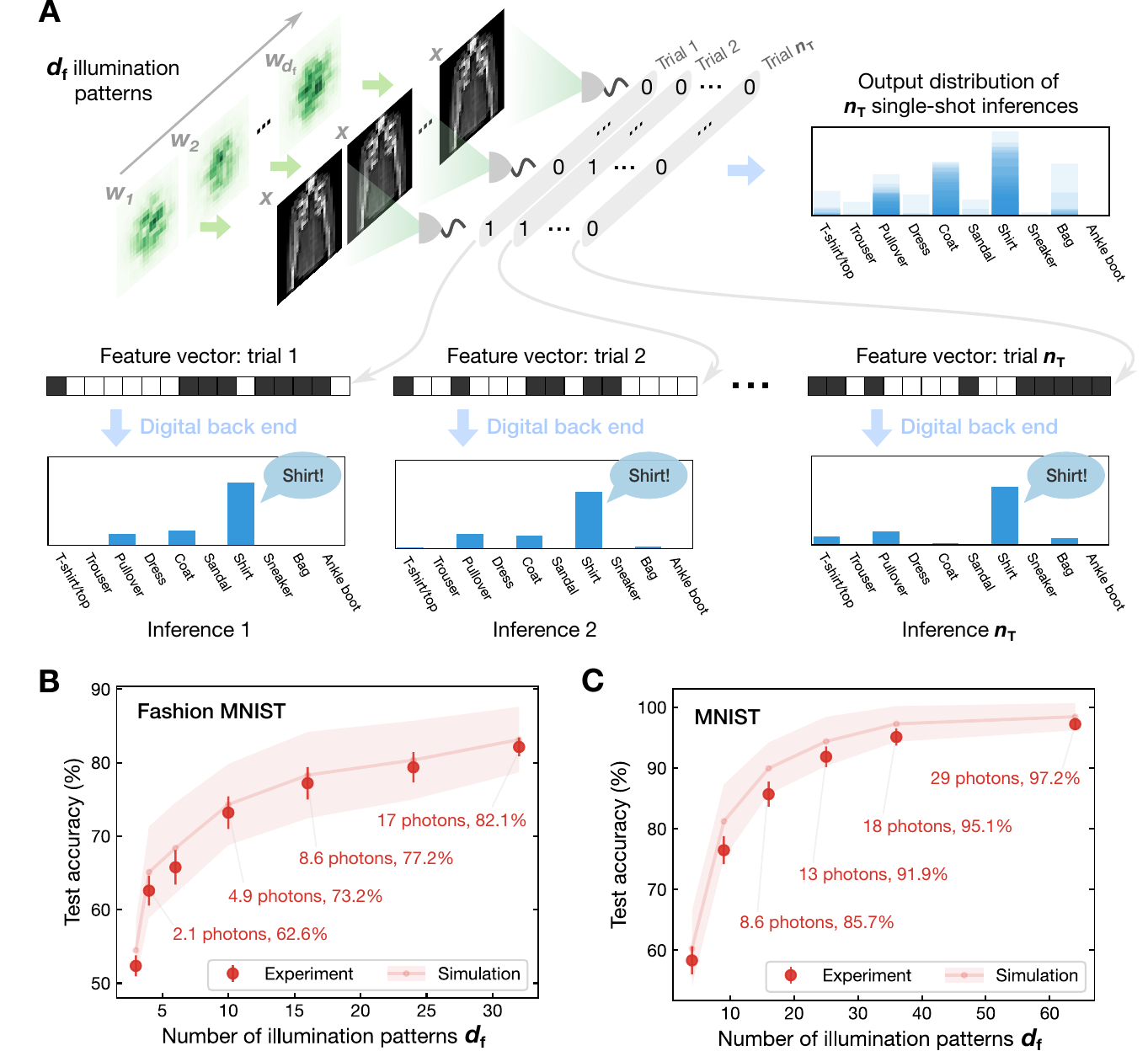}
\caption{\textbf{Stochastic single-shot inference and experimental validation of active PANS.}
\textbf{A}, Stochastic single-shot inference under extreme photon constraints. $d_\mathrm{f}$ learned illumination patterns $\{\vec{w}_i\}_{i=1}^{d_\mathrm{f}}$ are sequentially projected onto an object with transmission $\vec{x}$, each producing a binary single-photon detection (SPD) readout. Because detection is highly stochastic at these photon levels, $n_\mathrm{T}$ independent trials on the same object yield different feature vectors (bottom), each processed by the digital back end. Despite this trial-to-trial variability, the system consistently identifies the correct class across trials. The aggregate output distribution over $n_\mathrm{T}$ inferences (right) reflects the classification confidence (Appendix~13).
\textbf{B--C}, Experimental classification accuracy (red, mean $\pm$ std over $n_\mathrm{T}=30$ trials) versus number of illumination patterns $d_\mathrm{f}$ for FashionMNIST (\textbf{B}) and MNIST (\textbf{C}) under single-shot operation. The FashionMNIST data correspond to the red markers in Fig.~\ref{fig:fashion}E, here plotted against $d_\mathrm{f}$. Simulation results (light red band, mean $\pm\,3$ std) show close agreement with experiment. Annotations indicate the average total detected photon budget per inference $N_{\mathrm{det}}$ at selected $d_\mathrm{f}$ (Appendix~13).}
\label{fig:expresults}
\end{figure}

\subsection*{Case study: Experimental demonstration on FashionMNIST}

We tested this architecture experimentally on FashionMNIST classification \cite{xiao2017fashion}, training illumination patterns jointly with a digital classifier (Appendix~6) and deploying them to an OLED array calibrated to the ultra-low-light regime (Appendices~11--12). Fig.~\ref{fig:fashion}E reports results for $d_\mathrm{f}\in\{3,4,6,10,16,24,32\}$ illumination patterns. We report $N_{\mathrm{det}}$ as the total number of detected photons per inference, summed over all measurements and averaged over the test set, and $N_{\mathrm{illu}}$ as the corresponding total illumination energy incident on the object (Appendices~5,~8A). Active PANS achieves 73\% accuracy with only $N_\mathrm{det}=4.9$ total detected photons and 82\% with $N_\mathrm{det}=17$ (Appendix~13)---performance levels comparable to recent optical neural network demonstrations operating at substantially higher optical power \cite{chen2023all,xia2024nonlinear}, and orders of magnitude more photon-efficient than conventional approaches.

Active PANS substantially outperforms direct imaging (blue curve) at matched photon budgets. To isolate the contribution of photon-aware modeling (Fig.~A1), we evaluate a conventional end-to-end baseline (non-PA E2E; green curve) that uses the same architecture but does \emph{not} model the stochastic SPD process, instead training with approximated expected values and quantization-aware training \cite{wright2022deep,wang2022optical,sludds2022delocalized} (Appendix~8C). At each $N_\mathrm{det}$, the green curve shows the best accuracy achieved among separately trained models with different $d_\mathrm{f}$ (Appendix~8C; Fig.~A15). While this baseline improves over direct imaging, a clear gap remains in the extreme few-photon regime, highlighting the benefit of optimizing end-to-end \emph{through} the stochastic detection bottleneck. Compressive sensing baselines, which use fixed non-task-specific patterns, perform worse than the E2E baseline and are reported separately (Appendices~2D,~8B).

The t-SNE visualization \cite{van2008visualizing} (Fig.~\ref{fig:fashion}F; Figs.~A12,~A24) further illustrates how active PANS preserves task-relevant structure under photon starvation. Compared with direct imaging at similar photon budgets, PANS yields markedly improved class separation, confirming that the optimized front end retains more information through the detection bottleneck.

\subsection*{Stochastic inference and experimental validation}

Fig.~\ref{fig:expresults} details the experimental implementation (Appendices~11--13). At these photon levels, each single-shot inference is highly stochastic: the same object probed under identical conditions yields different detected feature vectors across independent trials (Fig.~\ref{fig:expresults}A, bottom; Fig.~A25). Despite this trial-to-trial variability, the optimized system consistently identifies the correct class, as reflected in the aggregate output distribution over $n_\mathrm{T}$ trials (Fig.~\ref{fig:expresults}A, right; Fig.~A26). The accuracy values reported in Fig.~\ref{fig:fashion}E correspond to the mean ($\pm$ standard deviation) over $n_\mathrm{T}=30$ trials on 200 test images.

Fig.~\ref{fig:expresults}B replots the experimental FashionMNIST results from Fig.~\ref{fig:fashion}E as a function of $d_\mathrm{f}$, and Fig.~\ref{fig:expresults}C extends the evaluation to MNIST, both alongside simulations of the same photon-aware models.
The close agreement across the full range of pattern counts confirms that the stochastic models are robust to realistic experimental imperfections (Appendix~7), validating the PANS framework. For MNIST, active PANS achieves 95.1\% (85.7\%) accuracy using only 18 (8.6) detected photons \emph{in total}---comparable to state-of-the-art works \cite{wang2022optical,sludds2022delocalized} operating at photon budgets $3$--$4$ orders of magnitude higher (see Discussion).

\subsection*{Simulation: Real-time sensing applications}

\begin{figure}[htp]
\centering
\includegraphics [width=.999\textwidth]{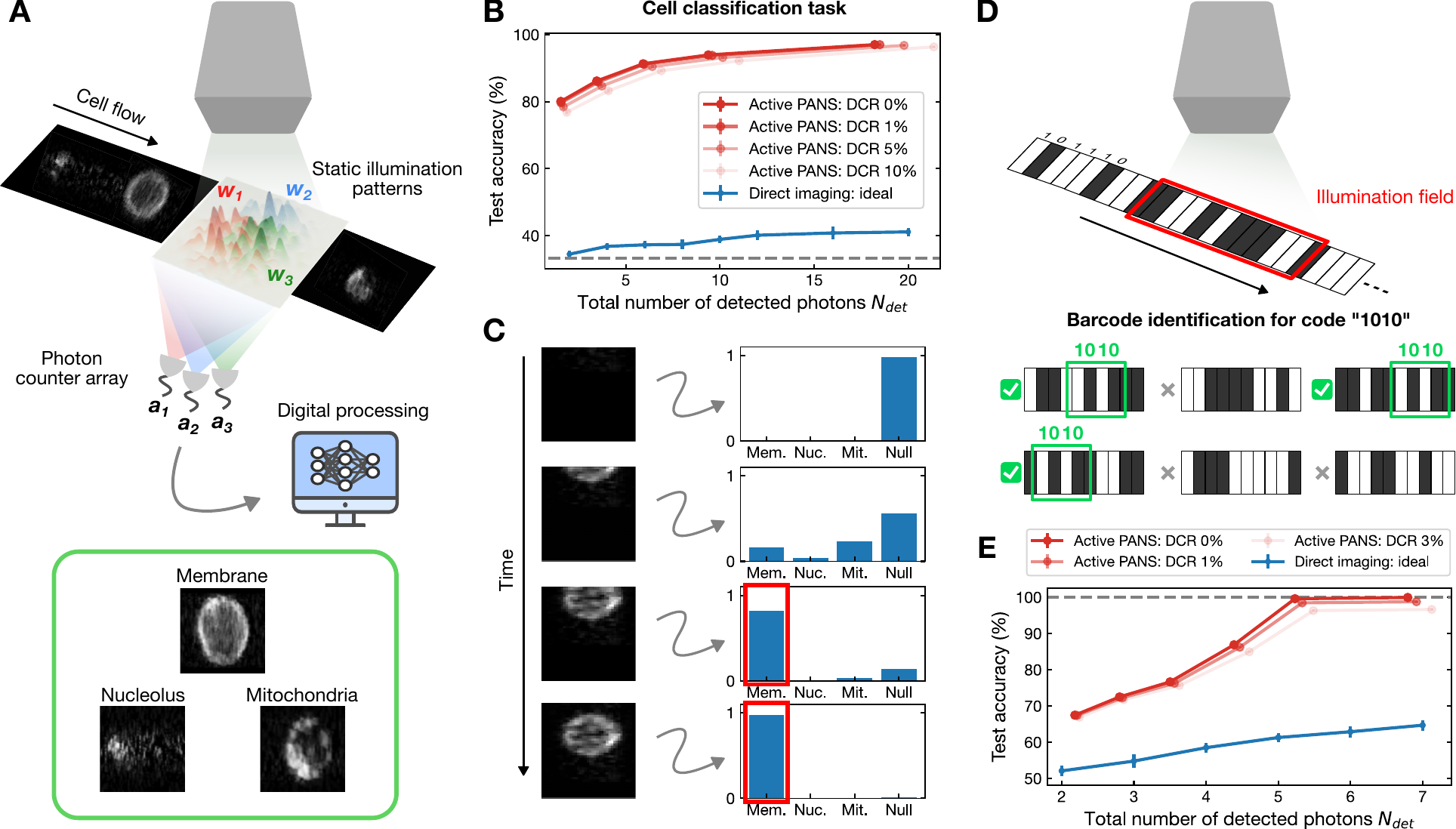}
\caption{\textbf{Proposed real-time image sensing with active photon-aware neuromorphic sensing (PANS) in simulation.}
\textbf{A}, Conceptual wavelength-multiplexed implementation for flow-cytometric cell sorting. Multiple static illumination patterns at distinct optical wavelengths (illustrated as $\vec{w}_1,\vec{w}_2,\vec{w}_3$ with different colors) are applied simultaneously; wavelength demultiplexing routes each channel to a dedicated photon counter, producing activations $(a_1,a_2,a_3)$ in parallel for real-time digital processing.
\textbf{B}, Simulated test accuracy for cell-organelle classification versus total detected photons $N_{\mathrm{det}}$ under different detector dark-count rates (DCRs), compared with an ideal direct-imaging baseline.
\textbf{C}, Example real-time sequence. Left: representative frames as a cell traverses the illumination field (top to bottom). Right: corresponding model outputs over time across classes (Mem.: membrane; Nuc.: nucleolus; Mit.: mitochondria; Null: no cell present).
\textbf{D}, Barcode identification task. The illumination field spans a 10-bar window (red box); the goal is to decide whether the target subsequence ``1010'' appears at any position.
\textbf{E}, Simulated test accuracy for barcode identification versus $N_{\mathrm{det}}$ under multiple DCR values. Direct imaging accounts only for ideal shot noise (no dark counts or additional detector noise), highlighting the robustness of active PANS under realistic counting noise.}
\label{fig:realtime}
\end{figure}

Our experimental demonstration uses time-multiplexing: patterns are projected sequentially and measured with a single detector. In applications where pattern switching limits throughput, active PANS can be implemented in parallel. We therefore propose a wavelength-multiplexed scheme (Fig.~\ref{fig:realtime}A), where different illumination patterns are encoded on distinct optical frequencies and separated by dispersive optics (e.g., gratings) for simultaneous detection on multiple detectors. 
Recent demonstrations of single-photon spectrometers with $>$400 modes spanning 580--660~nm \cite{presutti2024highly} suggest feasibility. With $d_\mathrm{f} \sim 10$ patterns spanning only a few nanometers, dispersion is negligible. This enables static illumination fields with throughput limited only by detector rates, reaching $>$GHz speeds with existing technology \cite{hadfield2009single}.

We evaluate this parallelized active PANS concept in simulation on two real-time tasks (Appendix~9). First, we consider flow cytometric cell classification \cite{schraivogel2022high} (Fig.~A16), where both low illumination (to reduce photodamage \cite{dixit2003cell,bernas2004minimizing}) and rapid decision-making are essential. PANS achieves $\sim 90\%$ accuracy with $\sim5$ detected photons and remains robust to realistic dark count rates (DCRs; Fig.~\ref{fig:realtime}B). We further simulate continuous operation in which cells enter and exit the illumination field; Fig.~\ref{fig:realtime}C shows the evolving activation vectors and corresponding predictions, illustrating real-time readout without reconstructing full image frames.

Additionally, we study a barcode identification task that requires recognizing a target bit pattern at arbitrary locations (Fig.~\ref{fig:realtime}D; Fig.~A17). The goal is to decide whether a 10-bar segment contains the sequence ``1010,'' like a Turing machine tape reader. Active PANS achieves near-unity accuracy with $\sim5$ detected photons (Fig.~\ref{fig:realtime}E), consistently across multiple noise conditions including realistic DCRs. In contrast, direct imaging (simulated with ideal shot noise only) performs poorly at these photon levels.

\begin{figure}
\centering
\includegraphics [width=.92\textwidth]{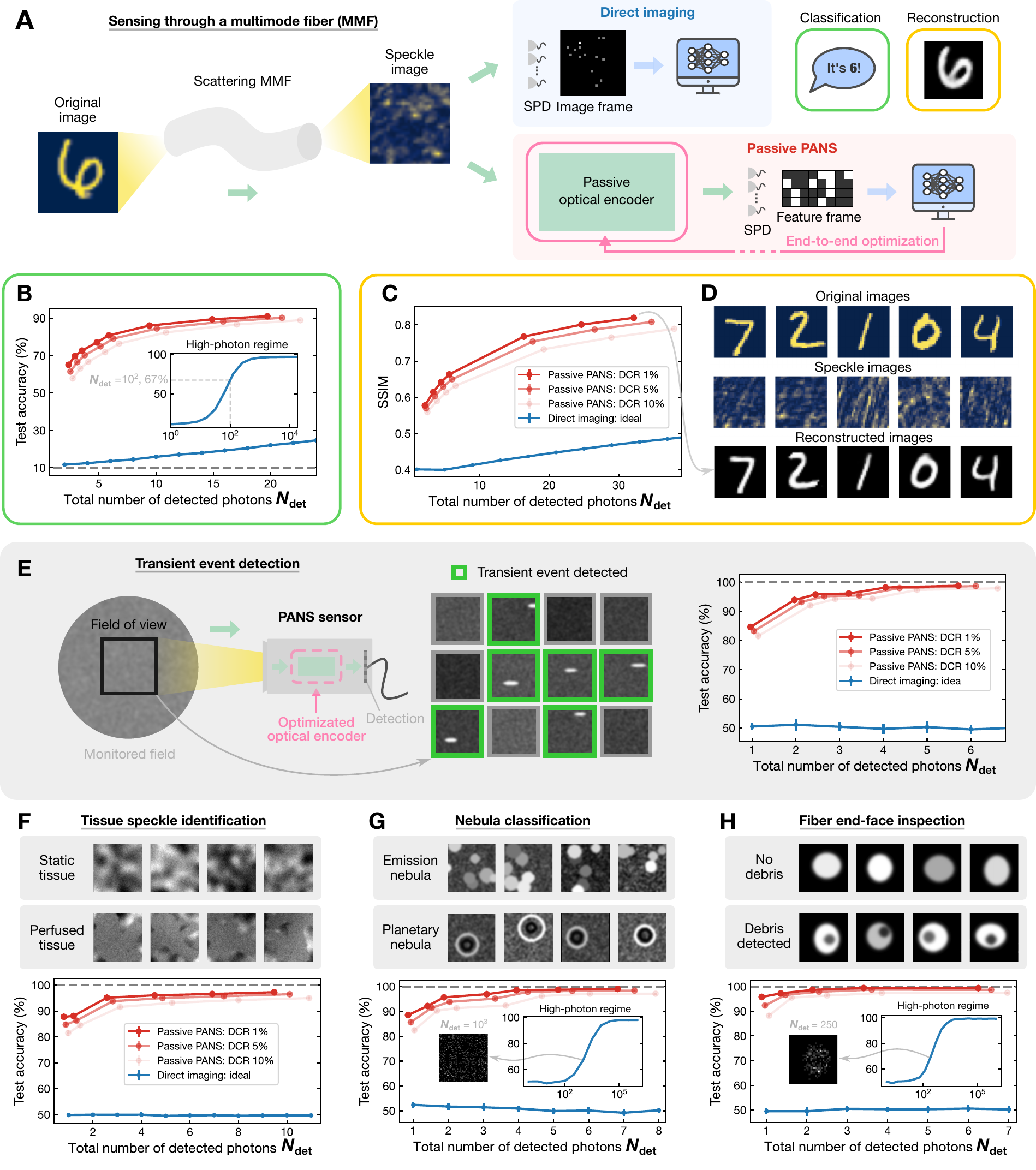}
\caption{
\textbf{Diagram and applications of passive photon-aware neuromorphic sensing (PANS) in simulation.}
\textbf{A}, Passive PANS vs. direct imaging for sensing images transmitted through a scattering multimode fiber (MMF). Input images propagate through the MMF, emerging as speckle patterns that scramble spatial information. In passive PANS, speckles pass through a passive optical encoder before detection; in direct imaging, speckle frames are captured directly at the image plane with $N_\text{det}$ photons per frame. 
Both schemes use post-processing neural networks for classification or reconstruction.
\textbf{B}, Classification accuracy on MNIST speckle images versus $N_\text{det}$, with dark count rates (DCRs) of 1\%, 5\%, and 10\% for passive PANS. Direct imaging (blue curve) simulated using only ideal shot noise.
Inset: direct imaging accuracy at higher $N_\text{det}$. Red markers show passive PANS with $d_\mathrm{f}=4,5,6,8,10,16,24,32$.
\textbf{C}, Average structural similarity index (SSIM) of reconstructed images from scattered speckles, evaluated with different DCRs. Passive PANS data points: $d_\mathrm{f}=4,6,8,10,32,48,64$.
\textbf{D}, Example images showing original MNIST digits (top), corresponding speckle patterns (middle), and reconstructed images (bottom) from passive PANS ($d_\mathrm{f} = 64$, DCR = 1\%). 
\textbf{E}, Transient event detection: fleeting objects in noisy backgrounds are identified in a monitored scene (left). Right: test accuracy vs. $N_\text{det}$. Passive PANS: $d_\mathrm{f}=2,4,5,6,8,10$.
\textbf{F}, Tissue blood flow detection via speckle contrast imaging.
\textbf{G}, Compact nebula classification.
\textbf{H}, Optical fiber end-face contamination inspection.
Insets show direct imaging performance at higher photon counts.
}
\label{fig:passive}
\end{figure}

\section{Passive PANS with optical linear operations}
\label{sec:passive}

Optical linear processors have driven decades of advancement in optical neural networks \cite{wetzstein2020inference}, with mature implementations in both free-space \cite{lin2018all,gigan2022imaging,spall2020fully,bernstein2023single,wang2024large} and integrated platforms \cite{reck1994experimental,shen2017deep,bandyopadhyay2024single}. Here we show in simulation that PANS extends naturally to \emph{passive} optical encoders that apply learnable transformations to incident optical fields.
In contrast to active PANS, passive PANS does not require illumination control: it processes existing optical signals under observation, making it applicable to a broad range of sensing settings.

As in active PANS, we model the optical front end as a linear operator $W$ representing the transmission matrix of an optical processor \cite{popoff2011controlling}. For the demonstrations below, we focus on coherent inputs and real-valued $W$ \cite{lin2018all,spall2020fully,gigan2022imaging}, a standard regime for many established linear optical processors; passive processing of incoherent signals \cite{rahman2023universal} is also possible and exhibits qualitatively similar behavior. After the optical transformation, the detected optical energy is given by the squared norm of the transformed field amplitudes, followed by single-photon detection (Appendix~4). 

\subsection*{Simulation: MMF-based image sensing---classification and reconstruction}

We first demonstrate passive PANS for image sensing through multimode fibers (MMFs), which scramble spatial information into speckle patterns \cite{cao2022shaping} (Appendix~10). Recent work demonstrated image reconstruction through MMFs using diffractive optical elements \cite{yu2025all}; we extend this setup to the extreme few-photon regime. Input images propagate through an MMF, emerging as random speckles (Fig.~\ref{fig:passive}A; Fig.~A18). In passive PANS, an optimized optical encoder transforms these speckles before single-photon detection; in direct imaging, speckle frames are captured directly under the same photon budget $N_\text{det}$.

For MNIST digit classification through an MMF, passive PANS with a two-layer MLP achieves $\sim$90\% accuracy at $N_\text{det} \sim 10$ photons, while direct imaging requires hundreds of photons to exceed 50\% (Fig.~\ref{fig:passive}B, inset). We also evaluated image reconstruction using structural similarity index measure (SSIM) as the quality metric, averaged over 10,000 test images (Fig.~\ref{fig:passive}C--D; Fig.~A18; Appendix~10A). Passive PANS achieves SSIM $\sim$0.7 at $N_\text{det} \sim$10 photons (Fig.~\ref{fig:passive}C). As expected, reconstruction demands more photons than classification since it requires retaining more complete information of the objects. Both tasks maintain robustness against realistic dark count rates (1--10\%), demonstrating practical viability.

\subsection*{Simulation: Transient event detection and diverse applications}

Passive PANS is particularly effective when weak and fleeting signals must be detected against noisy backgrounds---a common challenge in biomedical imaging, astronomy, security monitoring, and industrial inspection. Fig.~\ref{fig:passive}E demonstrates transient event detection (Appendix~10B) where brief objects appear in a noisy scene under uniform coherent illumination; here, the transient contribution is small compared with background fluctuations (Fig.~A19), and direct imaging degrades sharply at low photon budgets. Passive PANS reliably detects transient events in this regime, achieving $>95\%$ accuracy where direct imaging struggles.

To illustrate broad applicability, we further validate passive PANS across biomedical, astronomical, and industrial domains (Fig.~\ref{fig:passive}F--H; Figs.~A20--A22). Speckle-contrast imaging (Fig.~\ref{fig:passive}F) detects blood flow by identifying reduced speckle contrast in perfused versus ischemic tissue \cite{boas2010laser,konovalov2023laser}, enabling low-dose perfusion monitoring during surgery and endoscopy (Appendix~10C). Nebula classification (Fig.~\ref{fig:passive}G) separates planetary from emission nebulae using narrow-band imaging (e.g., H$\alpha$, [OIII] \cite{arnaboldi2003narrowband,galera2018deep}), where quasi-monochromatic emission and compact angular sizes support coherence---highlighting performance on inherently faint astronomical targets (Appendix~10D). Fiber end-face inspection (Fig.~\ref{fig:passive}H) detects surface contamination relevant to telecommunications and laboratory optics; passive PANS enables continuous monitoring by tapping only a small fraction of the signal, without disrupting primary operation (Appendix~10E).

Across these tasks, passive PANS achieves high performance with orders of magnitude lower optical energy than direct imaging in comparable photon-limited regimes, highlighting a task-agnostic strategy for preserving information through the detection bottleneck when optical power or acquisition time is severely constrained. Together with active PANS (Section~\ref{sec:active}), these results demonstrate a general paradigm: programmable physical transformations \emph{before} detection, optimized end-to-end under photon-aware modeling, to maximize task-relevant information flow under extreme resource limits.

\section{Discussion}
\label{sec:disc}

In this work, we have reported high accuracy on machine-vision tasks, including image recognition and reconstruction, even when only a handful of photons in total are detected---a situation in which accuracy would ordinarily be very low. One can think of information about an object as encountering a bottleneck before classification occurs due to there being a limited number of photons conveying information about the object, either because there were few photons illuminating the object to begin with, or because few of them successfully arrived at a detector and were detected, or both.
Our proposed photon-aware neuromorphic sensing (PANS) framework enables optimization considering the actual physical constraints rather than coarse approximations.
This allows the system to maximize task-relevant information \cite{neifeld2007task} flow through the bottleneck under extreme resource limitations.
We restricted our sensing setup to a few photon detectors, each performing only a \textit{single-shot} measurement per inference, and demonstrated high performance with only $\sim$1--10 detected photons.
Our results demonstrate that accurately modeling physical processes in sensing can enhance performance in the highly photon-starved regime.

\subsubsection*{Relation to previous work}
Existing approaches to photon-limited sensing can be categorized by how they address the detection bottleneck. Many methods operate exclusively through digital post-processing without altering the physical front end \cite{wernick1986image,morris1989pattern,saaf1995photon,kirmani2014first,chen2017seeing,gnanasambandam2020image,li2021photon,goyal2021photon}. While such approaches can extract information from noisy measurements, they are fundamentally constrained by only operating on post-detection data and cannot recover information already lost at the detection bottleneck \cite{beaudry2011intuitive} (Fig.~\ref{fig:intro}A, Fig.~\ref{fig:fashion}C).

Other approaches employ various modifications in the optical hardware---including structured illumination \cite{duarte2008single,jin2014light,butow2024generating}, meta-optical elements \cite{yu2014flat,mansouree2020multifunctional,wu2022integrated,tseng2021neural,lee2024mapping,huang2024photonic}, and diffractive elements \cite{popoff2011controlling,lin2018all,li2021spectrally,hu2024diffractive,yu2025all}. 
Many of these demonstrations use strategies that are not task-specific \cite{candes2006near,jin2014light}.
In works that optimize the physical front end for specific tasks \cite{sitzmann2018end,metzler2020deep,tseng2021neural,li2021spectrally,zheng2022meta,deb2022fouriernets,pinkard2024information,martel2020neural,mennel2020ultrafast,pad2020efficient,wang2023image,zhang2024high}, the highly stochastic nature of photon detection of weak light is typically not incorporated explicitly into the optimization, despite being central in the few-photon regime (see the taxonomy in Fig.~A1 and Table~A1). Meanwhile, many end-to-end optimization demonstrations \cite{horstmeyer2017convolutional,kellman2019physics,zhang2020image,mansouree2020multifunctional,bacca2021deep,hohenester2025optimizing} focus on photon budgets where optimization remains well-conditioned through the detection stage. 

Works targeting low-light scenarios while optimizing the front end generally focus on noise resilience of detection hardware rather than photon-aware modeling of the information encoding itself.
In the optical-neural-network (ONN) community, Ref.~\cite{hamerly2019large} proposed low-optical-energy ONNs in simulation, projecting $<1$ photon per multiply-accumulate (MAC) operation---equivalent to roughly $10^4$--$10^6$ photons per layer for reasonable accuracy. Even the most photon-efficient experimental ONN implementations \cite{wang2022optical, sludds2022delocalized} require $\sim 5 \times 10^4$ to $8 \times 10^4$ photons to achieve 90\% accuracy on MNIST when only the first optical layer is considered. This represents 3 orders of magnitude more photons than our approach, which achieves 91.9\% test accuracy with only \textit{13} detected photons (Fig.~3C, Table~A7).

The closest related result we could find in the literature is in Ref.~\cite{zhu2020photon}, which reported $\sim 90\%$ MNIST classification accuracy with $\sim 10^3$ detected photons (two orders of magnitude larger than the light levels in our experiments). Their approach relies on first-photon imaging \cite{kirmani2014first}, detecting photon arrival times rather than intensities via time-correlated single-photon counting. This effectively trades photon counts for temporal resolution: using $\sim 10^4$ detection time bins makes their system $\sim 10^4$ times slower, making it unsuitable for sensing tasks where the signal is weak and transient. Furthermore, first-photon timing measurements are highly sensitive to dark counts, whereas PANS maintains high robustness across realistic experimental conditions (Figs.~\ref{fig:realtime}B, \ref{fig:realtime}E, \ref{fig:passive}B--H; Fig.~A10).

The key distinction of our approach is the modeling of the stochastic detection process, which enables effective end-to-end optimization. By modeling the physical process we fully optimize the entire system against the actual constraints to maximize performance under extreme resource limitations.

\subsubsection*{Key factors of our approach}

Building on this distinction, we identify two major factors that empower PANS to function effectively at such low photon levels: information compression and accurate photon-aware modeling.

Information compression, which has been explored in image sensing \cite{li2021spectrally,wang2023image,huang2024photonic,huang2024pre,xia2024nonlinear} and compressive sensing \cite{takhar2006new,duarte2008single}, leverages the fact that task-specific information \cite{neifeld2007task} is often significantly lower-dimensional than the full image data. For a given task, fewer detectors may suffice; by projecting onto this lower-dimensional subspace, the limited photon budget is concentrated onto task-relevant dimensions rather than spread across the full image space. This insight underlies the effectiveness of many end-to-end approaches. However, compression alone does \textit{not} tell the entire story. For instance, in the barcode identification task (Fig.~\ref{fig:realtime}D), the number of detectors ($d_\mathrm{f}$ ranging from 2 to 16) was not always smaller than the dimension of the original input ($d_\text{obj}=10$), yet PANS still achieved superior performance even with a higher-dimensional feature space. Moreover, illumination patterns optimized through conventional methods---which do not model the stochastic detection process---exhibited significantly poorer performance than PANS even with smaller $d_\mathrm{f}$ (Fig.~\ref{fig:fashion}E; Appendices~8B--8D).

The second and more critical factor is the probabilistic model we developed to faithfully represent the single-photon detection process through photon-aware forward propagation~ \cite{wright2022deep,ma2025quantum}. This model simulates the actual physical process as a noisy channel, explicitly linking uncertainty to photon budgets. During optimization, this forces the system to prioritize photon allocation to features that maximally extract task-relevant information under the given constraints. This resembles the search for an optimal receiver in quantum sensing, where a unitary operation transforms the sensor's measurement basis to maximize state discrimination efficiency~\cite{helstrom1969quantum}. Although our models here are based on straightforward Poissonian photon statistics, our use of them already yielded significant gains in ultra-low-light sensing.

\subsubsection*{Distinction between sensing and ML acceleration}

Much work in optical neural networks (ONNs) aims to accelerate machine-learning inference or training by moving parts of the computation from electronics into optics for speed and energy gains. Our goal is related but fundamentally different: \emph{sensing under stringent detection-energy constraints}. This shift changes what should be optimized and where the bottlenecks lie.

First, the optimization target is different. In ONN accelerators, \emph{computation itself} is the scarce resource: the entire pipeline is judged by end-to-end latency and energy, so every multiplication, memory access, and conversion (between electronics and optics, and between the digital and analog domains) matters. In our setting, the scarce resource are the detected photons. We do not treat downstream digital processing of the digital record of the detected photons as the primary constraint (compute-limited sensing is a valid but different topic of study, which is outside the scope of our work). The key question is therefore: given a fixed detection budget (e.g., photons, detector throughput, temporal bandwidth), how should the optical front end shape and allocate measurements to capture the maximum task-relevant information---even if the subsequent reconstruction or inference is computationally expensive. 

Second, the detection stage becomes the dominant bottleneck. Unlike ONN accelerators, which can often buffer intermediate activations and repeatedly access stored digital values, many sensing signals are transient and cannot be replayed. The detection stage therefore serves as the sole interface between the analog physical process and downstream digital processing, and this interface is lossy \cite{beaudry2011intuitive}: any information not captured at detection is permanently lost. Consequently, the front end must be designed to preserve as much task-relevant information as possible through this one-shot transition, rather than to minimize digital-post-processing operations.

Finally, the evaluation baseline differs. Optical accelerators must justify themselves against highly optimized digital electronic hardware, where overheads such as analog-to-digital and digital-to-analog conversions can erode gains from analog optical processing. In contrast, sensing tasks inherently originate in the analog domain, which digital systems cannot access directly. This creates a natural opportunity for analog physical computing to provide value: by directly interfacing with physical signals and allocating scarce detection resources (e.g., photon budgets, detector throughput, temporal bandwidth) toward task-relevant features.

These considerations motivate PANS as a co-design problem: jointly optimizing the optical front end and the digital back end to maximize information retention under physical constraints at detection. 
In this sense, PANS can leverage the same programmable photonic hardware and optimization techniques developed in the ONN community, but repurpose them from compute acceleration to a sensing-centric goal: optimizing the optical front end to retain maximal task-relevant information under detection constraints, aligning with a broader trend toward task-driven optical sensing \cite{martel2020neural,mennel2020ultrafast,pad2020efficient,wang2023image,zhang2024high,wang2024non,choi2025free}.

\subsubsection*{Robustness and broad applications}
Our results demonstrate that PANS is robust to several non-idealities commonly encountered in practice, including detector and background noise (collectively modeled as dark count rate, DCR), source intensity fluctuations, and imperfections in optical operations (Appendix~7). This robustness is important for translating PANS from controlled demonstrations to realistic deployed sensing systems.

Because energy scales with both optical power and integration time, operating at an ultra-low photon budget benefits not only low-light settings where available power is limited, but also regimes that demand short time windows and high temporal resolution. For example, our transient-event-detection task (Fig.~\ref{fig:passive}E; Fig.~A19) captures a broadly useful morphology---rare, brief deviations from a noisy background---and could be extended to applications such as airborne contamination monitoring in cleanrooms, quality assurance in pharmaceutical manufacturing, conveyor-belt inspection for surface defects, pest detection in food-processing lines, and security perimeter monitoring.

PANS is also attractive when the measurement must minimally disturb the original system. In such settings, a passive optical tap (e.g., a weak beam splitter) can route only a negligible fraction of the light to the PANS front end while leaving the primary optical path essentially unchanged. As a concrete example, for continuous monitoring of contamination on optical fiber ends (Fig.~\ref{fig:passive}H; Fig.~A22), PANS could analyze a weakly tapped signal to detect stains or debris without disrupting normal operation, aligning with broader goals of low-disturbance sensing in security and monitoring scenarios \cite{sulimany2025quantum}.

\subsubsection*{Limitation: Expressivity of the programmable optical front end}
The performance of this framework heavily depends on the optical front end. 
As shown in Fig.~\ref{fig:intro}A, the optical front end is the only component before the lossy detection bottleneck that is configurable. 
In this work, we only demonstrated linear operations in the optical domain. 

A practical rule of thumb to determine if low-light sensing with linear PANS may be able to yield good results for a given task, and what optical front end may be required, is: if the task is solvable with a shallow network on low-dimensional features, then a linear optical encoder + photodetection + a digital back end could be sufficient; tasks requiring more sophisticated processing likely need more expressive optical front ends.

Recent studies have explored nonlinearity in ONN implementations \cite{wang2023image,huang2024pre,xia2024nonlinear,yildirim2024nonlinear,wanjura2024fully,li2024nonlinear}, demonstrating that hybrid optical--electronic neural networks that involve some nonlinearity outperform those with only linear operations.
From a computation perspective, ONNs featuring nonlinearity are more expressive than those featuring only linear operations.
From a sensing perspective, nonlinear encoders better preserve task-relevant information through the detection bottleneck, enabling more effective extraction by the digital back end. In the optical experiments we report, we only used linear operations in the optical front end. A natural extension of our work would be to incorporate a more expressive optical front-end architecture.

\subsubsection*{Outlook: Sensing when the background noise is dominant}
While PANS is robust to realistic noise levels as discussed above (see also Appendix~7), this robustness is not unlimited. In many practical sensing scenarios, additive noise can far exceed the signal---for example, when detecting faint sources against strong thermal backgrounds, or when using photon detectors whose dark counts dominate at low signal levels \cite{zhang2015advances}. 
When such noise buries the signal at detection, information about the object is effectively lost. To sense under these conditions, one can exploit physical properties in which the signal remains distinguishable from the background, such as wavelength \cite{lichtman2005fluorescence}, arrival time \cite{heide2013low}, or photon correlations \cite{morris2015imaging,shapiro2020quantum,defienne2024advances,zhang2015entanglement}. These represent natural extensions of the physical front end that could further broaden the range of sensing scenarios accessible at extreme photon budgets.

\subsubsection*{Outlook: Quantum and other systems in the physical front end}
Beyond classical light states and detection processes, our approach could be extended to other physical settings, such as those involving nonclassical optical states \cite{morris2015imaging,shapiro2020quantum,defienne2024advances} and novel light-matter interactions that can be exploited for sensing tasks, including in 2D materials \cite{mennel2020ultrafast,ma2022intelligent,zhang2024high}, spintronics \cite{grollier2020neuromorphic,islam2024hardware}, and other physical processes \cite{yeon2020alloying,yao2020protonic,roques2023biasing}.
The field of quantum-optical sensing has long focused on maximizing sensing efficiency with limited photon counts, achieving significant developments \cite{morris2015imaging,shapiro2020quantum,agarwal2022quantifying,mukamel2020roadmap,defienne2024advances}.  
Crucially, the detection processes in these systems are often inherently discrete and stochastic---precisely the regime where PANS's photon-aware modeling can provide a principled path to end-to-end optimization across the physical--digital interface.

\subsubsection*{Outlook: Optimization beyond gradient-based algorithms}
This work adopted gradient-based optimization that is compatible with state-of-the-art deep learning techniques. However, for general physical systems, alternative approaches such as in-situ training, backpropagation-free or label-free methods \cite{huo2023optical,momeni2023backpropagation,pinkard2024information} could further improve training efficiency. While these non-gradient methods have yet to match the performance of gradient-based optimizers, future developments in optimization techniques remain critical for advancing configurable systems, especially for stochastic, physical ones.

\subsubsection*{Outlook: Rigorous bounds for information propagation with limited physical resources}
Although our heuristic optimization approach has demonstrated effective practical performance, determining a tight lower bound on the number of photons required for a given task remains an open question.
Quantum theorists have successfully derived bounds for simpler tasks, such as binary hypothesis testing in quantum illumination \cite{shapiro2020quantum}, and rigorous bounds can guide experimentalists towards optimal operations \cite{zhang2015entanglement}. Metrics like Fisher information per photon have also been studied \cite{qin2023unconditional,hohenester2025optimizing}. However, questions like \textit{what is the minimum detected photon count required to classify FashionMNIST with 80\% test accuracy?} are far more challenging due to the high dimensionality of such tasks and the fact that class boundaries are only implicitly defined through finite training data, making a clean information-theoretic formulation difficult. Addressing these questions remains an important and compelling direction for future research \cite{refregier2019bhattacharyya,zhuang2019physical,ding2022bounds}.

\section*{Data and code availability}

All simulation and experimental data, trained model weights, and analysis code needed to reproduce the results presented in this paper are available at \url{https://doi.org/10.5281/zenodo.19210131}. 

\section*{Author contributions}

S.-Y.M., L.G.W., T.W., and P.L.M. conceived the project. S.-Y.M., T.W. and L.G.W. designed the experiments and built the experimental setup. S.-Y.M. developed the theoretical framework. S.-Y.M. performed the numerical simulations, experimental data collection, and data analysis, with assistance from T.W., J.L., M.M.S. and L.G.W. S.-Y.M. wrote the manuscript with input from all authors. P.L.M. supervised the project.

\section*{Acknowledgements}

We thank NTT Research for their financial and technical support (S.-Y.M., P.L.M., T.W. and L.G.W.). Portions of this work were supported by the National Science Foundation (award no. CCF-1918549; J.L., P.L.M. and T.W.) and a David and Lucile Packard Foundation Fellowship (P.L.M.).
We acknowledge discussions with Xingjian~Bai, Saumil~Bandyopadhyay, Chaohan~Cui, Dirk~Englund, Ryan~Hamerly, Mahmoud~Jalali~Mehrabad and Tatsuhiro~Onodera.

\section*{Competing interests}
S.-Y.M., T.W. and P.L.M. are listed as inventors on a U.S. provisional patent application (No. 63/974,312) on the techniques to optimize and implement a hybrid optical sensing system. 

\bibliographystyle{mcmahonlab}
\bibliography{references}

\end{document}


\clearpage
Appendices for 

\title{Machine vision with small numbers of detected photons per inference}

\author{Shi-Yuan~Ma}
\email{sm2725@cornell.edu}
\email{mashiyua@mit.edu}
\thanks{Present address: Research Laboratory of Electronics, Massachusetts Institute of Technology, Cambridge, MA 02139, USA.}
\affiliation{School of Applied and Engineering Physics, Cornell University, Ithaca, NY 14853, USA}

\author{Jérémie~Laydevant}
\affiliation{School of Applied and Engineering Physics, Cornell University, Ithaca, NY 14853, USA}

\author{Mandar~M.~Sohoni}
\affiliation{School of Applied and Engineering Physics, Cornell University, Ithaca, NY 14853, USA}

\author{Logan~G.~Wright} 
\thanks{Present address: Department of Applied Physics, Yale University, New Haven, CT 06511, USA}
\affiliation{School of Applied and Engineering Physics, Cornell University, Ithaca, NY 14853, USA}
\affiliation{NTT Physics and Informatics Laboratories, NTT Research, Inc., Sunnyvale, CA 94085, USA}

\author{Tianyu~Wang}
\thanks{Present address: Department of Electrical and Computer Engineering, Boston University, Boston, MA 02215, USA.}
\affiliation{School of Applied and Engineering Physics, Cornell University, Ithaca, NY 14853, USA}

\author{Peter~L.~McMahon}
\email{pmcmahon@cornell.edu}
\affiliation{School of Applied and Engineering Physics, Cornell University, Ithaca, NY 14853, USA}

\maketitle

\begin{spacing}{1.5}
\tableofcontents 
\end{spacing}

\clearpage
\newpage

\setcounter{page}{1}

\part*{Part I: \\Framework}
\addcontentsline{toc}{part}{I\quad Framework}
\label{part:framework}

\vspace{24pt}
\section{Photon-aware neuromorphic sensing framework}
\label{sec:pans_framework}

This section presents the photon-aware neuromorphic sensing (PANS) framework. The central idea is to explicitly model the stochastic single-photon detection (SPD) process during training, enabling end-to-end optimization \emph{through} the detection bottleneck in the few-photon regime.

\subsection{Modeling of single-photon detection}
\label{subsec:pans_spd}

The SPD process in a PANS model is modeled as a binary probabilistic neuron \cite{tang2013learning,peters2018probabilistic}, corresponding physically to a single readout from a photon detector. For each measurement, the single-shot output is either 0 or 1, with probabilities determined by the incident optical energy. The activation function takes its form directly from the physics of photon detection. We focus on Poissonian statistics, which govern photon arrivals for classical light sources, though the PANS framework generalizes to other detection statistics \cite{helstrom1969quantum,morris2015imaging,zhuang2018distributed,defienne2024advances}. In practice, we found the actual form of the statistic expression does \emph{not} affect the realistic numerical simulation significantly.

\subsubsection*{Derivation from Poisson statistics}
Consider a detector illuminated by classical light with expected photon number $\lambda$ representing the mean photon flux integrated over the measurement window. Assuming photon arrivals follow Poisson statistics:
\begin{equation}
P(n \mid \lambda) = \frac{\lambda^n e^{-\lambda}}{n!},
\end{equation}
where $n$ is the number of detected photons.

In the single-shot SPD regime, each detector makes a binary measurement: either at least one photon is detected ($a = 1$) or no photons are detected ($a = 0$). The probability of detecting a click is:
\begin{equation}
P_{\text{SPD}}(\lambda) = P(a = 1 \mid \lambda) = 1 - P(n = 0 \mid \lambda) = 1 - e^{-\lambda}.
\label{eq:pspd}
\end{equation}

The SPD activation can thus be regarded as a Bernoulli random variable:
\begin{equation}
a(\lambda) = \mathbf{1}_{t < P_{\text{SPD}}(\lambda)},
\end{equation}
where $t \sim \text{Uniform}[0,1]$ and $\mathbf{1}_{\{\cdot\}}$ is the indicator function.

\subsubsection*{Relationship between $\lambda$ and trainable parameters}

The expected photon number $\lambda$ on the optical front end and the chosen configuration. Two common examples are:
\begin{itemize}
    \item \textbf{Intensity encoding}: $\lambda = [\vec{w}^\top \vec{x}]_+$, where $\vec{w}$ represents optical weights and $\vec{x}$ is the input signal (with $[\cdot]_+$ enforcing non-negativity).
    \item \textbf{Field-amplitude encoding}: $\lambda = \left|\vec{w}^\top \vec{E}\right|^2$, where $\vec{E}$ is the input field amplitude and $\vec{w}$ represents a linear optical transformation.
\end{itemize}
PANS is agnostic to the specific optical front end form; it requires only that the trainable parameters in the optical front end can be effectively optimized.

\subsection{Gradient estimation through stochastic neurons}
\label{subsec:pans_gradient}

The discrete, stochastic nature of single-photon detection poses a challenge for gradient-based optimization. The sampling operation $a(\lambda) = \mathbf{1}_{t < P_{\text{SPD}}(\lambda)}$ is non-differentiable, precluding standard backpropagation.

We employ a straight-through estimator (STE)~\cite{bengio2013estimating,hubara2016binarized, ma2025quantum},  a technique proven effective for training networks with discrete activations. The standard STE would set $\partial a/\partial\lambda=1$, passing gradients through the non-differentiable operation unchanged (``straight-through''). However, this approach does not perform efficiently with the actual SPD process. Instead, we incorporate a damping factor $e^{-\lambda}$ that modulates the gradient during backpropagation. The gradient of the loss $\mathcal{L}$ with respect to $\lambda$ is estimated as:
\begin{equation}
\frac{\partial \mathcal{L}}{\partial \lambda} = \frac{\partial \mathcal{L}}{\partial a} \cdot \frac{\partial a}{\partial \lambda} \approx \frac{\partial \mathcal{L}}{\partial a} \cdot e^{-\lambda}.
\label{eq:ste_gradient}
\end{equation}
The factor $e^{-\lambda}$ naturally implements adaptive gradient scaling: gradients flow freely in the low-flux regime where detection is informative, and are naturally suppressed when detection probability saturates.

In practice, the precise form of the damping function matters less than its qualitative behavior: any function that vanishes for large $\lambda$ effectively regularizes against excessive photon counts. However, choosing $e^{-\lambda}$ specifically has a notable effect. Since $e^{-\lambda}$ is also the derivative $\partial P_{\mathrm{SPD}}(\lambda)/\partial \lambda$ and that $P_{\text{SPD}}$ is the expectation value of $a$, $P_{\text{SPD}}(\lambda) = \mathbb{E}[a]$, our estimator becomes \emph{unbiased}:  
\begin{equation}
\frac{\partial \mathcal{L}}{\partial a}\cdot e^{-\lambda} = \frac{\partial \mathcal{L}}{\partial a}\cdot \frac{\partial P_{\mathrm{SPD}}(\lambda)}{\partial \lambda}
= \frac{\partial \mathcal{L}}{\partial a} \cdot \frac{\partial \mathbb{E}[a]}{\partial \lambda}.
\end{equation}
That is, the estimated gradient equals the true gradient of the loss with respect to the expected activation, providing an unbiased signal for optimization.
While unbiasedness is not essential for effective training, this correspondence provides a satisfying theoretical grounding, and we adopt this form throughout.

The STE decouples the forward and backward passes:
\begin{itemize}
    \item \textbf{Forward pass}: Sample $a$ using the true stochastic process.
    \item \textbf{Backward pass}: Propagate gradients through the deterministic function $P_{\text{SPD}}$.
\end{itemize}

This approach connects the physical front end and digital back end despite the highly stochastic detection process. Given the gradient with respect to activation, $g_a = \partial \mathcal{L} / \partial a$, computed via backpropagation through the digital network, we obtain the gradient with respect to expected photon number as $g_\lambda = e^{-\lambda} \cdot g_a$. Gradients with respect to optical front-end parameters (e.g., weight matrix $W$) then follows from the chain rule applied to the specific encoding scheme.

\subsubsection*{Properties and implications in the few-photon regime}

The detection probability $P_{\mathrm{SPD}}(\lambda)=1-e^{-\lambda}$ saturates as $\lambda\rightarrow\infty$, with derivative
$P'_{\mathrm{SPD}}(\lambda)=e^{-\lambda}$ that
decays exponentially. In the high-flux regime ($\lambda\gg 1$), the detection probability approaches unity and becomes insensitive to changes in $\lambda$. In contrast, in the few-photon regime ($\lambda\sim 1$), the measurement is intrinsically discrete and strongly non-deterministic.

From an information-theoretic perspective, the binary entropy of $a$ is maximized when $P_{\mathrm{SPD}}(\lambda)=0.5$, corresponding to $\lambda=\ln2\approx0.69$ photons.
Although task-specific losses do not reduce to maximizing per-neuron entropy, this observation provides intuition: operating near $\lambda\sim 1$ balances informative activations against well-conditioned gradients.

\subsection{Training methods}
\label{subsec:pans_training}

Training jointly optimizes the optical front end and digital back end using gradient-based methods, with our customized STE providing gradient estimates through the stochastic detection layer. Encoding-specific details, including loss functions and implementation considerations, are described in \ref{sec:active_pans} and~\ref{sec:passive_pans}. Here we discuss training techniques common to all PANS implementations.

\subsubsection*{Initialization strategy}

Proper initialization is crucial for training stochastic physical models. If initial weights produce expected photon counts $\lambda$ that are too large, the detection probability saturates ($P_{\text{SPD}} \approx 1$) and gradients vanish ($P_{\text{SPD}}'(\lambda) = e^{-\lambda} \approx 0$), preventing effective learning. Conversely, if initial $\lambda$ values are too small, the activations are nearly always zero, providing weak learning signals.

We employ Kaiming uniform initialization~\cite{he2015delving} with a modified scaling factor. Standard Kaiming initialization includes a parameter controlling weight magnitude; we adjust this to ensure initial $\lambda$ values fall in the range where gradients are well-conditioned The specific choice of $a$ depends on the encoding scheme and is detailed in subsequent sections.
This initialization strategy ensures training begins in a regime where the stochastic detection process provides informative gradients.

\subsubsection*{Slope annealing and intensity clamping}

To further enhance training effectiveness, we introduce a slope parameter $\eta$ that modifies the effective intensity within the SPD activation function:
\begin{equation}
P_{\text{SPD}}^\eta(\lambda) = P_{\text{SPD}}(\eta \lambda) = 1 - e^{-\eta \lambda}.
\end{equation}
The slope annealing technique~\cite{chung2016hierarchical} enables controlled alteration of the gradients, facilitating more efficient navigation of the parameter space. The slope is updated after each epoch by a multiplicative factor $\theta$: $\eta \leftarrow \theta \eta$. In the optical implementation, the annealing factor can be absorbed into the mapping from trained weights to experimental parameters.

We can also clamp intensities to a maximum value $\lambda_{\text{max}}$:
\begin{equation}
\lambda_{\text{clamp}} = \min(\lambda, \lambda_{\text{max}}).
\end{equation}
This prevents vanishing gradients resulting from excessively large intensity values where the probability function plateaus. Both slope annealing and intensity clamping are applied during training only.

\subsubsection*{Batch normalization after detection}

We employ batch normalization on the $d_\text{f}$-dimensional feature vector immediately after the detection stage. This choice is enabled precisely because we are operating in the digital domain after the bottleneck, where such transformations are computationally trivial. Batch normalization serves several important functions:
\begin{itemize}
    \item \textit{Stabilization}: The raw detection outputs exhibit significant variance due to photon shot noise, particularly at low photon counts. Normalization stabilizes the distribution of activations across training batches, improving optimization dynamics.
    \item \textit{Scale invariance}: Normalization provides implicit regularization by making the network invariant to the absolute scale of photon counts. This is valuable when testing across different photon budgets than those used during training.
    \item \textit{Adaptation}: The learned affine parameters of batch normalization allow the network to adapt the feature distribution to the optimal range for subsequent layers.
    \item \textit{Gradient flow}: Normalized activations help maintain healthy gradient magnitudes throughout training, preventing vanishing or exploding gradients in the digital back end.
\end{itemize}

Notably, batch normalization cannot be implemented optically before detection---it requires computing statistics over a batch of measurements and applying learned affine transformations, operations that have no simple optical analog. This exemplifies a key design principle: digital processing should maximally exploit its flexibility to compensate for constraints at the detection bottleneck. Once photons are converted to digital signals, the objective is simply to extract information as effectively as possible.

\subsection{Inference procedure}
\label{subsec:pans_inference}

During inference of a trained model, the forward pass is similar to training except that intensity clamping is not applied. An effective scaling factor (slope) $\eta$ is calibrated with the physical setup to map the trained parameters to actual optical intensity in the setup. To reduce stochastic uncertainty, multiple detection shots can be averaged per inference.

In $K$-shot inference, we use $K$ independent binary SPD readouts and average them to obtain the activation:
\begin{equation}
a^{[K]} = \frac{1}{K} \sum_{k=1}^{K} a^{(k)}, \quad a^{(k)} \sim \text{Bernoulli}(p).
\end{equation}

This yields $a^{[K]} \in \{0, 1/K, 2/K, \ldots, 1\}$. The process reduces uncertainty: $\text{Var}[a^{[K]}] = p(1-p)/K$. In the limit $K \to \infty$, the activation converges to the expectation value without stochasticity: $a^{[\infty]} = \mathbb{E}[a] = P_{\text{SPD}}(\lambda)$.

In practice, models trained with $K=1$ generalize well to inference with larger $K$, achieving higher accuracy as $K$ increases. For the sensing applications presented in this work, we report results with $K=1$ (\textit{single-shot inference}) to demonstrate performance under the most stringent photon constraints, where each detected photon contributes maximally to the sensing task.

\begin{algorithm}[H]
\caption{Inference of PANS}
\label{alg:inference}
\begin{algorithmic}[1]
\Require Test input $\vec{x}$, trainable encoder $W$, digital back-end model $f_\theta$, number of shots $K$, slope $\eta$
\Ensure Output prediction
\State $\vec{\lambda} \gets \text{Encode}(W, \vec{x})$
\Comment{Mapping to $\vec{\lambda}$ depends on the encoding scheme}
\State $\vec{p} \gets 1 - \exp(-\eta \cdot \vec{\lambda})$
\Comment{Virtual probabilities based on input intensity}
\For{$k = 1$ to $K$} \Comment{$K$ shots per inference}
    \State $\vec{a}^{(k)} \gets \text{Bernoulli}(\vec{p})$ \Comment{Single-shot SPD measurement}
\EndFor
\State $\vec{a} \gets \frac{1}{K} \sum_{k=1}^{K} \vec{a}^{(k)}$ \Comment{Average over $K$ shots}
\State \Return $f_\theta(\vec{a})$ \Comment{Digital back-end processing}
\end{algorithmic}
\end{algorithm}

\subsection{The optical front end and the detection bottleneck}
\label{subsec:pans_frontend}

In a general optical sensing system, the object of interest interacts with light, which carries the object's information to the detector. This analog optical signal is then converted to digital data and processed by algorithms to extract insights. Conceptually, this represents an encoder--decoder architecture: the physical optical system acts as an ``encoder'' that determines how information is captured and packaged, and the digital system acts as a ``decoder'' that processes the captured data to complete the sensing task. Critically, these two stages are separated by the \textit{detection bottleneck}---the physical process where analog optical signals are irreversibly converted to digital measurements.

When optical resources are abundant, the detection bottleneck poses little concern; high signal-to-noise ratios ensure that most information survives the analog-to-digital transition. However, under photon-starved conditions, this bottleneck becomes the critical limiting factor. The stochastic nature of photon detection means that information is irreversibly lost at this stage---information that cannot be recovered by any subsequent digital processing, no matter how sophisticated. This asymmetry between pre- and post-detection processing motivates the core principle of PANS: \emph{optimize the optical front end to maximize information flow through the detection bottleneck}.

The optical front end is the only configurable component \textit{before} the lossy detection stage. Its role is to transform the incoming optical signal into a representation that preserves task-relevant information as effectively as possible given the photon budget constraints. Once detection occurs, the physical signal becomes digital data, and we can apply arbitrarily powerful computational methods---but we are limited to working with whatever information has survived the bottleneck. The encoder thus plays a decisive role, determining what information reaches the digital domain

Mathematically, we express the optical encoder as a parameterized mapping:
\begin{equation}
\vec{\lambda} = \mathcal{F}_W(\vec{x}),
\label{eq:general_encoder}
\end{equation}
where $\vec{x}$ denotes the object representation, $W$ represents the trainable parameters of the optical system, and $\vec{\lambda} \in \mathbb{R}_{\geq 0}^{d_\text{f}}$ is the vector of expected photon numbers at each of the $d_\text{f}$ detectors. The transformation $\mathcal{F}_W$ encapsulates all optical physics between the object and the detector array. Different encoder architectures correspond to different functional forms of $\mathcal{F}_W$, but all share the same downstream pipeline: detection samples from $\vec{\lambda}$ to produce digital outputs, followed by digital post-processing.

The key insight of PANS is that by faithfully modeling the detection process as it actually occurs---sampling from the distribution defined by $\vec{\lambda}$ according to the photon statistics---we can optimize the encoder parameters $W$ against the true physical constraints at the bottleneck. This photon-aware approach allows the system to learn optimal allocation of its limited photon budget across feature dimensions, maximizing retention of task-relevant information. Conventional end-to-end optimization methods, which do not explicitly model the non-deterministic detection process, cannot achieve this because their optimization landscapes do not reflect the actual information loss that occurs under the physical constraints of photon-limited detection.

This encoder-agnostic formulation reveals the broad applicability of the PANS framework. Any programmable optical system---whether based on structured illumination, linear transformations, nonlinear dynamics, scattering media, diffractive elements, or quantum states---can serve as the encoder $\mathcal{F}_W$, provided it can be modeled and optimized. The specific encoders presented in \ref{sec:active_pans} (active illumination) and~\ref{sec:passive_pans} (passive light-field transformation) represent particular instantiations chosen for their experimental accessibility and physical clarity. As optical processing platforms continue to advance, this framework provides a principled foundation for optimizing increasingly sophisticated architectures under extreme physical constraints where the detection bottleneck dominates system performance.

\newpage
\section{Sensing paradigms and algorithmic approaches}
\label{sec:paradigms}

Before presenting specific implementations of active and passive PANS, we establish a general framework to situate PANS among different algorithmic approaches to optical sensing, from direct imaging to conventional end-to-end optimization. 
We first describe the general sensing pipeline without assuming any particular operating regime, then discuss how the detection bottleneck---when it occurs---changes which approaches remain effective.
This taxonomy clarifies what distinguishes photon-aware methods and under what conditions they become essential.

\subsection{General framework for optical sensing}
\label{subsec:sensing_framework}

\subsubsection*{The physical object in optical sensing}

Consider a physical object characterized by some spatially-varying optical property---transmission, reflectance, fluorescence distribution, or similar. While such properties are fundamentally continuous, in practice we represent them as a finite-dimensional vector $\vec{x} \in \mathbb{R}^{d_{\text{obj}}}$, where $d_{\text{obj}}$ denotes the number of elements used to describe the object. This discretization is a modeling choice determined by the resolution of the system and computational considerations, not an intrinsic property of the physical object itself.

This representation is standard in computer vision, where $d_{\text{obj}}$ typically corresponds to the number of spatial pixels. More broadly, the $d_{\text{obj}}$ dimensions can represent frequency bins for polychromatic signals, time stamps for temporal measurements, or other bases appropriate for the sensing modality. The framework thus provides a tractable finite-dimensional description applicable across diverse optical sensing applications. The sensing task is then to extract information about $\vec{x}$---whether reconstructing it entirely, classifying it into categories, or estimating parameters of interest.

\subsubsection*{The sensing pipeline}

Any optical sensing system can be decomposed into three stages:

\begin{enumerate}
    \item \textbf{Optical front-end}: Physical operations applied before detection that transform the object representation $\vec{x}$ into optical signals at the detector plane. In active sensing, this includes the illumination that probes the object; in passive sensing, it includes optical elements (lenses, diffractive components, spatial light modulators, scattering media) that process incoming light. Mathematically, the optical front-end implements a mapping from the $d_{\text{obj}}$-dimensional object space to a $d_\text{f}$-dimensional detection space, producing expected optical intensities $\vec{\lambda} \in \mathbb{R}_{\geq 0}^{d_\text{f}}$ at the $d_\text{f}$ detector elements.
    
    \item \textbf{Detection}: Conversion of optical signals to digital data. Sampling from the expected intensities $\vec{\lambda}$, detectors produce digital readouts for subsequent processing. When optical energy is sufficient, the detected values faithfully reflect $\vec{\lambda}$, and this stage introduces negligible distortion---the common situation in standard computer vision applications. However, when optical energy is limited, detection becomes non-deterministic due to photon statistics: outcomes occur in discrete units of photon energy, introducing uncertainty governed by the measurement statistics of the light source.
    
    \item \textbf{Digital back-end}: Computational processing that operates on the detected digital data to accomplish the sensing task. This may range from simple algorithms to sophisticated neural networks, implementing classification, regression, reconstruction, or other inference tasks.
\end{enumerate}

\subsection{The detection bottleneck}

When optical resources are abundant, detection functions as a reliable channel: the detected values accurately represent the expected intensities $\vec{\lambda}$, and information encoded in the optical signal passes through to the digital domain with minimal loss. Standard cameras operating under typical illumination exemplify this regime, where the detection stage poses no fundamental limitation.

However, when optical energy becomes severely limited (e.g. due to weak sources, short integration times, low reflectivity, or the need to minimize optical exposure), detection can become the dominant source of information loss in the sensing pipeline. In this regime, the stochastic and discrete nature of photon detection introduces substantial uncertainty: measurements deviate significantly from their expected values, and outcomes become increasingly unpredictable as expected photon counts per detection decrease. The detection process then acts as a \emph{lossy information channel} that discards a significant portion of the information encoded in $\vec{\lambda}$.

We call this phenomenon the \emph{detection bottleneck}. Its key characteristic is \emph{irreversibility}: information lost during detection cannot be recovered by any subsequent digital processing, regardless of its sophistication. This asymmetry has a critical implication for system design---when the detection bottleneck is the limiting factor, optimization efforts should focus primarily on the optical front-end, which is the only configurable component before the lossy detection stage.

\begin{figure}[htp]
\includegraphics [width=0.8\textwidth] {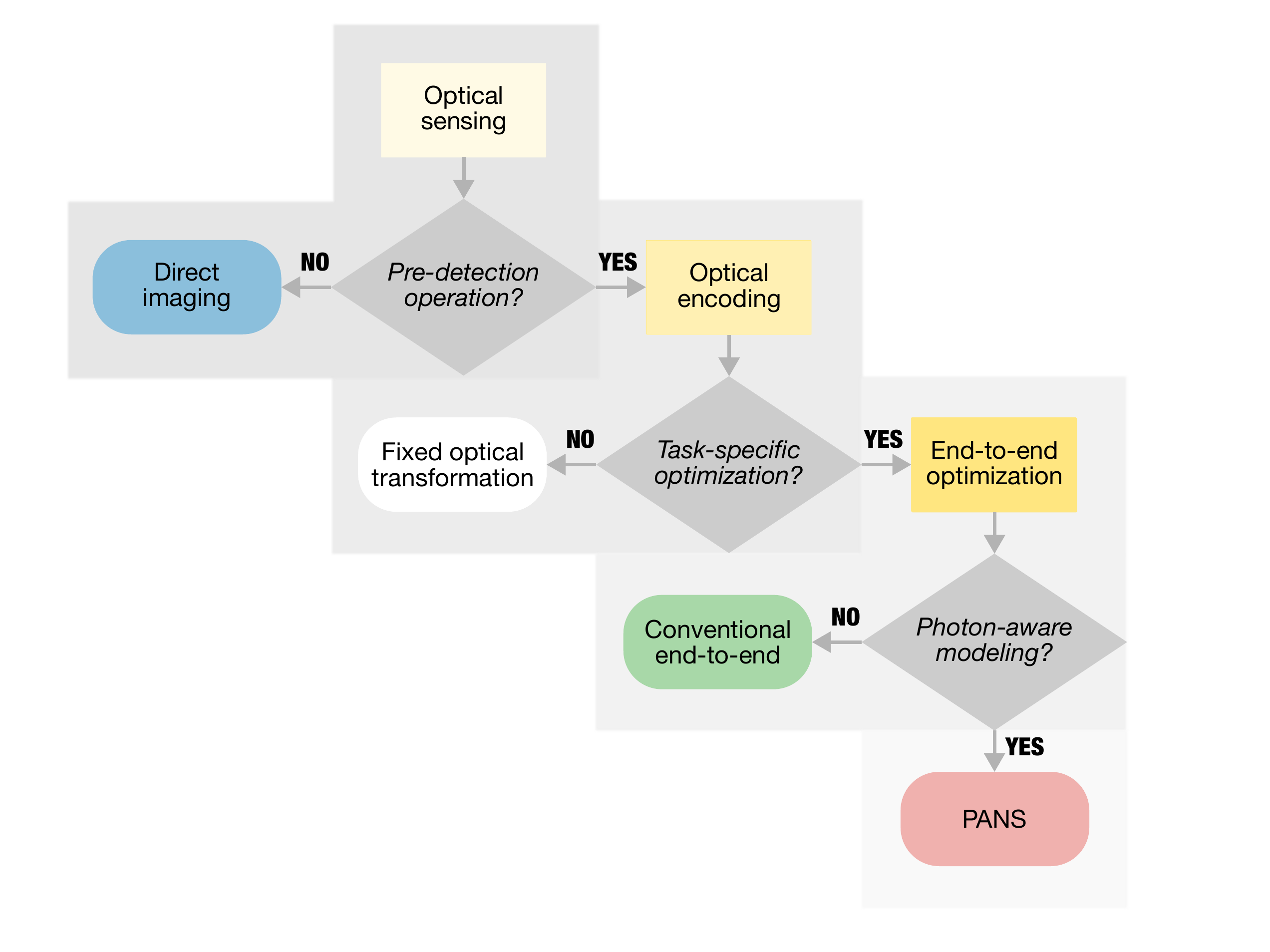}
\caption{\textbf{Taxonomy of optical sensing approaches.}
A flowchart illustrating the hierarchy of algorithmic approaches in optical sensing. Starting from the top, if no pre-detection operation is applied, the system performs direct imaging (blue). If an optical encoding operation is applied, the approach branches based on whether task-specific training is used: fixed transformations (white) use predetermined configurations such as downsampling or compressive sensing patterns, while end-to-end optimization learns task-specific optical parameters jointly with the digital back-end. End-to-end methods further divide based on whether photon-aware modeling is employed: conventional end-to-end methods (green) do not explicitly model stochastic photon detection, whereas photon-aware neuromorphic sensing (PANS, red) faithfully represents the probabilistic, discrete nature of the detection process during optimization. The color coding agrees with Fig.~2 in the main text. Shaded background regions group approaches by the level of sophistication in the optical front-end design.}
\label{fig:paradigms}
\end{figure}

\subsection{A taxonomy of sensing approaches}
\label{subsec:paradigm_taxonomy}

We organize sensing approaches according to three decisions regarding the optical front-end:

\subsubsection*{Decision 1: Pre-detection optical transformation?}

The first distinction is whether any optical transformation is applied before detection beyond simply imaging the object onto the detector array. If not, the system performs \textbf{direct imaging}: each detector element corresponds directly to a spatial location on the object, so that $d_\text{f} = d_{\text{obj}}$ and the mapping from $\vec{x}$ to $\vec{\lambda}$ is essentially an identity (up to scaling). If yes, an optical transformation maps the object from its native $d_{\text{obj}}$-dimensional space to a $d_\text{f}$-dimensional feature space before detection, where typically $d_\text{f} \neq d_{\text{obj}}$.

Examples of such optical transformations span a wide range: spatial binning or downsampling, coded apertures, structured illumination patterns used in compressive sensing, transmission through scattering media, and diffractive optical elements. The key point is that any pre-detection optical transformation fundamentally changes \emph{how information is presented to the detectors}.

\subsubsection*{Decision 2: Task-specific optimization?}

Given a pre-detection optical transformation, the second distinction is whether its parameters are optimized for the specific sensing task. \textbf{Fixed optical transformations} use predetermined configurations---downsampling grids, random projection matrices, or illumination patterns designed according to general principles (e.g., incoherence properties in compressive sensing). These are selected without reference to the particular task or object distribution.

In contrast, learned transformations optimize the optical parameters jointly with the digital back-end to maximize task performance. This \emph{end-to-end (E2E) optimization} discovers task-specific transformations that may be far more efficient than generic designs. For classification, the learned transformation projects objects onto directions that maximize class separability; for reconstruction, it preserves information most relevant to recovering the original signal.

\subsubsection*{Decision 3: Photon-aware forward modeling?}

Given learned optical transformations through E2E optimization, the third distinction concerns how the detection process is modeled during optimization. In many applications---particularly in computer vision where detection is reliable---\textbf{conventional end-to-end} methods work effectively. These typically model the forward pass as deterministic, incorporate additive noise, or employ robustness techniques such as quantization-aware training~\cite{wang2022optical,wright2022deep}. Such approximations are effective when detection introduces only minor perturbations to the optical signals.

However, when the detection bottleneck becomes significant---when detection is unreliable, highly non-deterministic, and produces discrete outcomes---these approximations break down. \textbf{Photon-aware} methods address this by faithfully modeling the actual stochastic photon-detection process during optimization. Rather than treating detection as deterministic-plus-noise, photon-aware methods represent detection as the inherently probabilistic process it physically is, enabling optimization against the true physical constraints at the bottleneck.

\subsection{Summary of sensing paradigms}
\label{subsec:paradigm_summary}

These three decisions yield four distinct stages of increasing sophistication:

\begin{enumerate}
    \item \textbf{Direct imaging}: No pre-detection optical transformation. Detectors capture an image of the object directly, with $\vec{\lambda}$ proportional to $\vec{x}$ (up to system response) and $d_\text{f} = d_{\text{obj}}$. The digital back-end processes these direct measurements.
    
    \item \textbf{Fixed optical transformation}: A predetermined operation (e.g., downsampling, pixel binning, fixed coded aperture, illumination patterns from compressive sensing) maps $\vec{x}$ to $\vec{\lambda}$. The transformation is not optimized for the task; only the digital back-end is trained.
    
    \item \textbf{Conventional (``non-photon-aware'', non-PA) E2E}: The optical transformation parameters are optimized jointly with the digital back-end through end-to-end training. The forward model does not explicitly represent the non-deterministic nature of photon detection---it treats detection as deterministic or subject to approximate noise models.
    
    \item \textbf{Photon-aware neuromorphic sensing (PANS)}: The optical transformation and digital back-end are jointly optimized with a forward model that faithfully represents the stochastic photon-detection process. Training explicitly accounts for the probabilistic, discrete nature of detection.
\end{enumerate}

\ref{tab:approach_comparison} summarizes these stages.

\begin{table}[h]
\centering
\caption{Taxonomy of sensing approaches based on optical front-end design choices.}
\label{tab:approach_comparison}
\begin{tabular}{l|ccc}
\hline
\textbf{Approach} & \textbf{  Optical operation} & \textbf{  Task-optimized} & \textbf{  Photon-aware} \\
\hline
Direct imaging & \texttimes & — & — \\
Fixed optical transformation & \checkmark & \texttimes & — \\
Conventional (non-PA) E2E & \checkmark & \checkmark & \texttimes \\
Photon-aware neuromorphic sensing (PANS) & \checkmark & \checkmark & \checkmark \\
\hline
\end{tabular}
\end{table}

When detection is reliable and introduces negligible information loss---as in standard computer vision and many optical sensing applications---direct imaging is often sufficient, assuming adequate detector availability. All approaches can be effective in this regime, with the choice depending on practical considerations such as hardware constraints and computational resources.

However, when optical energy at detection is severely limited, detection becomes unreliable and loses most information. This motivates the search for more efficient information encoding at the detection bottleneck. Fixed optical transformations may help by concentrating measurements into fewer dimensions, but cannot adapt to the task or to the detection statistics. Conventional E2E methods optimize the optical transformation for specific tasks, but at a few photons per detection, signals become highly non-deterministic and the learned parameters no longer apply.

Photon-aware methods address precisely this challenge. By modeling the stochastic detection process during training, PANS learns optical transformations that remain discriminative even when individual sampled measurements are highly uncertain. The system learns to allocate its limited photon budget across feature dimensions to maximize retention of task-relevant information through the lossy detection channel, with respect to the actual physical constraints imposed by photon statistics.

\newpage
\section{Active PANS with structured illumination}
\label{sec:active_pans}

\subsection{Optical encoding via structured illumination}
\label{subsec:active_encoding}

Active optical sensing involves designing illumination patterns that are projected onto an object to efficiently extract task-relevant information. This strategy has been widely used in computational optics, with applications in compressive sensing~\cite{duarte2008single}, LiDAR~\cite{gao2018object}, and biomedical imaging~\cite{jin2014light,tian2015computational}.

\subsubsection*{Physical setup}

Consider an object characterized by a spatially-varying transmission (or reflection) profile $\vec{x} \in [0, 1]^{d_{\text{obj}}}$, where $d_{\text{obj}}$ is the number of spatial modes (pixels) and each element $x_j$ represents the local transmission coefficient. A structured illumination pattern with intensity profile $\vec{w} \in \mathbb{R}_{\geq 0}^{d_{\text{obj}}}$ is projected onto the object. The transmitted (or reflected) light is collected by a bucket detector (single-photon counter), which integrates the total optical power without spatial resolution.

\subsubsection*{Mathematical formulation}

The expected photon count at the detector is proportional to the overlap between the illumination pattern and the object transmission:
\begin{equation}
\lambda = \vec{w} \cdot \vec{x} = \sum_{j=1}^{d_{\text{obj}}} w_j x_j.
\label{eq:active_lambda}
\end{equation}
Here $\lambda$ directly represents the expected number of photons reaching the detector, determined by the element-wise product of the illumination intensity and object transmission, summed over all spatial modes.

\begin{figure}[htpb]
    \centering
    \includegraphics[width=0.6\linewidth]{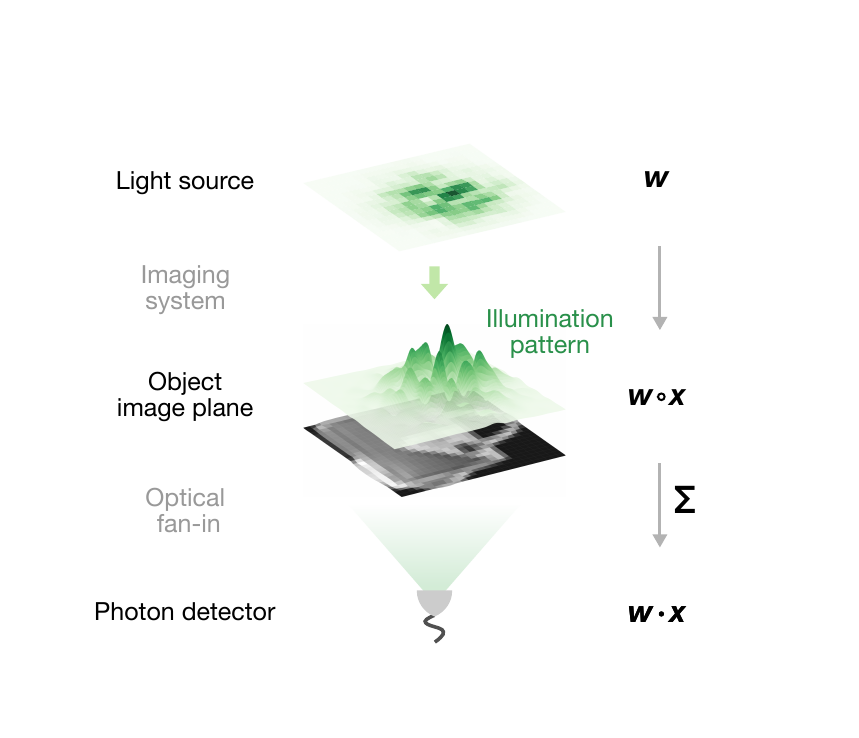}
    \caption{\textbf{Diagram of the optical front end operation in active sensing.} The light source provide a spatially distributed illumination pattern and project to the object, the object then attentuate the illuminatino light, equivalent to element-wise multiplication. then the optical energy was focused on to a detector to be measured. the output is equivalent to a dot product process.}
    \label{fig:vvm}
\end{figure}

By projecting $d_\text{f}$ different illumination patterns $\{\vec{w}_1, \vec{w}_2, \ldots, \vec{w}_{d_\text{f}}\}$ onto the object, we obtain a $d_\text{f}$-dimensional feature vector $\vec{\lambda} = (\lambda_1, \lambda_2, \ldots, \lambda_{d_\text{f}})$. Collectively, these patterns define a linear transformation:
\begin{equation}
\vec{\lambda} = W \vec{x}, \quad W \in \mathbb{R}_{\geq 0}^{d_\text{f} \times d_{\text{obj}}},
\end{equation}
where each row of $W$ corresponds to one illumination pattern. This linear transformation occurs \textit{before} the detection bottleneck, allowing the optical encoder to concentrate task-relevant information into a compact feature representation.

\subsubsection*{Compatibility with incoherent light}

This scheme operates on the modulation and redistribution of light intensity, without requiring phase stability or coherent interference. This makes active PANS fully compatible with spatially incoherent illumination sources, where the optical intensity at each spatial mode can be controlled independently. While coherent light sources would also work within this framework, the use of incoherent light offers significant practical advantages:
\begin{itemize}
    \item \textit{Robustness}: Incoherent systems are inherently immune to environmental perturbations such as vibrations, thermal fluctuations, and air turbulence that would degrade coherent optical setups.
    \item \textit{Simplicity}: The mathematical model (Eq.~\ref{eq:active_lambda}) directly describes the physical process without requiring interferometric alignment or phase stabilization.
    \item \textit{Accessibility}: Implementation is straightforward using widely available intensity-modulating devices.
\end{itemize}
By operating in the intensity domain, active PANS minimizes hardware requirements and maximizes compatibility with real-world deployment scenarios.

\subsection{Non-negativity constraint and squared parameterization}
\label{subsec:nonneg}

Because we encode information directly in light intensity, the values must be non-negative. This constraint must be respected during optimization to ensure that trained patterns are physically realizable.

We enforce non-negativity through a squared parameterization: the trainable parameters $\tilde{W} \in \mathbb{R}^{d_\text{f} \times d_{\text{obj}}}$ are unconstrained real values, and the physical illumination weights are computed as:
\begin{equation}
W = \tilde{W}^2 \quad \text{(element-wise squaring)},
\end{equation}
so that $w_{ij} = \tilde{w}_{ij}^2 \geq 0$ for all elements.

This approach offers several advantages over alternatives such as clamped value ranges~\cite{wang2023image,ma2025quantum} or projected gradient descent:
\begin{itemize}
    \item \textit{Smooth gradient flow}: Gradients flow smoothly through the squaring operation, avoiding the discontinuities and zero-gradient regions that arise with projection-based methods. Common methods like clamping $W \leftarrow \max(W, 0)$ \cite{wang2022optical,ma2025quantum} create a hard boundary where gradients vanish for negative values, stalling optimization.
    \item \textit{Automatic gradient scaling}: The gradient $\frac{\partial w_{ij}}{\partial \tilde{w}_{ij}} = 2\tilde{w}_{ij}$ provides natural scaling---gradients for small weights are automatically suppressed, stabilizing training and preventing oscillations near zero.
    \item \textit{Sparse solutions}: The squared parameterization naturally encourages sparsity, as small parameter values $|\tilde{w}_{ij}| \ll 1$ are mapped to very small (nearly zero) physical weights $w_{ij} \ll 1$. This promotes efficient photon allocation by concentrating illumination on informative spatial modes.
    \item \textit{No projection step}: Unlike projected gradient descent, which requires an explicit projection onto the non-negative orthant after each update, the squared parameterization inherently satisfies the constraint without additional computational overhead.
\end{itemize}

The gradient with respect to the trainable parameters follows from the chain rule:
\begin{equation}
\frac{\partial \mathcal{L}}{\partial \tilde{w}_{ij}} = \frac{\partial \mathcal{L}}{\partial w_{ij}} \cdot \frac{\partial w_{ij}}{\partial \tilde{w}_{ij}} = \frac{\partial \mathcal{L}}{\partial w_{ij}} \cdot 2\tilde{w}_{ij}.
\end{equation}

\subsection{Photon budget control}
\label{subsec:active_budget}

A key advantage of photon-aware modeling is that all parameters maintain precise physical meanings. The total number of illumination photons across all patterns is:
\begin{equation}
N_{\text{illu}} = \sum_{i,j} w_{ij}.
\end{equation}

We incorporate photon budget control through L2 regularization on the parameterization $\tilde{W}$:
\begin{equation}
\mathcal{L}_{\text{total}} = \mathcal{L}_{\text{task}}(f_\theta(\vec{a}), y) + \alpha \cdot \|\tilde{W}\|_2^2,
\label{eq:loss_active}
\end{equation}
where $\mathcal{L}_{\text{task}}$ is the task-specific loss (e.g., cross-entropy for classification), $f_\theta$ is the digital decoder, and $\alpha \geq 0$ controls the trade-off between task performance and photon budget.

The key observation is that the L2 norm of the parameterization equals the total illumination:
\begin{equation}
\|\tilde{W}\|_2^2 = \sum_{i,j} \tilde{w}_{ij}^2 = \sum_{i,j} w_{ij} = N_{\text{illu}}.
\end{equation}
Thus, the regularization term directly penalizes the total illumination energy. By varying $\alpha$, we train models optimized for different photon budget regimes (Fig.~2E). Without photon-aware modeling, such direct optimization of the physical photon budget would not be possible.

\subsection{Training procedure}
\label{subsec:active_training}

Training follows the general framework described in \ref{sec:pans_framework}, with the encoding $\vec{\lambda} = \tilde{W}^2 \vec{x}$ and regularization term $\alpha \|\tilde{W}\|_2^2$.

For initialization, we use Kaiming uniform initialization on $\tilde{W}$ with scaling factor $a = 1.2$ (\ref{subsec:training_procedure}). This produces moderate initial $\lambda$ values after the squaring operation, ensuring that training begins in a regime where gradients are well-behaved.

The complete training algorithm is summarized in Algorithm~\ref{alg:active_training}.

\begin{algorithm}[H]
\caption{Training of active PANS}
\label{alg:active_training}
\begin{algorithmic}[1]
\Require Training data $\{(\vec{x}^{(n)}, y^{(n)})\}$, regularization coefficient $\alpha$, initial slope $\eta_0$, annealing factor $\theta_{\text{anneal}}$, intensity clamp $\lambda_{\text{max}}$
\Ensure Trained parameters $\tilde{W}^*$, $\theta^*$
\State Initialize $\tilde{W}$ and $\theta$
\State $\eta \gets \eta_0$
\For{epoch $= 1$ to num\_epochs}
    \For{each minibatch $\{(\vec{x}, y)\}$}
        \State \underline{\textbf{Forward pass}}
        \State $W \gets \tilde{W}^2$ \Comment{Squared parameterization for non-negativity}
        \State $\vec{\lambda} \gets W \vec{x}$ \Comment{Expected detected optical energy}
        \State $\vec{\lambda} \gets \min(\vec{\lambda}, \lambda_{\text{max}})$ \Comment{Maximum intensity clamping}
        \State $\vec{p} \gets 1 - \exp(-\eta \cdot \vec{\lambda})$ \Comment{Detection probabilities from photon statistics}
        \State $\vec{a} \sim \text{Bernoulli}(\vec{p})$ \Comment{Stochastic SPD activation}
        \State $\vec{a} \gets \text{BatchNorm}(\vec{a})$ \Comment{Batch normalization}
        \State $\hat{y} \gets f_\theta(\vec{a})$ \Comment{Digital back-end output}
        \State $\mathcal{L} \gets \mathcal{L}_{\text{task}}(\hat{y}, y) + \alpha \|\tilde{W}\|_2^2$ \Comment{Total loss}
        \State
        \State \underline{\textbf{Backward pass}}
        \State Compute $g_a = \partial \mathcal{L}_{\text{task}} / \partial \vec{a}$ and $g_\theta = \partial \mathcal{L}_{\text{task}} / \partial \theta$ via backprop through $f_\theta$
        \State $g_p \gets g_a$ \Comment{``Straight-through'' here (STE)}
        \State $g_\lambda \gets \eta \cdot \exp(-\eta \vec{\lambda}) \odot g_p$ \Comment{$\partial p / \partial \lambda = \eta e^{-\eta\lambda}$}
        \State $g_W \gets g_\lambda \vec{x}^\top$ \Comment{$\partial \lambda / \partial W$}
        \State $g_{\tilde{W}} \gets 2\tilde{W} \odot g_W + 2\alpha \tilde{W}$ \Comment{$\partial W / \partial \tilde{W}$ + regularization}
        \State
        \State \underline{\textbf{Parameter update}}
        \State $\tilde{W} \gets \text{Update}(\tilde{W}, g_{\tilde{W}})$ \Comment{Update optical front end}
        \State $\theta \gets \text{Update}(\theta, g_\theta)$ \Comment{Update digital back end}
    \EndFor
    \State $\eta \gets \theta_{\text{anneal}} \cdot \eta$ \Comment{Slope annealing}
\EndFor
\end{algorithmic}
\end{algorithm}

\newpage
\section{Passive PANS with linear optical processors}
\label{sec:passive_pans}

Active PANS (\ref{sec:active_pans}) controls structured illumination to encode object information before detection. However, many sensing scenarios involve optical signals from sources that cannot be controlled---astronomical observations, surveillance, fiber-optic sensing, scattered light through turbid media, and similar applications. Passive PANS addresses these scenarios by applying learnable transformations to the \textit{incoming} optical field before detection, rather than shaping the outgoing illumination.

From an algorithmic perspective, both active and passive PANS instantiate the same photon-aware framework: a parameterized optical encoder produces expected photon counts $\vec{\lambda}$, which pass through stochastic single-photon detection and digital post-processing. The distinction lies entirely in the physical deployment---whether the encoder acts on light we generate (active) or light we receive (passive). This flexibility makes passive PANS applicable wherever the signal originates from the environment or system under observation rather than from controlled illumination.

Optical linear processors have driven decades of advancement in optical neural networks~\cite{wetzstein2020inference}, with established implementations in both free-space~\cite{lin2018all,gigan2022imaging,spall2020fully,bernstein2023single,wang2024large} and on-chip~\cite{reck1994experimental,shen2017deep,bandyopadhyay2024single} platforms. In this section, we demonstrate through simulations that PANS extends naturally to passive optical encoders that apply learnable linear transformations to incoming optical fields.

\subsection{Optical encoding via field transformation}
\label{subsec:passive_coherent}

We focus on linear optical processors with real-valued transmission matrices as our primary demonstration of passive PANS. This choice reflects the prevalence of such transformations across optical computing architectures---from spatial light modulation to integrated photonic meshes~\cite{shen2017deep,bernstein2023single}---and their experimental maturity developed over decades of research in optical matrix-vector multiplication. 

The transmission matrix $W$ of a general linear optical processor can have complex-valued elements. For our demonstrations, we restrict to real-valued $W \in \mathbb{R}^{d_\text{f} \times d_{\text{in}}}$, primarily because real-valued matrix-vector multiplication is already well-established in optical hardware---multiple platforms have demonstrated programmable real-valued linear transformations in both free-space~\cite{spall2020fully,bernstein2023single} and integrated photonic~\cite{shen2017deep,wang2024large} implementations. Since our goal is to demonstrate that photon-aware modeling provides advantages for photon-limited sensing rather than to optimize the encoder architecture itself, real-valued linear encoders provide a concrete, experimentally accessible choice for this purpose.

\subsubsection*{Mathematical formulation}

The optical encoder is a linear optical processor characterized by a transmission matrix $W \in \mathbb{R}^{d_\text{f} \times d_{\text{in}}}$. For an input field with amplitude vector $\vec{E}$, the transformed field is:
\begin{equation}
\vec{E}' = W \vec{E}.
\end{equation}
The detected intensity at each output port $i$ is the squared magnitude of the corresponding field amplitude:
\begin{equation}
\lambda_i = |E'_i|^2 = \left| \sum_j w_{ij} E_j \right|^2.
\label{eq:passive_coherent_lambda}
\end{equation}

In vector notation:
\begin{equation}
\vec{\lambda} = (W \vec{E})^2 \quad \text{(element-wise squaring)}.
\end{equation}

This intensity-detection model differs from active PANS in several ways. First, the mapping from weights to expected photon counts is \textit{quadratic}, creating a more complex optimization landscape. Second, interference between input modes contributes to each output, enabling the system to exploit phase relationships in the input field. Third, total detected power is bounded by input power—photons can only be redistributed among output ports, not created.

Because many optical platforms support both positive and negative real-valued weights through phase control, no squared parameterization is needed---the weights $W$ are trained directly as unconstrained real values.

\subsubsection*{Gradient computation}

The squared relationship between field amplitude and intensity introduces an additional factor in the gradient. Defining the intermediate variable $\vec{z} = W\vec{E}$ so that $\vec{\lambda} = \vec{z}^2$, the gradient for the $i$-th detector is:
\begin{equation}
\frac{\partial \lambda_i}{\partial w_{ij}} = 2 z_i \cdot E_j = 2 \left(\sum_k w_{ik} E_k\right) E_j.
\end{equation}

In the backward pass, we first compute the gradient with respect to the intermediate field amplitudes, $g_z = 2\vec{z} \odot g_\lambda$, where $g_\lambda = \partial \mathcal{L} / \partial \vec{\lambda}$ is backpropagated through the detection and digital layers. The gradient then propagates to the weights via $g_W = g_z \vec{E}^\top$, or equivalently in matrix form:
\begin{equation}
\frac{\partial \mathcal{L}}{\partial W} = 2 \, \text{diag}(\vec{z}) \cdot g_\lambda \, \vec{E}^\top.
\end{equation}

\subsection{Intensity-based encoding}
\label{subsec:passive_intensity}

As another example, we explore passive intensity-based linear transformations~\cite{rahman2023universal} that are compatible with spatially incoherent light sources. This encoding can be understood as a reallocation of the input intensity distribution across the detector array. For incoherent light characterized by intensity distribution $\vec{I} \in \mathbb{R}_{\geq 0}^{d_{\text{in}}}$, intensities add directly without interference:
\begin{equation}
\lambda_i = \sum_j w_{ij} I_j, \quad w_{ij} \geq 0.
\label{eq:passive_intensity_lambda}
\end{equation}

In matrix form:
\begin{equation}
\vec{\lambda} = W \vec{I}, \quad W \in \mathbb{R}_{\geq 0}^{d_\text{f} \times d_{\text{in}}}.
\end{equation}

This linear mapping $\vec{\lambda} = W\vec{I}$ is mathematically identical to the active PANS encoding, but the intensity originates from an external source rather than controlled illumination. Since intensities cannot interfere destructively, the transformation weights must be non-negative, enforced through squared parameterization $W = \tilde{W}^2$, similar to \ref{subsec:nonneg}.

\subsection{Training procedure}
\label{subsec:passive_training}

Training follows the general framework described in \ref{sec:pans_framework}. For passive encoders processing external light sources, there is no illumination budget to regularize, so the loss function is simply:
\begin{equation}
\mathcal{L}_{\text{total}} = \mathcal{L}_{\text{task}}(f_\theta(\vec{a}), y).
\end{equation}

The optical encoder learns to optimally redistribute the available photons across feature dimensions to maximize task performance. 
The training algorithm for coherent passive PANS is summarized in Algorithm~\ref{alg:passive_training_coh}.

\begin{algorithm}[H]
\caption{Training of passive PANS (coherent field encoding)}
\label{alg:passive_training_coh}
\begin{algorithmic}[1]
\Require Training data $\{(\vec{E}^{(n)}, y^{(n)})\}$, initial slope $\eta_0$, annealing factor $\theta_{\text{anneal}}$, intensity clamp $\lambda_{\text{max}}$
\Ensure Trained parameters $W^*$, $\theta^*$
\State Initialize $W$ and $\theta$
\State $\eta \gets \eta_0$
\For{epoch $= 1$ to num\_epochs}
    \For{each minibatch $\{(\vec{E}, y)\}$}
        \State \underline{\textbf{Forward pass}}
        \State $\vec{z} \gets W \vec{E}$ \Comment{Linear transformation of field amplitudes}
        \State $\vec{\lambda} \gets |\vec{z}|^2$ \Comment{Intensity from squared amplitude}
        \State $\vec{\lambda} \gets \min(\vec{\lambda}, \lambda_{\text{max}})$ \Comment{Maximum intensity clamping}
        \State $\vec{p} \gets 1 - \exp(-\eta \cdot \vec{\lambda})$ \Comment{Detection probabilities from photon statistics}
        \State $\vec{a} \sim \text{Bernoulli}(\vec{p})$ \Comment{Stochastic SPD activation}
        \State $\vec{a} \gets \text{BatchNorm}(\vec{a})$ \Comment{Normalize activations}
        \State $\hat{y} \gets f_\theta(\vec{a})$ \Comment{Digital back-end output}
        \State $\mathcal{L} \gets \mathcal{L}_{\text{task}}(\hat{y}, y)$ \Comment{Task loss}
        \State
        \State \underline{\textbf{Backward pass}}
        \State Compute $g_a = \partial \mathcal{L} / \partial \vec{a}$ and $g_\theta = \partial \mathcal{L} / \partial \theta$ via backprop through $f_\theta$
        \State $g_p \gets g_a$ \Comment{``Straight-through'' here (STE)}
        \State $g_\lambda \gets \eta \cdot \exp(-\eta \vec{\lambda}) \odot g_p$ \Comment{$\partial p / \partial \lambda = \eta e^{-\eta\lambda}$}
        \State $g_z \gets 2 \vec{z} \odot g_\lambda$ \Comment{$\partial \lambda / \partial z = 2z$}
        \State $g_W \gets g_z \vec{E}^\top$ \Comment{$\partial z / \partial W$}
        \State
        \State \underline{\textbf{Parameter update}}
        \State $W \gets \text{Update}(W, g_W)$ \Comment{Update optical front end}
        \State $\theta \gets \text{Update}(\theta, g_\theta)$ \Comment{Update digital back end}
    \EndFor
    \State $\eta \gets \theta_{\text{anneal}} \cdot \eta$ \Comment{Slope annealing}
\EndFor
\end{algorithmic}
\end{algorithm}

\newpage
\section{Photon budget scaling}
\label{sec:photon_budget}

Evaluating and comparing sensing performance under photon-limited conditions requires systematically varying the photon budget. This section introduces the relevant quantities and describes the principal methods for scaling photon budget, clarifying when each method applies and the considerations specific to PANS.

\subsection{Detected photons $N_{\text{det}}$}

The photon budget for sensing is characterized by $N_{\text{det}}$, the total number of detected photons per inference. Given an object $\vec{x}$, recall from \ref{sec:pans_framework} that the optical front-end maps it to expected photon numbers $\vec{\lambda} \in \mathbb{R}_{\geq 0}^{d_\text{f}}$ at the $d_\text{f}$ detectors. The total expected detected photons is simply the sum:
\begin{equation}
    \mathbb{E}[N_{\text{det}}] = \sum_{i=1}^{d_\text{f}} \lambda_i.
    \label{eq:Ndet_expectation}
\end{equation}
The actual detected photon count $N_{\text{det}}$ is a random variable sampled from this expectation according to the photon statistics at each detector.

\subsection{Scaling average photon level}
\label{subsec:eta_scaling}

The most straightforward way to vary $N_{\text{det}}$ is to scale the average photon level at detection. The detected optical energy per measurement is determined by both the optical power reaching the detector and the integration time. Denoting the combined effect as a global scaling factor $\eta > 0$ applied to the object signal, we have $\vec{x} \to \eta \cdot \vec{x}$, and consequently $\vec{\lambda} \to \eta \cdot \vec{\lambda}$ and $\mathbb{E}[N_{\text{det}}] \to \eta \cdot \mathbb{E}[N_{\text{det}}]$.

In practice, this scaling can be achieved by attenuating or amplifying the optical signal (e.g., using neutral density filters, adjusting the light source power, or modifying transmission through the system), or equivalently by adjusting the detector integration time. Both approaches allow \textit{continuous} scaling of the photon budget to any desired level.

This scaling does not affect the fundamental ability to extract information from the signal. Consider everyday experience: we recognize a familiar face whether seen under bright midday sun or in the soft light of dusk, and putting on sunglasses uniformly attenuates all incoming light yet we perceive and recognize the same scene. Our perception depends on the relative contrasts and spatial patterns, not on the absolute brightness. The same principle applies broadly to sensing and machine learning: pattern recognition tasks depend on \emph{relative} relationships in the input, not absolute values. Uniformly scaling $\eta$ preserves these relative relationships, so information extraction is unaffected.

This is why $\eta$ scaling is the standard method for evaluating sensing performance across different photon regimes: it provides continuous control over photon budget while preserving the information content of the signal. This applies universally to direct imaging and optical encoding methods where the detected signal is treated as a continuous quantity.

However, scaling $\eta$ is \textit{not} well-suited for PANS. PANS fundamentally differs because it explicitly models the stochastic photon detection process during training. In this framework, each parameter acquires a precise physical meaning: the elements of the optical transformation directly determine expected photon numbers $\vec{\lambda}$, which govern the detection statistics. Once the model is optimized, the trained parameters are optimal with respect to a specific value of $\eta$ (i.e., specific optical power and integration time settings). Tuning $\eta$ after training---whether increasing or decreasing it---pushes the system away from its optimized operating point.

This distinction highlights a key aspect of PANS: the photon-aware modeling ensures that trained parameters have definite physical interpretations, but this also means the system needs to be calibrated to meet these specific operating conditions in experimental implementation (\ref{sec:implementation}). Fortunately, our trained models exhibit robustness to $\eta$ within a reasonable range (\ref{fig:eta_sweep}), so the calibration requirement is not overly stringent and the system remains stable during deployment, including robustness to typical intensity fluctuations.

\subsection{$K$-shot scaling}
\label{subsec:K_scaling}

$K$-shot scaling provides an alternative way to vary photon budget by taking $K$ repeated measurements per detector. Recall from \ref{sec:pans_framework} that PANS uses single-photon detection (SPD) measurements, where each measurement produces a binary outcome $a_i^{(k)} \in \{0, 1\}$. With $d_\text{f}$ detectors and $K$ shots each, the total detected photons is:
\begin{equation}
    N_{\text{det}} = \sum_{i=1}^{d_\text{f}} \sum_{k=1}^{K} a_i^{(k)}.
\end{equation}
This provides discrete scaling of the photon budget.

$K$-shot scaling is essentially equivalent to $\eta$ scaling via integration time, but in discrete increments. This discretization is not merely a theoretical construct---it reflects how real photon-counting detectors operate. Photon detection devices general have fixed integration windows determined by their recovery time or readout clock, and they output binary or discrete count signals. Even detectors that appear to provide ``continuous'' exposure operate with discrete clock cycles at the hardware level. The notion of truly continuous integration time is an idealization; in practice, photon detection always involves some fundamental time discretization. Thus, $K$-shot scaling---accumulating $K$ discrete measurements---is a natural and realistic way to increase photon budget.

For conventional sensing approaches that treat detection as continuous, $K$-shot scaling is equivalent to $\eta$ scaling via integration time: taking $K$ repeated measurements is analogous to integrating for $K$ times longer. The two approaches yield the same expected $N_{\text{det}}$ and, in principle, the same information content (neglecting practical considerations such as per-readout noise).

For PANS, $K$-shot scaling is a natural extension once a system is trained and deployed. The training process optimizes the optical front-end for a specific configuration, including the average photon level per detection (corresponding to fixed input power, transmission, and integration time settings). Given this fixed configuration, a straightforward way to collect more information is simply to perform additional measurements under identical conditions. Once experimental parameters are set, the detector continuously operates and can accumulate data over multiple shots. 

At first glance, $K$-shot scaling appears to be an appropriate way to vary photon budget for PANS while respecting the trained configuration. However, this is \textit{not} how we report results in the main text, because $K$-shot scaling does not fully demonstrate the algorithmic capability of PANS.

The key observation is this: with $K$-shot scaling, the total number of detection events is $d_\text{f} \times K$, but during training we only optimized the mapping for $d_\text{f}$ individual features and simply repeat sampling from the same expected feature values. In principle, if we could optimize the mapping for every single feature we detect---meaning all $d_\text{f} \times K$ features independently, each designed to extract complementary information about the object---the system would have greater capacity to retain information through the detection bottleneck for a given total number of detections (or equivalently, a given photon budget).

To illustrate, consider $d_\text{f} = 10$ with $K = 2$ versus $d_\text{f} = 20$ with $K = 1$. Both configurations yield 20 total detection events with similar expected $N_{\text{det}}$. However, the $d_\text{f} = 20$ configuration has 20 independently optimized features, while the $d_\text{f} = 10$ configuration optimizes only 10 features that are each sampled twice. The former should retain at least as much---and typically more---task-relevant information because every feature can be tailored to extract unique information about the object. Simulation results in \ref{subsec:testing_K} confirm this intuition.

That said, $K$-shot scaling remains practically valuable for deployment. Once a model is trained and the experimental setup is configured, it is often easier to take samples continuously than to reconfigure the system. Real-world constraints such as speed requirements, detector availability, or optical system complexity may make $K > 1$ the pragmatic choice. The point is that for \textit{reporting} the fundamental scaling behavior of PANS---demonstrating what the algorithm can achieve given certain physical constraints---we should use $d_\text{f}$ scaling with $K = 1$ to show the true expressive capacity of the end-to-end optimization.

\subsection{Feature dimension ($d_\text{f}$) scaling}
\label{subsec:df_scaling}

Given the above considerations, $d_\text{f}$ scaling emerges as the principled approach for characterizing PANS performance: we vary $d_\text{f}$ while fixing $K = 1$, training each configuration independently. This ensures that for any given photon budget, all available detection events correspond to independently optimized features, fully exploiting the optimization's capacity to maximize information retention through the detection bottleneck for the given $N_{\text{det}}$.

More generally, $d_\text{f}$ scaling applies to any end-to-end optimized approach where the number of output dimensions can be varied. Each $d_\text{f}$ configuration is trained separately, allowing the optimization to find the best solution given that specific constraint. Performance is then evaluated at the corresponding photon budget, providing a fair characterization of what each method can achieve.

\subsection{Summary of scaling methods}

\ref{tab:photon_scaling} summarizes the applicability of each scaling method across different sensing approaches.

\begin{table}[h]
\centering
\renewcommand{\arraystretch}{1.3}
\setlength{\tabcolsep}{12pt}
\caption{Photon budget scaling methods and their applicability across sensing approaches.}
\label{tab:photon_scaling}
\begin{tabular}{l|ccc}
\hline
\textbf{Approach} & \textbf{Photon level ($\eta$)} & \textbf{$K$-shot} & \textbf{$d_\text{f}$} \\
\hline
Direct imaging & Applicable & $\equiv \eta$ & N/A \\
Conventional optical encoding & Applicable & $\equiv \eta$ & Applicable \\
PANS & Limited & Practical & Optimized \\
\hline
\end{tabular}
\end{table}

Scaling the average photon level ($\eta$) is applicable to any approach where information extraction depends only on relative signal values, not absolute values. This includes direct imaging and conventional optical encoding methods, where the detected signal is treated as a continuous quantity. $K$-shot scaling is equivalent to $\eta$ scaling for these approaches (denoted ``$\equiv \eta$'' in the table), as repeated measurements are analogous to extended integration time.

For PANS, scaling $\eta$ has limited applicability because the trained parameters are optimized for specific absolute photon levels, as discussed in \ref{subsec:eta_scaling}. $K$-shot scaling is practical for deployment but suboptimal for characterizing algorithmic capability, since it does not allow independent optimization of every single feature. $d_\text{f}$ scaling with $K = 1$ is the optimized approach that fully demonstrates what PANS can achieve at each photon budget level; this is the method used throughout the main text.

For conventional optical encoding, $d_\text{f}$ scaling is applicable and provides a way to systematically evaluate performance, though full optimization at each $d_\text{f}$ is needed to observe meaningful scaling behavior.

\subsection{Reporting $N_{\text{det}}$ and performance variability}
\label{subsec:Ndet_variability}

When reporting results such as test accuracy for classification tasks, $N_{\text{det}}$ exhibits variability from two sources that must be properly accounted for.

\subsubsection*{Object-dependent variability.}
Different test objects $\vec{x}$ produce different expected photon numbers $\vec{\lambda}$, and hence different $\mathbb{E}[N_{\text{det}}]$ via Eq.~\eqref{eq:Ndet_expectation}. When reporting test accuracy with respect to a particular $N_{\text{det}}$ value, we average the detected photon counts over the test set to obtain a representative value. This averaging over test samples is standard practice in machine learning evaluation.

\subsubsection*{Shot-to-shot variability.}
Even for a fixed object, $N_{\text{det}}$ fluctuates across repeated measurements due to the stochastic nature of photon detection. This source of variability is less common in conventional machine learning, where models are typically deterministic or nearly deterministic---given the same input, the same or very similar output is produced. In photon-starved sensing, however, the intrinsic stochasticity of detection means that repeated inferences on the same object yield very different feature vectors and potentially different outcomes; we must consider the probability distribution in the feature space rather than fixed points.

In high signal-to-noise regimes, this variability is relatively small (the coefficient of variation scales as $1/\sqrt{N_{\text{det}}}$) and has minimal impact on performance. However, in the few-photon regime central to PANS, shot-to-shot variability becomes pronounced. For example, with Poisson statistics, detecting on average 1 photon per measurement ($\lambda = 1$) gives a ``signal-to-noise ratio'' of 1, meaning the standard deviation of the signal equals its mean value.

A key contribution of our work is demonstrating that despite this significant stochasticity, PANS achieves reliable high-accuracy sensing. To faithfully characterize this reliability, we report test accuracy as mean $\pm$ standard deviation over multiple repeated inferences.

\subsubsection*{Distinction between $K$ and repeated trials.}
It is important to distinguish between $K$-shot measurements and repeated trials for evaluating variability. When using $K > 1$ shots, all $K$ measurements contribute to a \textit{single} inference: the total $N_{\text{det}}$ for that inference is the sum over all $K \times d_\text{f}$ detection events, and the system produces one output (e.g., one classification decision) based on this combined information. In contrast, when we report mean $\pm$ standard deviation over multiple trials, we perform multiple independent \textit{single-shot} inferences, each with $K = 1$, to characterize the probability distribution of single-shot outcomes. Although the data collection process may appear similar---repeatedly acquiring measurements with fixed experimental setup---the conceptual distinction is crucial: $K > 1$ increases the photon budget per inference, while repeated trials characterize the statistical distribution of outcomes at a fixed photon budget ($K = 1$).

For example, reporting ``30 trials per test image'' means we perform 30 independent single-shot ($K = 1$) inferences on each test image, computing the mean and standard deviation of the resulting accuracy to characterize reliability. This is fundamentally different from using $K = 30$, which would constitute a single inference with 30 times more detected photons. Detailed evaluation protocols for specific tasks are described in subsequent sections.

\newpage
\part*{Part II: \\Numerical simulation details}
\addcontentsline{toc}{part}{II\quad Numerical simulation details}
\label{part:numerical}
\vspace{24pt}

This part provides detailed numerical simulation results and implementation specifics for the PANS framework introduced in Part~\ref{part:framework}. We begin with training considerations using active PANS on FashionMNIST as a case study, then present testing procedures and results for additional tasks.

\vspace{24pt}

\section{Training PANS models}
\label{sec:sim_training}

This section walks through the training process for photon-aware neuromorphic sensing (PANS) models using active PANS with FashionMNIST 10-class object classification as a representative example. The considerations discussed here---model architecture, hyperparameter selection, and photon budget control---apply broadly to other PANS configurations and tasks.

FashionMNIST serves as a suitable benchmark: it is widely used in the machine learning community, its classes (T-shirt, trouser, pullover, dress, coat, sandal, shirt, sneaker, bag, ankle boot) represent recognizable object shapes, and its grayscale 28$\times$28 images better approximate general object morphology than simple digit datasets. While relatively simple compared to real-world sensing tasks, it suffices to establish methodology and demonstrate the framework's capabilities.

\subsection{Model architecture}
\label{subsec:sim_model}

The active PANS model consists of three stages following the framework in \ref{sec:active_pans}: an optical front-end that encodes object information into expected photon counts, a stochastic photon detection layer, and a digital back-end that processes the detected features for classification.

The key components are:
\begin{itemize}
    \item \texttt{input\_weights}: trainable parameters $\tilde{W}$, squared in the forward pass to enforce non-negativity (\ref{subsec:nonneg}).
    \item \texttt{PhotonActivation}: stochastic SPD process, serving as neural activation with straight-through gradient estimation (\ref{subsec:pans_gradient}).
    \item \texttt{feature\_bn}: batch normalization of the feature vectors (dim.\ $d_\text{f}$) after detection.
    \item \texttt{conv\_layers}, \texttt{fc\_layers}: configurable digital back-end neural network.
\end{itemize}

Listing~\ref{lst:pans_active} shows the PyTorch implementation.

\vspace{200pt}

\begin{lstlisting}[language=Python, caption={PyTorch implementation of the active PANS model.}, label=lst:pans_active, basicstyle=\normalsize\ttfamily, lineskip=4pt]
import torch
import torch.nn as nn
import torch.nn.functional as F

class PANSActive(nn.Module):
    """Active PANS with structured illumination encoding.
    
    Args:
        n_input: Input dimension (d_obj in paper, e.g., 784 for 28x28)
        n_output: Number of output classes
        n_feature: Number of features/illumination patterns (d_\text{f} in paper)
        in_channels: Input channels for conv layers (default 1)
        conv_channels: Channel config for conv blocks ([] to disable)
        conv_kernel: Kernel size for conv layers
        pool_size: Pooling window size
        fc_units: Hidden units in fully connected layers
    """
    
    def __init__(self, n_input, n_output, n_feature, in_channels=1,
                 conv_channels=[[64,64], [128,128], [256,256,256]],
                 conv_kernel=15, pool_size=2, fc_units=[512]):
        super().__init__()
        
        # Optical front-end: trainable illumination patterns
        self.input_weights = nn.Parameter(torch.empty(n_feature, n_input))
        nn.init.kaiming_uniform_(self.input_weights, a=1.2)
        
        # Photon detection layer (defined in Section 1)
        self.photon_act = PhotonActivation()
        
        # Digital back-end: Batch normalization
        self.feature_bn = nn.BatchNorm1d(n_feature)
        
        # Digital back-end: Conv blocks
        conv_modules = []
        conv_output_size = n_feature
        for block in conv_channels:
            for out_ch in block:
                conv_modules.extend([
                    nn.Conv1d(in_channels, out_ch, conv_kernel,
                              padding=conv_kernel // 2),
                    nn.BatchNorm1d(out_ch), nn.ReLU()])
                in_channels = out_ch
            conv_modules.append(nn.AvgPool1d(pool_size))
            conv_output_size //= pool_size
        self.conv_layers = nn.Sequential(*conv_modules)
        
        # Digital back-end: Fully connected layers
        if len(conv_channels) > 0:
            fc_input_size = conv_output_size * conv_channels[-1][-1]
        else:
            fc_input_size = n_feature
        fc_modules = []
        for units in fc_units:
            fc_modules.extend([nn.Linear(fc_input_size, units), nn.ReLU()])
            fc_input_size = units
        fc_modules.append(nn.Linear(fc_input_size, n_output))
        self.fc_layers = nn.Sequential(*fc_modules)

    def forward(self, x, n_rep=1, slope=1.0):
        x = x.view(x.size(0), -1)
        # Optical encoding: squared weights ensure non-negativity
        x = F.linear(x, self.input_weights ** 2)
        # Stochastic photon detection
        x = self.photon_act(x, n_rep=n_rep, slope=slope)
        # Digital processing
        x = self.feature_bn(x)
        x = self.conv_layers(x.unsqueeze(1))
        x = self.fc_layers(x.view(x.size(0), -1))
        return F.log_softmax(x, dim=1)
        
\end{lstlisting}

\subsubsection*{Optical front end}

The optical front-end is parameterized by \texttt{input\_weights} $\tilde{W} \in \mathbb{R}^{d_\text{f} \times d_{\text{obj}}}$. As described in \ref{subsec:nonneg}, the physical illumination weights are $W = \tilde{W}^2$ (element-wise), ensuring non-negativity. Kaiming initialization with $a = 1.2$ produces moderate initial $\lambda_i$ values (\ref{subsec:active_training}).

\subsubsection*{Photon detection layer}

The \texttt{PhotonActivation} module implements the stochastic SPD process (\ref{subsec:pans_spd}) with straight-through gradient estimation (\ref{subsec:pans_gradient}). The \texttt{n\_rep} parameter controls $K$-shot averaging: \texttt{n\_rep=1} for single-shot mode. \texttt{slope} corresponds to the slope parameter $\eta$ discussed in \ref{subsec:pans_training}.

\subsubsection*{Digital back end}

The digital back-end processes the $d_\text{f}$-dimensional feature vector after detection. Its architecture should be expressive enough that the information bottleneck remains at detection, not at digital processing (cf.\ Fig.~1A in main text).

The back end consists of the following stages:
\begin{enumerate}
    \item \textbf{Batch normalization}: Normalizes the feature vector across the batch, stabilizing training given the high variance of photon-limited measurements.
    
    \item \textbf{Convolutional layers} (optional): 
    The \texttt{conv\_channels} parameter specifies the convolutional architecture as a nested list, where each inner list defines a block of consecutive convolutional layers, and average pooling is applied between blocks. 
    For example, \texttt{[[64,64], [128,128], [256,256,256]]} specifies three blocks: the first with two 64-channel layers, the second with two 128-channel layers, and the third with three 256-channel layers, with average pooling (size \texttt{pool\_size}) after each block. We found average pooling to work better than max pooling in the few-photon regime, likely because it better preserves the statistical properties of the sparse activations.
    Setting \texttt{conv\_channels=[]} disables convolutional layers entirely, leaving only the fully connected layers specified by \texttt{fc\_units}. This flexibility allows trading off model expressivity against training efficiency, as discussed below.
    
    \item \textbf{Fully connected layers}: The \texttt{fc\_units} parameter specifies hidden layer sizes with ReLU activations. For example, \texttt{fc\_units=[512,512]} creates two hidden layers with 512 neurons each.
    
    \item \textbf{Output layer}: A linear layer mapping to \texttt{n\_output} classes, followed by log-softmax for use with negative log-likelihood loss.
\end{enumerate}

The specific hyperparameters (kernel sizes, pooling sizes, number of channels, etc.) are configurable and may require tuning for different tasks.

\subsection{Training procedure}
\label{subsec:training_procedure}

Training PANS models involves navigating a complex hyperparameter space. The total loss combines task performance with photon budget regularization (\ref{subsec:active_budget}):
\begin{equation}
    \mathcal{L} = \mathcal{L}_{\text{CE}}(\hat{y}, y) + \alpha \cdot \|\tilde{W}\|_F^2,
    \label{eq:loss_alpha}
\end{equation}
where $\mathcal{L}_{\text{CE}}$ is the cross-entropy loss and $\alpha \geq 0$ controls the accuracy--photon budget trade-off.

The general workflow proceeds as follows:
\begin{enumerate}
    \item \textbf{Choose digital back-end architecture.} Select the depth and width of convolutional and FC layers based on task complexity.
    
    \item \textbf{For each feature dimension $d_\text{f}$ and regularization strength $\alpha$}, train models while fine-tuning hyperparameters (Algorithm~\ref{alg:active_training}). Many factors interact: learning rate distribution across model components, slope annealing schedule, intensity clamping threshold $\lambda_{\text{max}}$, batch size, initialization scale, and more.
    
    \item \textbf{Select models} achieving desired accuracy--photon trade-offs.
\end{enumerate}

Achieving optimal performance requires careful hyperparameter tuning for \emph{each} combination of $(d_\text{f}, \text{backend}, \alpha)$. This makes systematic exploration computationally expensive and labor-intensive.
\ref{tab:training_config} summarizes the baseline configuration. 

\begin{table}[h]
\centering
\renewcommand{\arraystretch}{1.3}
\setlength{\tabcolsep}{12pt}
\caption{Baseline training configuration for active PANS on FashionMNIST.}
\label{tab:training_config}
\begin{tabular}{ll}
\hline
\textbf{Parameter} & \textbf{Value} \\
\hline
Optimizer & Adam \\
Learning rate (optical front-end, conv layers) & $10^{-3}$ \\
Learning rate (FC layers) & $10^{-4}$ \\
Batch size & 128 \\
Maximum epochs & 500 \\
Training shots per activation ($K$) & 1 \\
Weight initialization & Kaiming uniform, $a = 1.2$ \\
\hline
\end{tabular}
\end{table}

\subsection{Choice of digital back-end architecture}

For the digital classifier, we compared several architectures on FashionMNIST with $d_\text{f} = 32$: a simple MLP with a single hidden layer of 512 units ($d_\text{f} \to 512 \to 10$), and a convolutional network with channel configuration $[[64,64], [128,128], [256,256,256]]$ (as stated in \ref{subsec:sim_model}), using kernel size 15 (which we found performed better than the conventional kernel size 3) and average pooling with pool size 2. The convolutional network achieves approximately 83\% test accuracy, while the MLP achieves approximately 82\%---a difference of roughly one percentage point.

Given this modest gap, we adopt the MLP architecture throughout this work. This choice keeps the digital back-end consistent across all experiments and, more importantly, makes hyperparameter optimization practical. As discussed below, achieving the desired balance between classification accuracy and photon budget requires careful tuning of the regularization strength $\alpha$ along with learning rate schedules and other training parameters. These interactions are delicate: small changes in $\alpha$ can shift the accuracy--photon operating point substantially, and the optimal settings depend on $d_\text{f}$, the target photon budget, and the specific task. Using a simple MLP architecture reduces the dimensionality of this search space, allowing us to explore the accuracy--photon tradeoff more thoroughly.

\subsection{Illumination regularization and the accuracy--photon tradeoff}
\label{subsec:alpha_tuning}

\begin{figure}[htp]
    \centering
    \includegraphics[width=.999\textwidth]{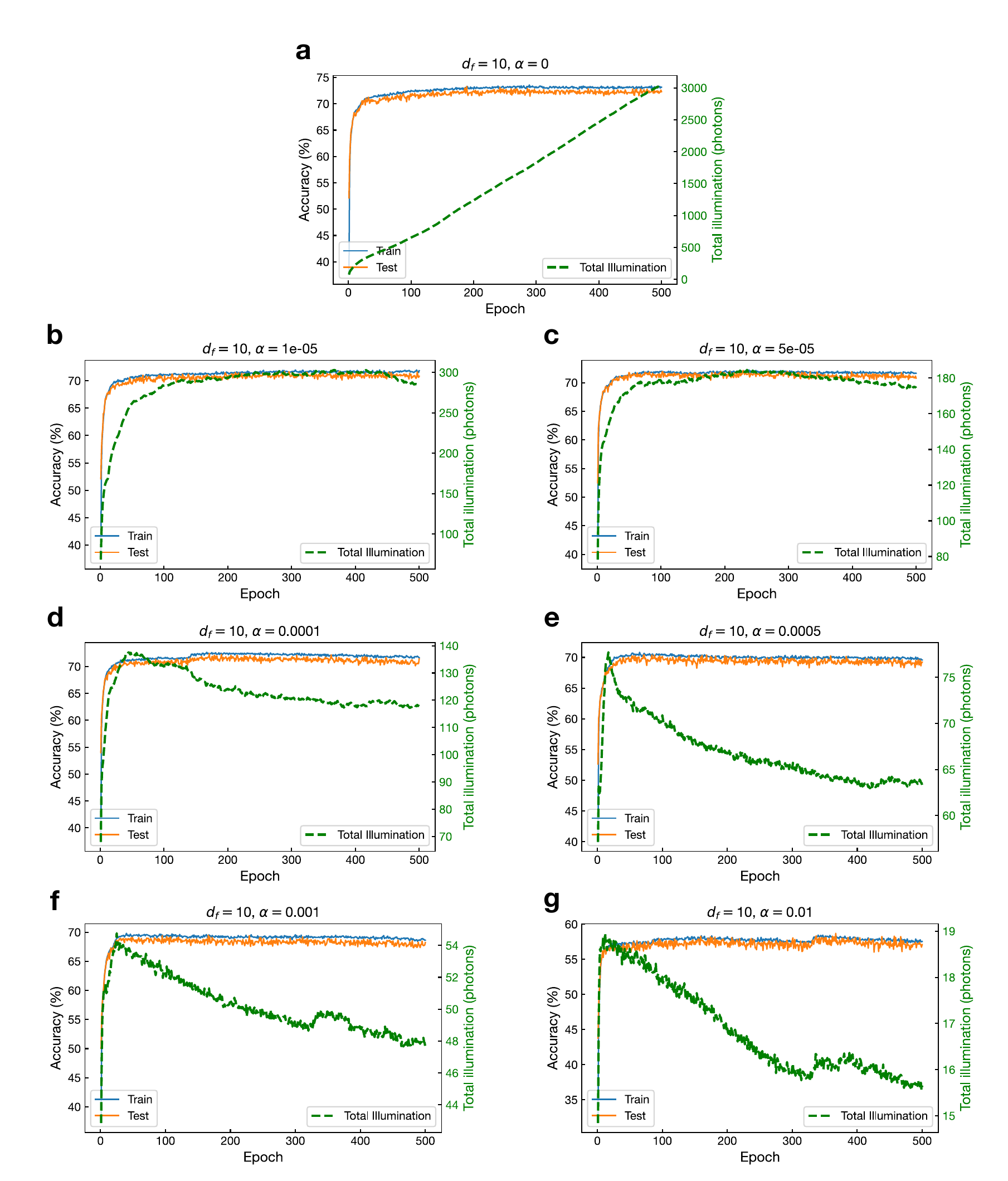}
    \caption{\textbf{Training dynamics under different illumination regularization $\alpha$ for $d_\text{f} = 10$.} Each panel shows training accuracy (blue), test accuracy (orange), and total illumination photon budget $N_{\text{illu}}$ (green dashed, right axis) versus training epoch. The $d_\text{f}$ and $\alpha$ values are indicated above each panel. All models use the training configuration in \ref{tab:training_config} with only $\alpha$ varied.}
    \label{fig:alpha_sweep_10}
\end{figure}

\begin{figure}[htp]
    \centering
    \includegraphics[width=.999\textwidth]{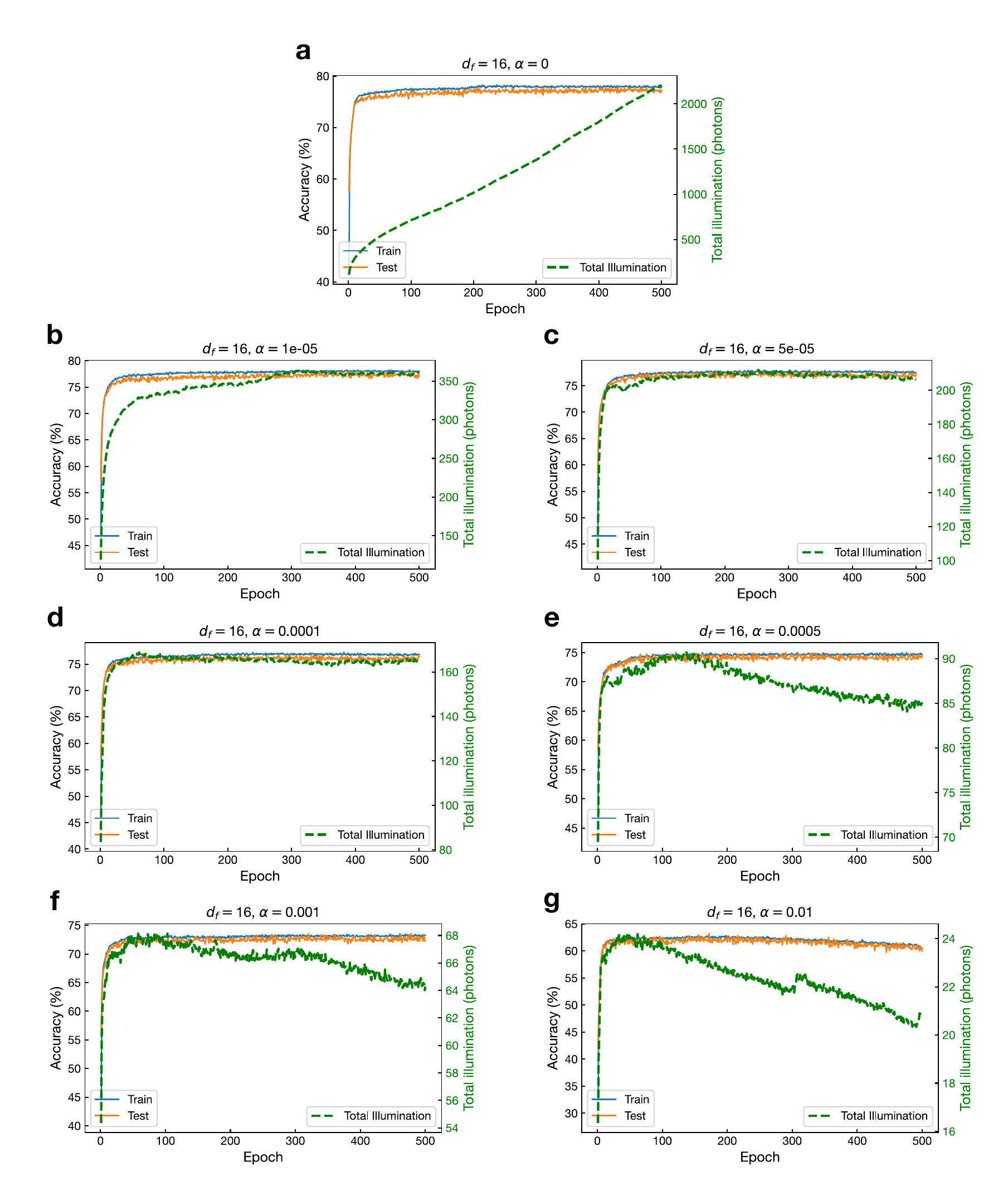}
    \caption{\textbf{Training dynamics under different illumination regularization $\alpha$ for $d_\text{f} = 16$.} Each panel shows training accuracy (blue), test accuracy (orange), and total illumination photon budget $N_{\text{illu}}$ (green dashed, right axis) versus training epoch. The $d_\text{f}$ and $\alpha$ values are indicated above each panel. All models use the training configuration in \ref{tab:training_config} with only $\alpha$ varied.}
    \label{fig:alpha_sweep_16}
\end{figure}

\begin{figure}[htp]
    \centering
    \includegraphics[width=.999\textwidth]{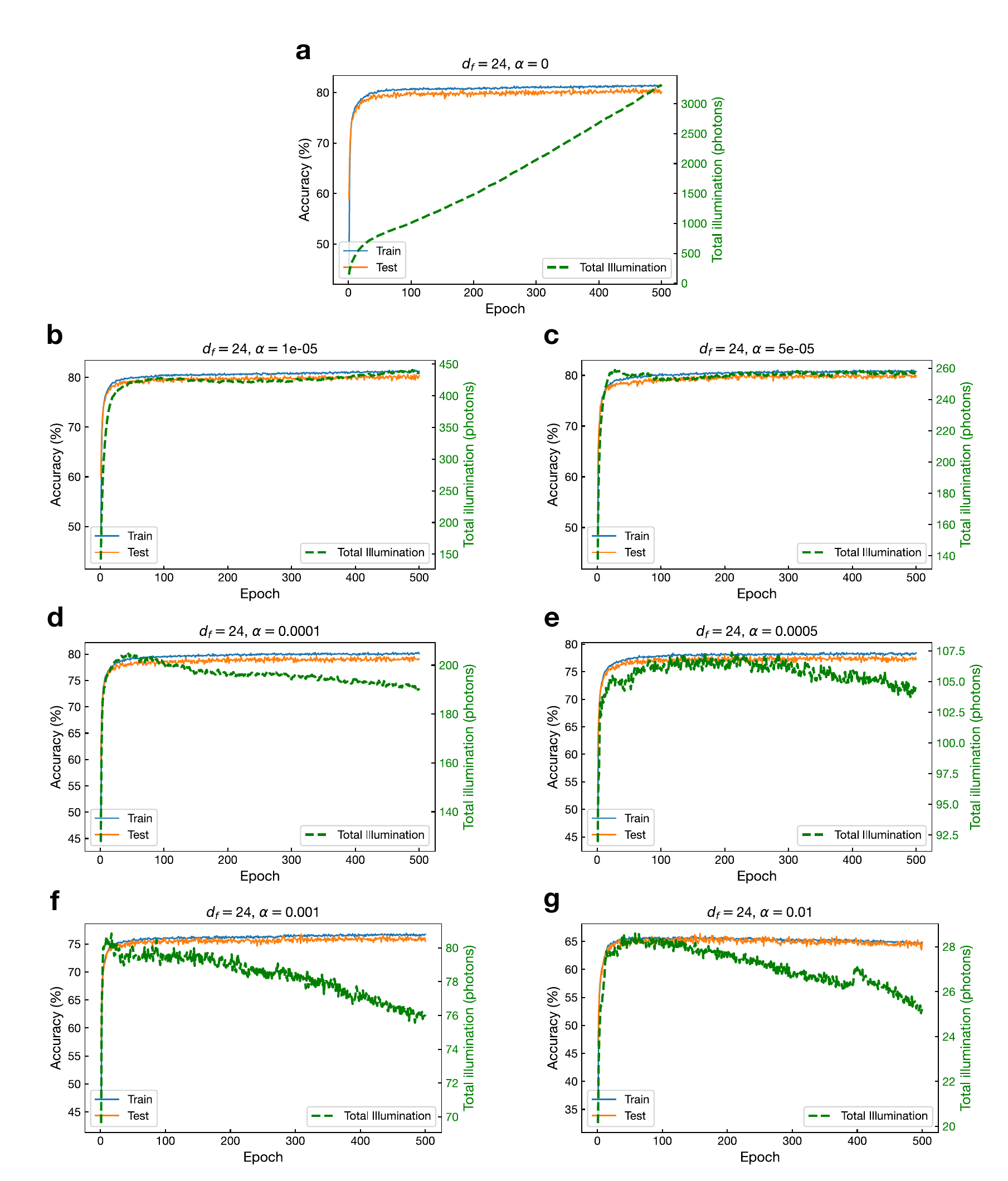}
    \caption{\textbf{Training dynamics under different illumination regularization $\alpha$ for $d_\text{f} = 24$.} Each panel shows training accuracy (blue), test accuracy (orange), and total illumination photon budget $N_{\text{illu}}$ (green dashed, right axis) versus training epoch. The $d_\text{f}$ and $\alpha$ values are indicated above each panel. All models use the training configuration in \ref{tab:training_config} with only $\alpha$ varied.}
    \label{fig:alpha_sweep_24}
\end{figure}

\begin{figure}[htp]
    \centering
    \includegraphics[width=.999\textwidth]{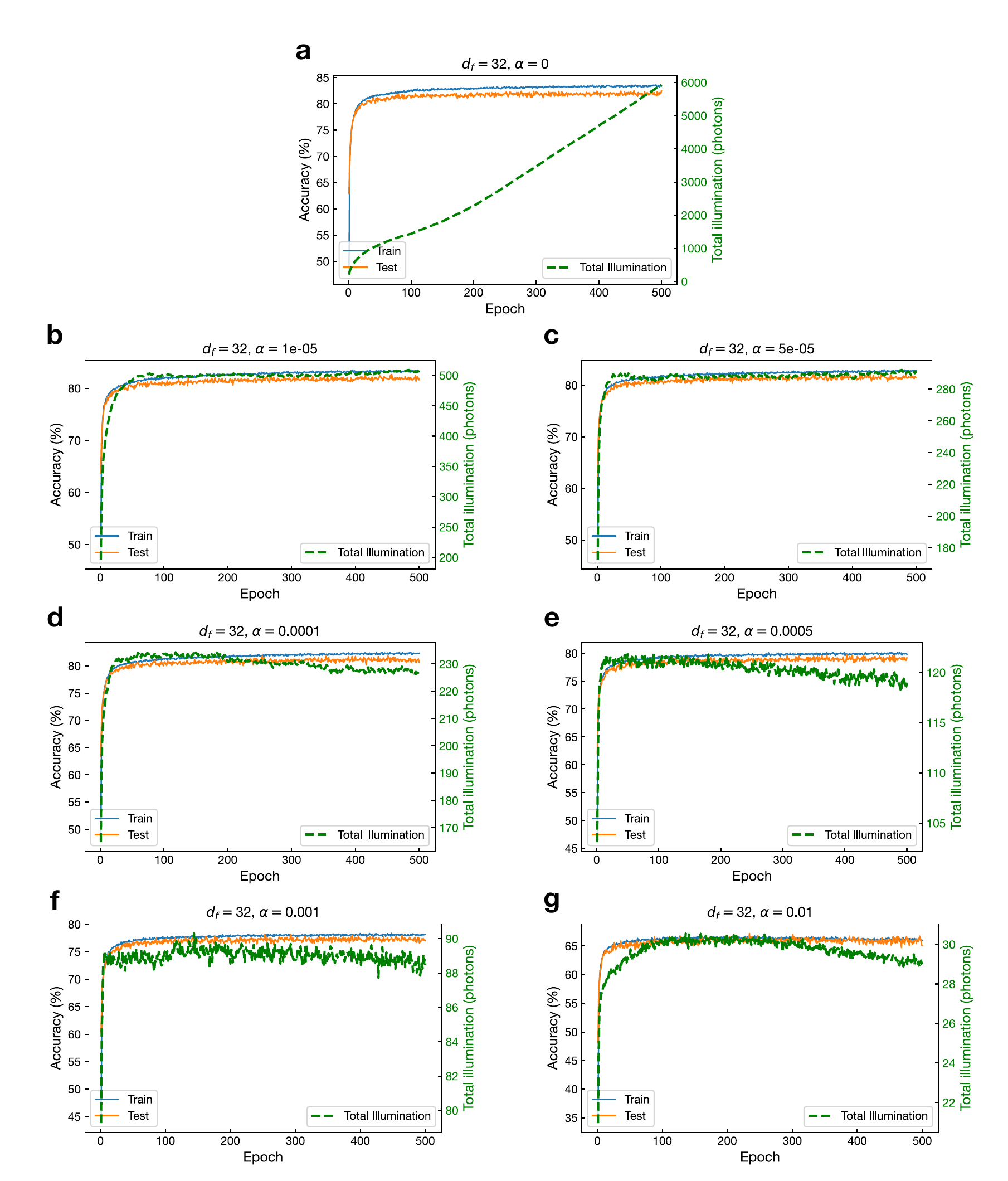}
    \caption{\textbf{Training dynamics under different illumination regularization $\alpha$ for $d_\text{f} = 32$.} Each panel shows training accuracy (blue), test accuracy (orange), and total illumination photon budget $N_{\text{illu}}$ (green dashed, right axis) versus training epoch. The $d_\text{f}$ and $\alpha$ values are indicated above each panel. All models use the training configuration in \ref{tab:training_config} with only $\alpha$ varied.}
    \label{fig:alpha_sweep_32}
\end{figure}

The regularization coefficient $\alpha$ in the loss function (Eq.~\ref{eq:loss_alpha}) controls the tradeoff between classification accuracy and total illumination photons $N_{\text{illu}}$. To illustrate how this tradeoff manifests during training, \ref{fig:alpha_sweep_10}--\ref{fig:alpha_sweep_32} show training dynamics across a range of $\alpha$ values for $d_\text{f} \in \{10, 16, 24, 32\}$ on FashionMNIST.

These training curves provide a general sense of how $\alpha$ affects the accuracy--photon tradeoff, helping to identify the parameter regime worth exploring for a given application. The $\alpha$ values shown span several orders of magnitude, from no regularization ($\alpha = 0$) to strong regularization ($\alpha = 10^{-2}$), illustrating the qualitative transitions in training behavior.

The specific $\alpha$ values in these figures are illustrative rather than prescriptive. The models deployed in our experiments (Tables~\ref{tab:exp_fmnist_K} and \ref{tab:exp_mnist_K}) were trained with $\alpha$ values chosen through additional fine-tuning for each $(d_\text{f})$ configuration. This fine-tuning involves adjusting not only $\alpha$ but also learning rate schedules, batch sizes, and training duration in ways that depend on the particular operating point. Such model-specific adjustments are difficult to systematize into general rules. The sweeps shown here serve as a starting point: they help locate the $\alpha$ range where accuracy and photon budget are both reasonable, after which more targeted exploration within that range yields the final model.

\subsubsection*{Qualitative regimes}

Despite the need for model-specific tuning, several qualitative observations apply broadly:

\paragraph*{No regularization ($\alpha = 0$).} Without illumination penalty, the optimizer freely increases pattern intensities to improve classification. Accuracy converges normally, but $N_{\text{illu}}$ grows without bound, reaching thousands of photons. Such models are unsuitable for photon-constrained sensing.

\paragraph*{Weak regularization.} The illumination penalty prevents $N_{\text{illu}}$ from diverging. In this regime, $N_{\text{illu}}$ converges to a stable value during training, but the photon level remains relatively high, providing limited benefit over the unregularized case.

\paragraph*{Moderate regularization.} The penalty significantly constrains $N_{\text{illu}}$ while causing only modest accuracy reduction. This is the practically useful regime: the system learns to allocate photons efficiently, and the accuracy--photon tradeoff is favorable. When selecting models for deployment, we search within this regime to find configurations that achieve target accuracy at minimal photon cost.

\paragraph*{Strong regularization.} When the penalty dominates, $N_{\text{illu}}$ is driven to extremely low levels---likely below the point where sufficient information remains for accurate classification. Accuracy degrades substantially as the optimizer prioritizes reducing illumination at the expense of learning discriminative features. Training may also become unstable, exhibiting persistent oscillations.

\subsection{Effect of regularization on learned patterns}

\begin{figure}[htp]
    \centering
    \includegraphics[width=0.67\textwidth]{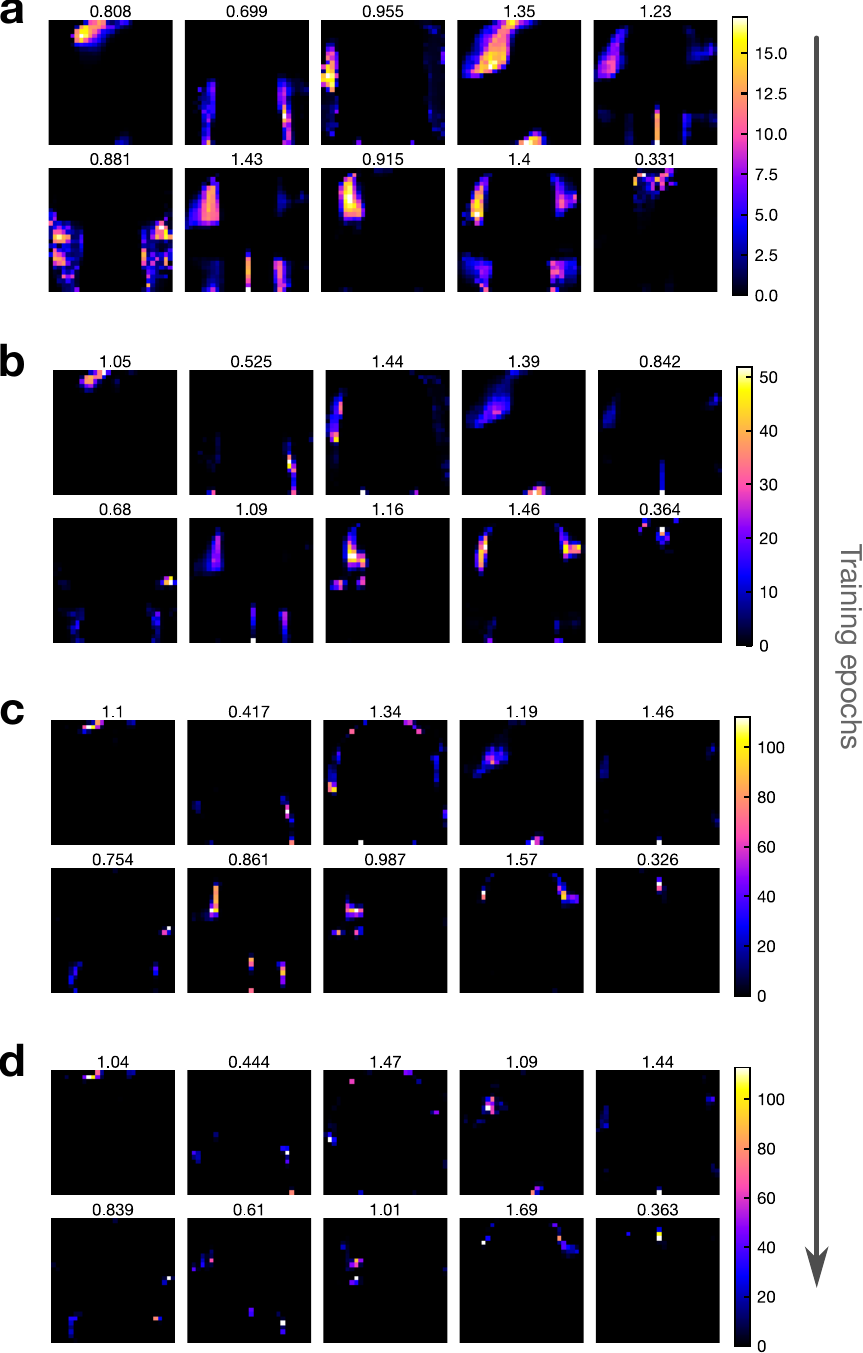}
    \caption{\textbf{Evolution of learned illumination patterns during training.} Each panel displays the $d_\text{f} = 10$ patterns from a model trained with $\alpha = 10^{-4}$, shown at four stages with training progressing from \textbf{a} to \textbf{d}. For visualization, patterns are normalized such that the mean pixel value across all patterns equals 1 (corresponding to a total of 784 per pattern); the number above each pattern shows its individual mean. The color bar indicates pixel value range. Early in training (\textbf{a}), patterns are spatially diffuse with moderate intensity variation. Through training (\textbf{b}--\textbf{d}), patterns become increasingly sparse and localized, concentrating intensity into small regions while the maximum value expands from $\sim$18 to $\sim$100 times the mean (color bar), reflecting efficient photon allocation to discriminative spatial locations.}
    \label{fig:w_train_epoch}
\end{figure}

The regularization affects not only the total photon budget but also the spatial structure of the learned patterns. To illustrate this, \ref{fig:w_train_epoch} displays the $d_\text{f} = 10$ illumination patterns in a PANS front end at four stages during training with $\alpha = 10^{-4}$.

To visualize how the weight distribution evolves during training regardless of the absolute intensity scale (which varies substantially as shown in \ref{fig:alpha_sweep_10}--\ref{fig:alpha_sweep_32}), we normalize patterns by their mean pixel value for display purposes---this normalization is not used during actual training or inference, where values retain their physical meaning in photon numbers. Specifically, for $d_\text{f} = 10$ patterns each containing $28 \times 28 = 784$ pixels, we divide all pixel values by the mean across all $10 \times 784$ values. After normalization, the mean pixel value equals 1, so the total across one pattern equals 784. This allows us to examine how the distribution of weights evolves independent of the overall intensity scale.

The number above each pattern indicates its individual mean pixel value after normalization, and the color bar shows the range of pixel values at each stage. We do not specify exact epoch numbers because the training dynamics depend heavily on configuration details such as slope scheduling, learning rate distribution across model components, and other hyperparameters discussed earlier in this section; the qualitative progression matters more than the specific epoch count.

Early in training (\ref{fig:w_train_epoch}a), patterns are spatially diffuse, with intensity spread across broad regions resembling edge-like or blob-like features. The value range spans 0--18 times the mean, indicating moderate variation. As training progresses (\ref{fig:w_train_epoch}b--d), the patterns become increasingly sparse and localized: intensity concentrates into small clusters of pixels while most of the image becomes dark. By the later stages, the maximum value extends to $\gtrsim$100 times the mean, reflecting highly uneven distributions where a few pixels carry most of the energy despite the fixed mean pixel value of 1.

This concentration reflects the PANS framework learning to allocate photons efficiently. Under the constraint of a detection bottleneck, the optimizer discovers that selectively probing specific spatial locations---rather than illuminating the object uniformly---better preserves task-relevant information under stringent photon budgets.

\newpage
\section{Testing trained PANS models}
\label{sec:sim_testing}

After training produces models across a range of $(d_\text{f}, \alpha)$ configurations (\ref{sec:sim_training}), we select representative models that achieve good accuracy--photon tradeoffs for systematic evaluation. This section examines three factors that affect testing performance beyond the trained model weights: the number of measurement shots $K$, the photon level scaling factor $\eta$, and the dark count rate (DCR). Understanding these factors is essential for bridging the gap between numerical simulation and practical sensing applications.

All testing results in this section use 200 test images from FashionMNIST, with each configuration repeated 100 times to capture the stochastic variability inherent to single-photon detection. This protocol mirrors our experimental implementation described in later sections.

\subsection{Multi-shot ($K > 1$) evaluation}
\label{subsec:testing_K}

Our models are trained with $K = 1$ (single-shot measurement per feature), but a natural question arises: how does performance scale when we allow multiple measurements? This directly tests the theoretical argument from \ref{subsec:K_scaling}, where we predicted that increasing $d_\text{f}$ with $K = 1$ should outperform increasing $K$ with fixed $d_\text{f}$, given the same total number of detection events (total feature size) $d_\text{f}\times K$.

In numerical simulation, evaluating at different $K$ values is straightforward---we simply specify the number of repeated measurements in the forward pass, and the model internally sums the stochastic binary outcomes to form the activation vector. 
In practice, this is equivalent to continuous collection on the photon detector with the same experimental setup. \ref{tab:sim_fmnist_K} presents the results across different $(d_\text{f}, K)$ configurations.

\begin{table}[h]
\centering
\renewcommand{\arraystretch}{1.3}
\setlength{\tabcolsep}{12pt}
\caption{Test accuracy (\%) on FashionMNIST for different feature dimensions $d_\text{f}$ and shot counts $K$. Each entry shows mean $\pm$ standard deviation over 100 trials with 200 test images.}
\label{tab:sim_fmnist_K}
\begin{tabular}{lcccc}
\hline
 & $K = 1$ & $K = 2$ & $K = 3$ & $K = 5$ \\
\hline
$d_\text{f} = 3$ & $54.46 \pm 1.31$ & $55.20 \pm 1.67$ & $55.88 \pm 1.35$ & $56.68 \pm 1.44$ \\
$d_\text{f} = 4$ & $65.11 \pm 2.07$ & $66.26 \pm 1.90$ & $67.20 \pm 1.55$ & $68.38 \pm 1.59$ \\
$d_\text{f} = 6$ & $68.37 \pm 2.06$ & $70.15 \pm 1.84$ & $71.33 \pm 1.50$ & $71.29 \pm 1.32$ \\
$d_\text{f} = 10$ & $74.31 \pm 1.84$ & $77.55 \pm 1.32$ & $77.53 \pm 1.32$ & $78.48 \pm 1.16$ \\
$d_\text{f} = 16$ & $78.30 \pm 1.96$ & $81.42 \pm 1.48$ & $81.85 \pm 1.36$ & $82.34 \pm 1.32$ \\
$d_\text{f} = 24$ & $80.35 \pm 1.78$ & $81.98 \pm 1.39$ & $82.22 \pm 1.26$ & $82.35 \pm 1.16$ \\
$d_\text{f} = 32$ & $83.13 \pm 1.48$ & $84.39 \pm 1.48$ & $85.02 \pm 1.15$ & $85.34 \pm 1.05$ \\
\hline
\end{tabular}
\end{table}

For a fixed $d_\text{f}$, increasing $K$ generally improves accuracy. However, the critical comparison is between configurations with equivalent total detection events but different $(d_\text{f}, K)$ allocations. This tests our theoretical prediction: when $K > 1$, the effective feature space of size $d_\text{f} \times K$ was not fully optimized during training, since only $d_\text{f}$ independent features were learned.

\begin{figure}[htp]
    \centering
    \includegraphics[width=0.6\textwidth]{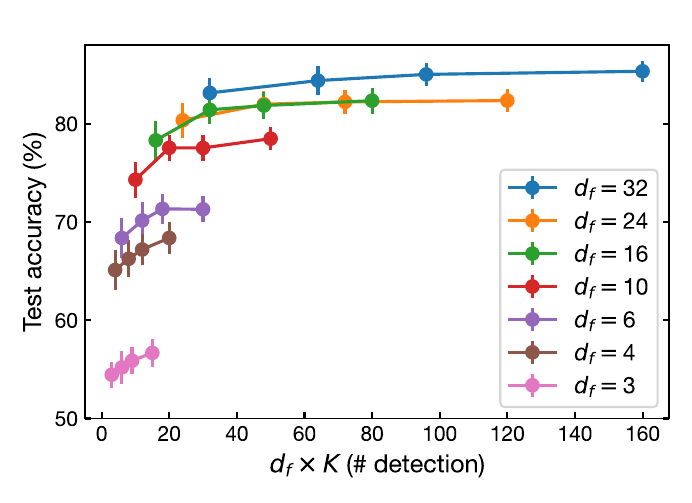}
    \caption{\textbf{Multi-shot evaluation validates $d_\text{f}$ scaling over $K$ scaling.} Test accuracy versus total number of detections ($d_\text{f} \times K$). Different curves correspond to models trained with different feature dimensions $d_\text{f} \in \{3, 4, 6, 10, 16, 24, 32\}$, as indicated in the legend. Points along each curve correspond to increasing $K$ (1, 2, 3, 5) from left to right. At the same total detection budget, larger $d_\text{f}$ with smaller $K$ consistently achieves higher accuracy than smaller $d_\text{f}$ with larger $K$. Dot points show mean values computed over 100 repetitions; error bars indicate standard deviation.}
    \label{fig:K_scaling_plot}
\end{figure}

\ref{fig:K_scaling_plot} presents this comparison directly, plotting test accuracy against total number of detections $d_\text{f} \times K$. The results validate our prediction: at the same total detection budget, configurations with larger $d_\text{f}$ (and correspondingly smaller $K$) consistently outperform those with smaller $d_\text{f}$ and repeated measurements. For example, at approximately 30 total detections, $(d_\text{f} = 10, K = 3)$ substantially outperforms $(d_\text{f} = 6, K = 5)$ and $(d_\text{f} = 3, K = 10)$. This confirms that each independently optimized feature extracts unique task-relevant information, whereas repeated measurements of the same feature merely refine an already-captured signal component.

One might note that the $d_\text{f} = 16$ curve appears to be close to the $d_\text{f} = 24$ curve in some regions. This reflects the practical challenge of training PANS models: achieving good accuracy--photon tradeoffs requires careful hyperparameter tuning, and some configurations may be better optimized than others. The $d_\text{f} = 16$ model happened to be particularly well-tuned in our experiments. Despite such variations, the overall trend clearly supports the theoretical prediction that $d_\text{f}$ scaling is more efficient than $K$ scaling.

These results justify our choice to train and evaluate primarily at $K = 1$, which represents the most photon-efficient operating regime (\ref{subsec:df_scaling}).

\subsection{Photon level scaling ($\eta$)}
\label{subsec:testing_eta}

During training, the optimization determines an optimal photon level (\ref{subsec:eta_scaling}) for each PANS model through the accuracy--photon budget tradeoff controlled by $\alpha$. Since this photon level has no absolute units, we define $\eta = 1$ as the trained operating point. We then vary $\eta$ to examine how robust the models are to photon level variations.

This analysis serves two purposes. First, it characterizes robustness: in practical sensing scenarios, illumination power may fluctuate due to source instability, objects may have varying reflectivity, or environmental conditions may differ from training assumptions. Second, it explores whether we can operate at lower photon budgets than training produced: by reducing $\eta$ below 1, we attenuate the signal before detection and reduce the detected photon budget $N_{\text{det}}$, potentially achieving acceptable performance with fewer photons than the training-optimized level.

\begin{figure}[htp]
    \centering
    \includegraphics[width=0.6\textwidth]{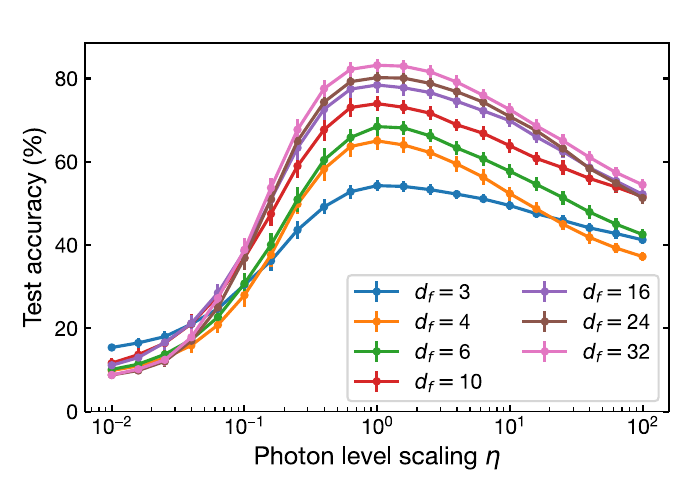}
    \caption{\textbf{Effect of photon level scaling $\eta$.} Test accuracy versus photon scaling parameter $\eta$ for trained PANS models with different feature dimensions $d_\text{f} \in \{3, 4, 6, 10, 16, 24, 32\}$, as indicated in the legend. The horizontal axis uses logarithmic scale spanning $\eta \in [10^{-2}, 10^2]$. All models achieve peak accuracy near $\eta = 1$ (the trained operating point) and degrade for both lower and higher $\eta$ values. Dot points show mean values computed over 100 repetitions; error bars indicate standard deviation.}
    \label{fig:eta_sweep}
\end{figure}

\ref{fig:eta_sweep} shows test accuracy versus $\eta$ for different $d_\text{f}$ configurations. The curves exhibit a peak near $\eta = 1$, confirming that the trained models are indeed optimized for this photon level. As $\eta$ decreases below 1, accuracy degrades gracefully over a substantial range before dropping sharply at very low $\eta$ values where detection probabilities become negligibly small. As $\eta$ increases above 1, accuracy also decreases---this may seem counterintuitive, but reflects that the learned illumination patterns were optimized for specific detection probability distributions, and simply increasing photon counts does not improve (and can degrade) the feature discrimination learned during training.

The graceful degradation for $\eta < 1$ has practical implications: it suggests that models trained at one photon level can maintain useful performance when operated at moderately lower photon budgets, providing flexibility without requiring retraining for every specific condition. However, this flexibility is limited---at very low $\eta$, all models converge toward random-guess performance as insufficient photons are available for any meaningful discrimination.

This robustness to photon level variation has practical implications beyond simply enabling lower-photon operation. In real sensing scenarios, illumination power may fluctuate due to source instability, objects may have varying reflectivity, or environmental conditions may differ from training assumptions. The ability to maintain performance across a range of effective photon levels provides important operational flexibility without requiring retraining for every specific condition.

\subsection{Dark count rate (DCR)}
\label{subsec:testing_dcr}

Dark counts are often the first technical concern for single-photon detection (SPD) applications---and potentially a showstopper for low-light sensing. If PANS cannot tolerate realistic dark count levels, its practical utility would be severely limited regardless of its photon efficiency.

Dark counts in single-photon detectors arise from thermal excitation, afterpulsing, and other detector-intrinsic noise sources. For sensing applications more broadly, functionally equivalent effects arise from ambient light leakage, thermal background radiation, stray reflections, and other environmental sources. We use ``dark count rate'' (DCR) to encompass all such contributions: any process that triggers detection events independently of the signal being measured.

\begin{figure}[h]
    \centering
    \includegraphics[width=0.6\textwidth]{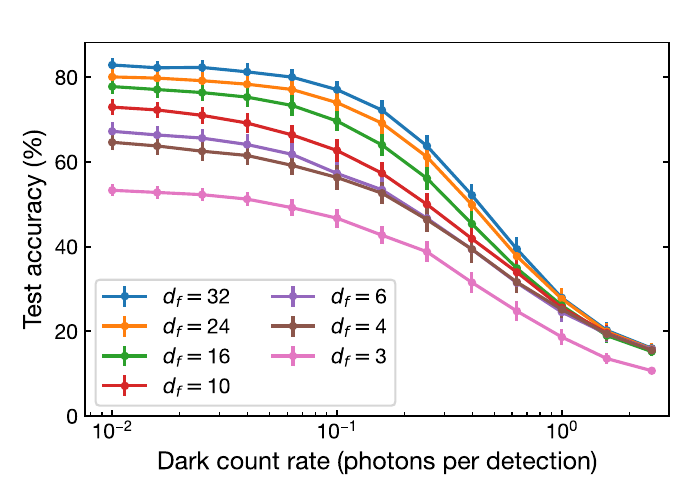}
    \caption{\textbf{Robustness to dark counts.} Test accuracy versus dark count rate (expected dark counts per detector per measurement) for different feature dimensions $d_\text{f} \in \{3, 4, 6, 10, 16, 24, 32\}$ at $\eta = 1$, as indicated in the legend. The horizontal axis uses logarithmic scale. Accuracy degrades gradually up to DCR $\sim 0.1$ and more rapidly at higher DCR values. Dot points show mean values computed over 100 repetitions; error bars indicate standard deviation.}
    \label{fig:dcr_sweep}
\end{figure}

We model DCR by adding independent Poisson-distributed counts to the signal detection. Specifically, for each detector, we sample from $\text{Poisson}(\lambda_{\text{dark}})$ and add the result to the detection outcome. \ref{fig:dcr_sweep} shows how test accuracy varies with DCR across different $d_\text{f}$ configurations. PANS maintains strong performance at DCR levels up to approximately 0.1 (10\% of the signal level), which exceeds typical values in well-designed single-photon detection systems. Beyond this point, accuracy degrades more rapidly as noise begins to overwhelm the signal.

This dark count tolerance is consistently observed across the various sensing tasks reported in this work, where we explicitly evaluate performance under multiple DCR conditions. The combination of $\eta$ robustness (to photon-level variations) and DCR tolerance (to background noise) establishes that PANS can operate reliably under realistic, non-ideal conditions---a prerequisite for any practical sensing system.

\newpage
\section{Paradigm comparison in numerical simulation}
\label{sec:sim_paradigm}

\ref{sec:paradigms} established a conceptual taxonomy of sensing approaches based on three design decisions: whether to apply pre-detection optical transformations, whether to optimize them for the task, and whether to model the stochastic detection process during training. That section argued that when the detection bottleneck becomes significant, photon-aware methods should outperform alternatives by optimizing against the true physical constraints rather than approximations.

Here we validate these arguments through numerical simulation on FashionMNIST, systematically comparing the four paradigms summarized in \ref{tab:approach_comparison}: direct imaging, fixed optical transformations, conventional end-to-end optimization (non-PA E2E), and PANS. For each approach, we evaluate classification accuracy across a range of photon budgets $N_{\text{det}}$, revealing how performance degrades as the detection bottleneck tightens.

\subsection{Direct imaging}
\label{subsec:direct_imaging_sim}

In direct imaging, the detector array captures the spatial intensity distribution without optical preprocessing. A digital classifier then processes the recorded frame. This paradigm is standard in computer vision, where photon counts are typically high enough that detection introduces negligible information loss.

\subsubsection*{Simulation protocol}

We apply a uniform scaling factor $\eta$ to all test images: for an image with pixel transmissions $\{x_j\}$, the expected photon count at pixel $j$ is $\lambda_j = \eta \cdot x_j$. Each pixel independently undergoes Poisson sampling to simulate shot noise, yielding a detected count $n_j \sim \text{Poisson}(\lambda_j)$. The resulting frame $\{n_j\}$ is normalized and fed to the classifier.

Since different images have different total transmissions $\sum_j x_j$, a given $\eta$ produces different total detected photon counts for different images. To characterize the operating point with a single number, we report $N_{\text{det}}$ as the average total detected photons across the test set. By sweeping $\eta$, we obtain performance curves parameterized by this average $N_{\text{det}}$. For the illumination budget $N_{\text{illu}}$, we assume uniform illumination across all pixels---the standard imaging configuration where each pixel receives equal incident optical power.

Our simulation considers \emph{only} fundamental photon shot noise, excluding detector dark counts, readout noise, ambient light, and other practical degradation mechanisms. The resulting frames represent an idealized \emph{best case}: the maximum information that direct imaging could preserve at any given $N_{\text{det}}$. Realistic implementations with practical considerations would perform worse. The gap between PANS and direct imaging shown in main text Fig.~2 is therefore conservative.

We use an AlexNet-style convolutional neural network (5 convolutional layers followed by fully-connected layers) trained on clean (high-$N_{\text{det}}$) images to report the test accuracy. This choice reflects a practical consideration: we cannot exhaustively search for the optimal digital classifier architecture at every $N_{\text{det}}$ value, nor is this our focus. Convolutional neural networks are well-understood and sufficient to extract class-discriminative information survives the detection process, although better models exist (e.g., Vision Transformers). When information is severely degraded at detection, more sophisticated architectures might yield modest improvements at any given operating point, but the fundamental trend---accuracy collapsing as $N_{\text{det}}$ decreases---would remain. We develop this insight in the following subsection.

\begin{figure}[htp]
    \centering
    \includegraphics[width=\textwidth]{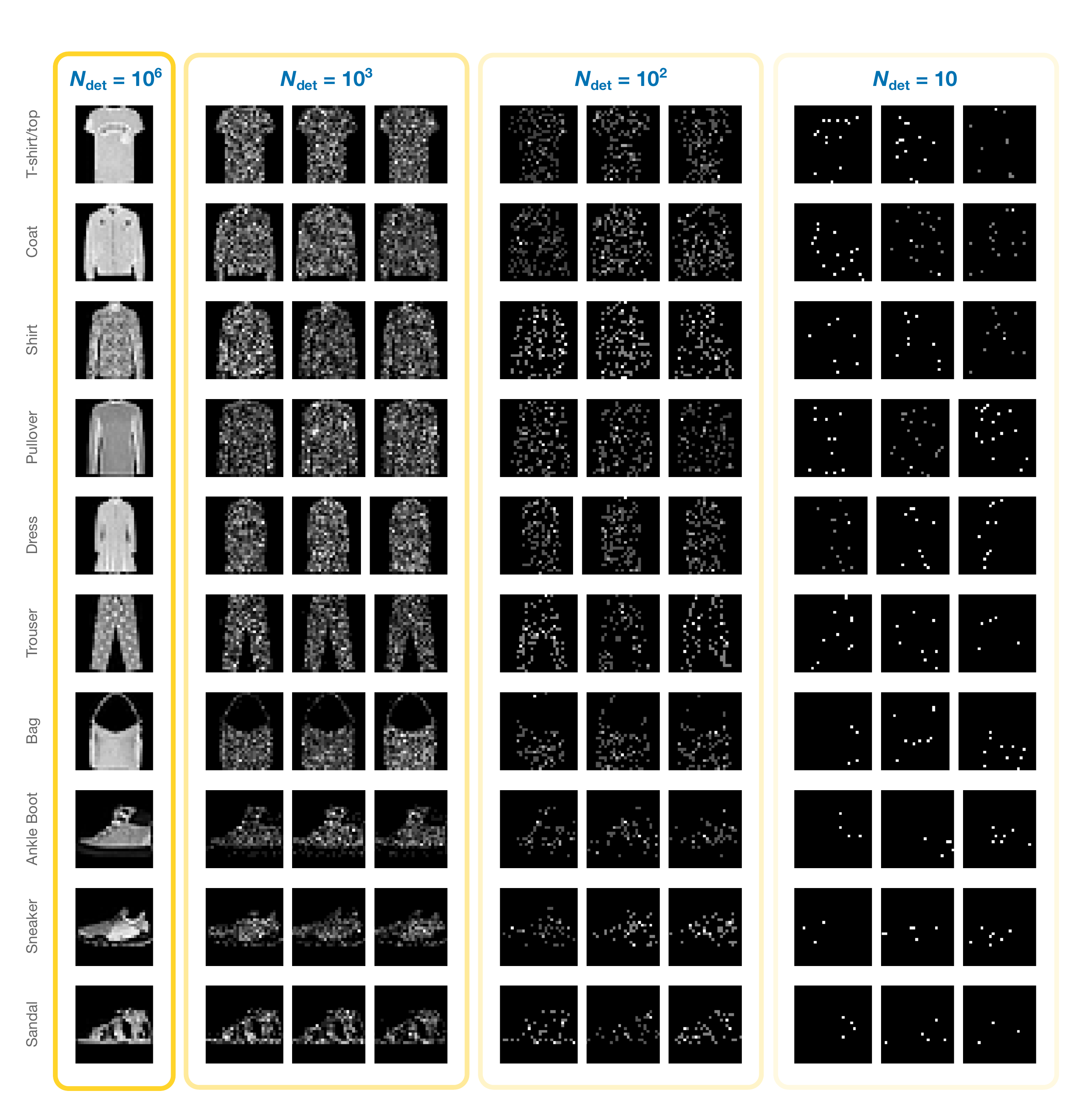}
    \caption{\textbf{Shot-noise-limited frames at different photon budgets.} Each row shows the same test image captured at decreasing average $N_{\text{det}}$ (left to right: $10^6$, $10^3$, $10^2$, 10). For $N_{\text{det}} \leq 10^3$, three independent samples are shown to illustrate stochastic variation. Class-discriminative features are progressively destroyed by shot noise. At low photon counts, even human observers cannot reliably identify the object category.}
    \label{fig:direct_imaging_frames}
\end{figure}

\subsubsection*{Visualizing information loss}

The claim that information is ``lost at detection'' may seem abstract. Here we make it concrete through direct visual inspection of the captured frames, leveraging a simple but powerful principle: if a human observer cannot distinguish between classes by looking at the frames, it is evident that the discriminative information is gone.

This principle deserves elaboration. The human visual system, evolved over millions of years, is remarkably adept at pattern recognition---often outperforming sophisticated algorithms on tasks involving noisy or degraded images. When trained observers consistently fail to identify objects in captured frames, this constitutes strong evidence that the frames genuinely lack the information needed for classification. The limitation is not in the observer's skill but in the data itself. Just as any digital classifier, no matter how advanced, can only extract information that exists in its input; it cannot recover information that was never captured.

\ref{fig:direct_imaging_frames} shows example frames at different photon budgets for all 10 FashionMNIST classes. At $N_{\text{det}} = 10^6$, frames are visually identical to ground truth---every detail is preserved, and classification is trivial. As $N_{\text{det}}$ decreases, shot noise progressively corrupts the spatial structure. At $N_{\text{det}} = 10^3$, fine details are lost, though overall shape remains discernible. The three independent samples per image shown in the figure illustrate the stochastic variation at this photon level. At $N_{\text{det}} = 10^2$, fine details are lost and the variation between samples of the same image becomes substantial---a ``pullover'' might plausibly be a ``shirt'' or ``coat,'', a ``sandal'' might be a ``sneaker'' or ``ankle boot.'' At $N_{\text{det}} = 10$, frames reduce to sparse, random-looking dot patterns bearing no visible resemblance to the original objects---one cannot tell whether the original was clothing, footwear, or a bag.

This visual inspection reveals something important: the difficulty is not uniform across classes. Categories with similar silhouettes---such as ``pullover'' versus ``shirt,'' or ``sneaker'' versus ``ankle boot''---become indistinguishable at higher $N_{\text{det}}$ than categories with distinct shapes like ``trouser'' or ``bag.'' This observation will resurface in our experimental results (\ref{sec:evaluation}), where confusion matrices show elevated error rates between precisely these visually similar categories.

The frames at low $N_{\text{det}}$ also illustrate why no classifier can rescue the situation. Consider the $N_{\text{det}} = 10$ frames: the detected photons are so sparse that the frames from different images of the \emph{same} class look as different from each other as frames from \emph{different} classes. There is simply no consistent pattern that a classifier could learn to exploit. The information has been destroyed by the stochastic detection process, not by any deficiency in the classification algorithm.

\begin{figure}[htp]
    \centering
    \includegraphics[width=\textwidth]{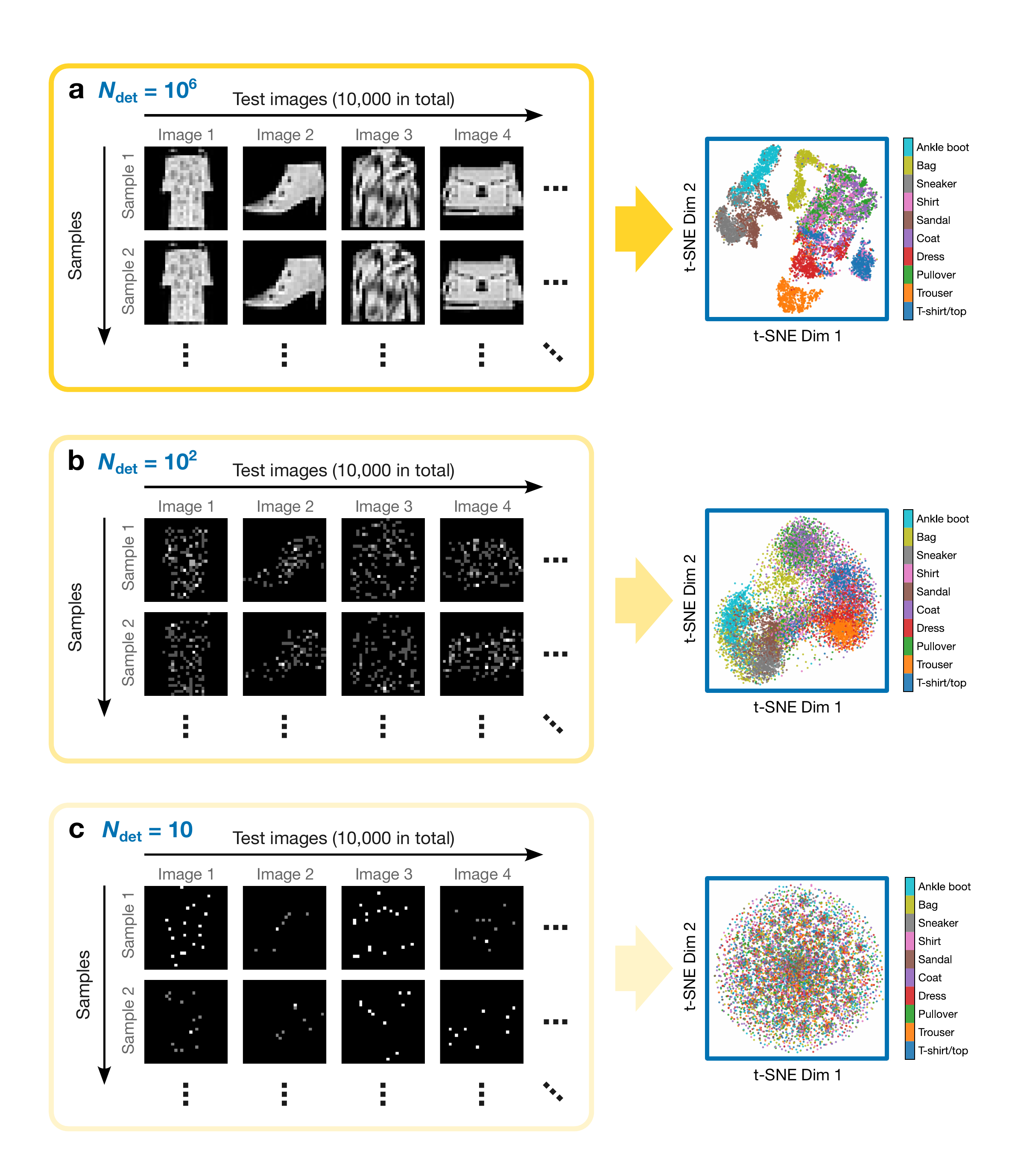}
    \caption{\textbf{t-SNE visualization of information loss.} Each panel shows the procedure and resulting embedding at a different $N_{\text{det}}$: each test image (10,000 total) is sampled 3 times at the target photon budget, and all 30,000 resulting vectors are projected via t-SNE. Points are colored by class label. \textbf{a,}~$N_{\text{det}} = 10^6$: well-separated clusters indicate preserved class structure. \textbf{b,}~$N_{\text{det}} = 10^2$: clusters overlap as shot noise corrupts discriminative features. \textbf{c,}~$N_{\text{det}} = 10$: class structure is entirely lost; the embedding collapses into an undifferentiated distribution.}
    \label{fig:tsne_direct_imaging}
\end{figure}

To move beyond simple visual intuition, we employ t-SNE dimensionality reduction~\cite{van2008visualizing}. For each of the 10,000 test images, we generate 3 independent shot-noise-limited frames at the target $N_{\text{det}}$, yielding 30,000 frame vectors of dimension 784. The t-SNE algorithm projects these high-dimensional vectors into 2D while preserving local neighborhood structure, allowing us to visualize whether frames from the same class cluster together or scatter indiscriminately.
\ref{fig:tsne_direct_imaging} illustrates this procedure and representative embeddings at three different photon budgets spanning the transition from well-preserved to completely destroyed class structure.

At $N_{\text{det}} = 10^6$, the 10 FashionMNIST classes form distinct, well-separated clusters---the frames preserve enough information that same-class frames are far more similar to each other than to frames from other classes. At $N_{\text{det}} = 10^2$, cluster boundaries blur and substantial overlap appears. By $N_{\text{det}} = 10$, the embedding collapses into a single undifferentiated blob: frames from all classes are statistically interchangeable. The t-SNE embedding cannot find structure because there is no structure to find.

This visualization reinforces the central point: the performance degradation we observe is not a failure of the classifier but a fundamental loss of information at the detection stage. The frames simply do not contain the information needed for classification. For quantitative information-theoretic metrics confirming this picture, we refer to main text Fig.~2C, which shows mutual information (MI) and Fisher discriminant ratio (FDR) as functions of $N_{\text{det}}$. Both metrics decline sharply below $N_{\text{det}} \sim 10^4$, consistent with these visualizations.
This confirms the intuition from \ref{sec:paradigms}: without any optical transformation to concentrate information, direct imaging is vulnerable to the detection bottleneck.

\subsection{Fixed optical transformations}
\label{subsec:fixed_transforms_sim}

Physical objects are spatially continuous: any measurement system necessarily projects this continuous distribution onto a finite set of detectors. Standard imaging performs this projection through pixelization---binning the continuous light field into discrete spatial regions. When pixel resolution is sufficiently fine, this binning is assumed to preserve all information relevant for downstream tasks.

The simplest modification to direct imaging is \emph{spatial binning}: summing adjacent pixels to reduce the measurement dimensionality from $d_{\text{obj}}$ to some smaller $d_\text{f}$. While straightforward, this approach discards high-frequency spatial information that may carry discriminative features.

More generally, one can apply arbitrary linear transformations before detection. Compressive sensing~\cite{candes2006robust,donoho2006compressed} formalizes this idea: rather than localized measurements (pixels), each detector integrates a weighted combination of the entire input field. With appropriate measurement patterns, compressive sensing can capture signal information more efficiently than direct imaging, particularly for signals with sparse structure.

As discussed in \ref{sec:paradigms}, fixed transformations are fundamentally limited because they cannot adapt to the specific sensing task. The patterns are chosen \emph{a priori} without knowledge of which features matter for the downstream objective. We therefore expect fixed transformations to underperform learned approaches, and indeed this gap is large enough that we did not include fixed transformations in the main text comparisons. Nevertheless, showing explicit results provides useful context.

\subsubsection*{Compressive sensing: implementation and results}

Compressive sensing requires random or structured measurement patterns. We implemented random binary patterns as a representative example: each element of the measurement matrix $W \in \mathbb{R}^{d_\text{f} \times d_{\text{obj}}}$ is independently set to 0 or 1 with equal probability. This is a standard choice in the compressive sensing literature; other pattern families (Hadamard, Gaussian, etc.) exhibit qualitatively similar behavior.

\begin{figure}[htbp]
    \centering  \includegraphics[width=0.9\textwidth]{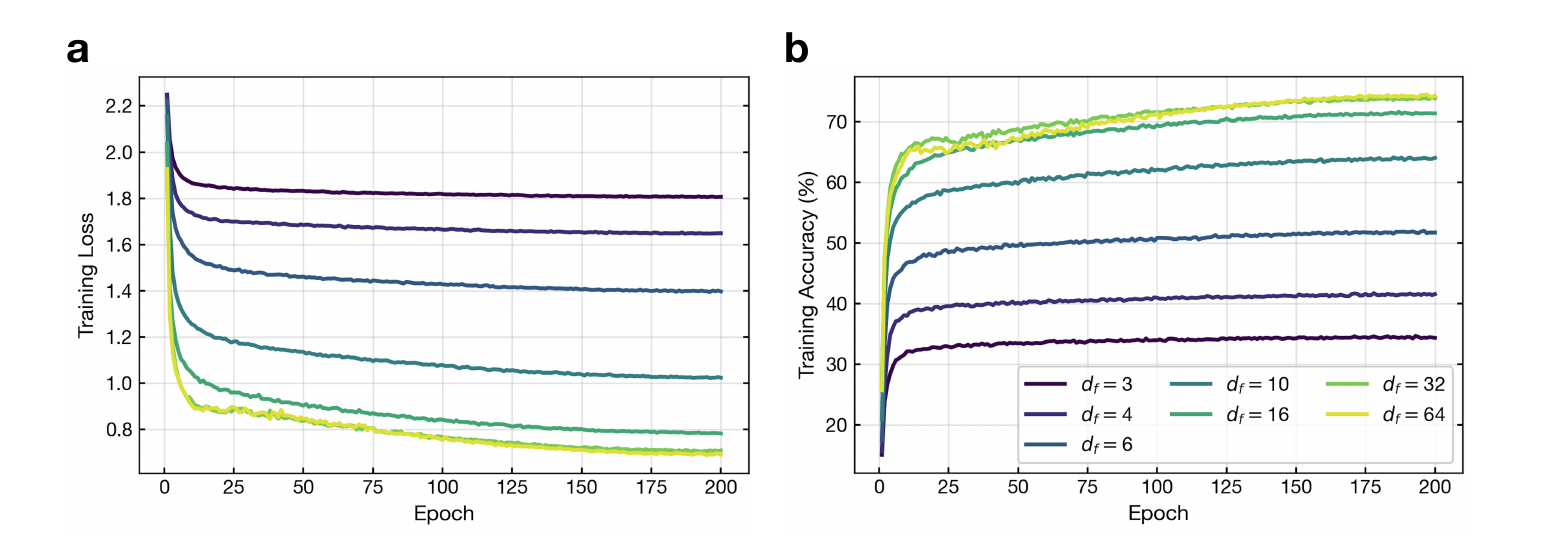}
    \caption{\textbf{Training compressive sensing classifiers.} Training loss (left) and accuracy (right) versus epoch for random binary patterns at different $d_\text{f}$ values. Models are trained on clean (high photon count) measurements.}
    \label{fig:cs_training}
\end{figure}

We trained the digital back end (CNN classifier) on \emph{clean} measurements (high photon budget) to isolate the representational capacity of the fixed patterns from photon noise effects. Testing then proceeds across different $N_{\text{det}}$ values using $\eta$ scaling, identical to the PANS evaluation protocol. \ref{fig:cs_training} shows training curves for different $d_\text{f}$ values, confirming that the classifiers successfully learn from the compressive measurements when photon counts are abundant.

\subsubsection*{Results}

\begin{figure}[htbp]
    \centering
    \includegraphics[width=0.7\textwidth]{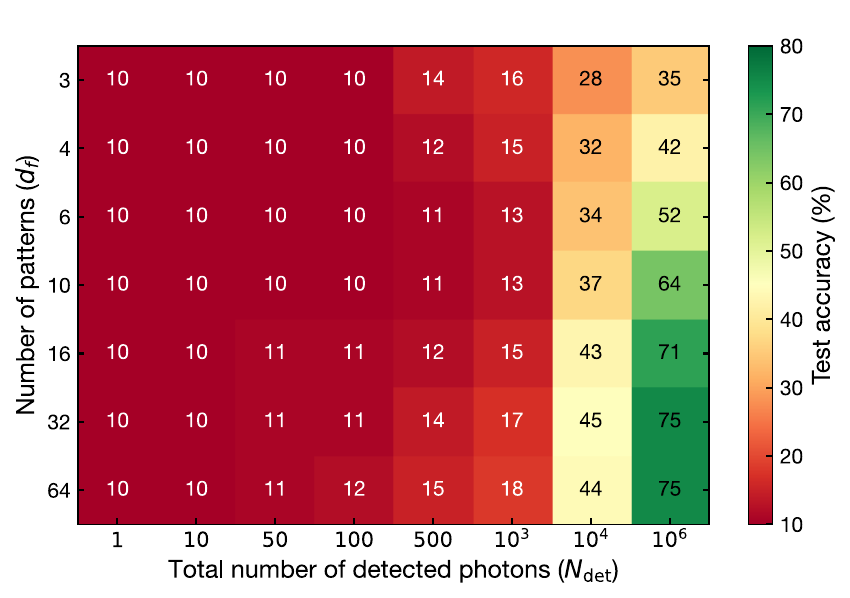}
    \caption{\textbf{Compressive sensing performance.} Test accuracy (\%) as a function of number of patterns ($d_\text{f}$, rows) and photon budget ($N_{\text{det}}$, columns). Performance collapses in the few-photon regime regardless of $d_\text{f}$.}
    \label{fig:cs_heatmap}
\end{figure}

\ref{fig:cs_heatmap} presents test accuracy as a function of both $d_\text{f}$ and $N_{\text{det}}$. At high photon budgets ($N_{\text{det}} = 10^6$), larger $d_\text{f}$ yields better accuracy, reaching $\sim$75\% with $d_\text{f} = 64$. This makes sense: more measurements capture more information about the object, and with abundant photons, detection noise is negligible.

However, performance collapses uniformly in the few-photon regime. At $N_{\text{det}} \leq 100$, accuracy remains near random guessing (10\%) for all $d_\text{f}$ values tested. Increasing $d_\text{f}$ does not help---in fact, spreading a fixed photon budget across more detectors reduces the signal per detector. The fundamental issue is that fixed patterns cannot adapt to concentrate photons on task-relevant features. Without knowing which spatial locations matter for classification, the measurement patterns waste photons on uninformative regions.

This result highlights the limitation of fixed transformations: compression alone is insufficient. Even though compressive sensing reduces the measurement dimensionality, the \emph{choice} of which linear combinations to measure critically determines performance. Random patterns, while mathematically convenient, have no mechanism to prioritize discriminative features over irrelevant ones. This motivates learning the measurement patterns through end-to-end optimization, which we examine next.

\subsection{Conventional end-to-end optimization}
\label{subsec:conventional_e2e_sim}

The natural progression from fixed transformations is to \emph{learn} the measurement patterns through end-to-end optimization, adapting them to the specific task. This approach---optimizing the optical front-end jointly with the digital back-end---has been extensively explored in computational imaging and optical neural networks~\cite{sitzmann2018end,metzler2020deep,deb2022fouriernets,wang2023image}.

To isolate the contribution of photon-aware modeling, we compare PANS against conventional end-to-end optimization where illumination patterns are learned but the stochastic single-photon detection process is \emph{not} explicitly modeled during training. This baseline uses the same network architecture as PANS: identical optical front end (learned illumination patterns) and digital back end (MLP with 512 hidden units). The key difference lies solely at the detection stage: conventional end-to-end optimization usually employs deterministic expected values rather than stochastic sampling from the actual photon detection distribution. To improve noise resilience despite this deterministic approximation, we applied quantization-aware training (QAT)~\cite{wright2022deep,wang2022optical}, a standard technique that trains the model to be robust against realistic noise at inference time.

\subsubsection*{Training and evaluation}

We trained conventional end-to-end models across $d_\text{f} \in \{3, 4, 6, 8, 10, 16, 24, 32, 64\}$. For each trained model, we evaluated performance by sweeping $\eta$ to trace out continuous accuracy--$N_{\text{det}}$ curves (\ref{sec:photon_budget}). The illumination photon budget $N_{\text{illu}}$ is computed identically to PANS: the sum of all illumination pattern elements.

\begin{figure}[h]
    \centering
    \includegraphics[width=.7\textwidth]{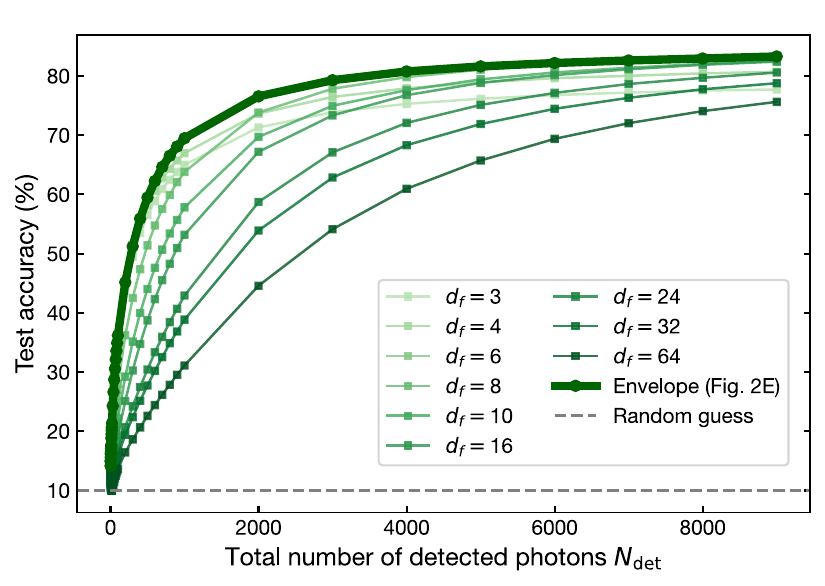}
    \caption{\textbf{Conventional end-to-end optimization results.} Test accuracy versus $N_{\text{det}}$ for different $d_\text{f}$ (light to dark curves). The thick curve shows the envelope---maximum accuracy at each $N_{\text{det}}$ across all $d_\text{f}$---corresponding to the green curve in main text Fig.~2E.}
    \label{fig:conv_e2e_results}
\end{figure}

\ref{fig:conv_e2e_results} shows test accuracy versus $N_{\text{det}}$ for different $d_\text{f}$ values, with the thick curve indicating the envelope (maximum accuracy across all $d_\text{f}$ at each $N_{\text{det}}$), corresponding to the green curve in main text Fig.~2E.

Conventional end-to-end optimization substantially outperforms both direct imaging and compressive sensing. At high photon budgets ($N_{\text{det}} \gtrsim 5000$), accuracy approaches $\sim$83\%, comparable to PANS. The learned patterns successfully capture task-relevant features, and with sufficient photons, the detection process introduces minimal information loss.

However, performance degrades as $N_{\text{det}}$ decreases. This degradation is more gradual than compressive sensing but still substantial, and a significant gap versus PANS emerges in the few-photon regime: at $N_{\text{det}} \sim 5$, conventional end-to-end achieves only $\sim$15--20\% accuracy while PANS achieves $\sim$70\% (Fig.~2E, Fig.~3A in the main text; \ref{fig:exp_Ks}). The learned patterns provide considerable benefit over untrained patterns, but without explicit modeling of the stochastic detection process, the optimization cannot fully account for how photon noise corrupts measurements at test time.

This gap emerges precisely in the regime where detection stochasticity dominates. 
This validates the third decision in our taxonomy (\ref{sec:paradigms}; \ref{fig:paradigms}): task-specific optimization helps, but photon-aware modeling becomes essential when detection is highly stochastic.
Both approaches learn task-adapted measurement patterns, but only PANS trains against the true physical detection process under extreme few-photon conditions, especially for $N_{\text{det}} \lesssim 10$. This difference becomes critical when every photon matters: PANS learns to allocate photons to maximize information flow through the exact photon detection bottleneck, while conventional approaches optimize against an approximation that diverges from reality at such low photon counts.

\subsection{Summary}
\label{subsec:sim_paradigm_summary}

The comparisons in this section establish a clear hierarchy. Direct imaging preserves full spatial information but suffers catastrophic degradation when shot noise corrupts pixel values (\ref{subsec:direct_imaging_sim}). Fixed transformations (used compressive sensing for demonstration) reduce dimensionality but cannot adapt to the task, collapsing to random guessing in the few-photon regime \ref{subsec:fixed_transforms_sim}. Conventional end-to-end optimization learns task-relevant patterns and outperforms fixed approaches, but without modeling the stochastic detection process, it cannot fully optimize in few-photon regime (\ref{subsec:conventional_e2e_sim}).

PANS addresses this gap by incorporating photon-aware modeling of single-photon detection directly into the training loop. The stochastic forward pass and gradient estimation enable the optimizer to learn patterns that maximize information preservation through the detection bottleneck---not just for an idealized model, but for the actual physical process that will actually occur at test time. 

\newpage
\section{Active PANS: Additional tasks}
\label{sec:active_pans_tasks}

The previous sections focused on FashionMNIST to develop the methodology in detail. Here we present numerical simulation results for the remaining active PANS tasks reported in the main text: MNIST digit classification, cell-organelle classification, and barcode identification. These tasks demonstrate the generality of the PANS framework across different sensing scenarios, from standard benchmarks to application-specific challenges.

\subsection{MNIST digit classification}
\label{subsec:sim_mnist}

MNIST handwritten digit recognition is a foundational benchmark in machine learning, widely used for validating new methods and hardware platforms due to its simplicity and interpretability. Reporting results for this task helps us to compare PANS to many existing work \cite{hamerly2019large,wang2022optical,sludds2022delocalized,zhu2020photon}. The dataset consists of 28$\times$28 grayscale images of digits 0--9, with 60,000 training and 10,000 test images. Compared to FashionMNIST, MNIST presents a simpler classification task with more distinct class boundaries, providing a complementary test of PANS performance.

Training followed the procedure detailed in \ref{sec:sim_training}. The digital back-end is a two-layer MLP: $d_\text{f} \to 512 \to 10$. We trained models with $d_\text{f} \in \{4, 9, 16, 25, 36, 64\}$.
This choice of squared numbers was motivated by potential spatial multiplexing arrangements in hardware implementation, though the specific values do not qualitatively affect the results.
\ref{tab:sim_mnist_k} shows test accuracy across different $(d_\text{f}, K)$ configurations, evaluated on 200 test images with 100 repetitions per configuration.

\begin{table}[h]
\centering
\renewcommand{\arraystretch}{1.3}
\setlength{\tabcolsep}{12pt}
\caption{Test accuracy (\%) on MNIST for different feature dimensions $d_\text{f}$ and shot counts $K$. Each entry shows mean $\pm$ standard deviation over 100 trials.}
\label{tab:sim_mnist_k}
\begin{tabular}{lcccc}
\hline
 & $K = 1$ & $K = 2$ & $K = 3$ & $K = 5$ \\
\hline
$d_\text{f} = 4$  & $60.28 \pm 2.08$ & $64.15 \pm 1.92$ & $65.38 \pm 1.49$ & $67.03 \pm 1.60$ \\
$d_\text{f} = 9$  & $81.23 \pm 2.03$ & $85.59 \pm 1.72$ & $86.93 \pm 1.27$ & $88.49 \pm 1.21$ \\
$d_\text{f} = 16$ & $89.96 \pm 1.42$ & $92.32 \pm 1.37$ & $93.11 \pm 1.11$ & $93.42 \pm 0.89$ \\
$d_\text{f} = 25$ & $94.39 \pm 1.35$ & $96.35 \pm 0.97$ & $96.88 \pm 0.69$ & $97.20 \pm 0.66$ \\
$d_\text{f} = 36$ & $97.28 \pm 0.98$ & $98.20 \pm 0.77$ & $98.47 \pm 0.68$ & $98.63 \pm 0.56$ \\
$d_\text{f} = 64$ & $98.46 \pm 0.75$ & $99.20 \pm 0.54$ & $99.40 \pm 0.45$ & $99.52 \pm 0.38$ \\
\hline
\end{tabular}
\end{table}

\subsection{Cell-organelle classification}
\label{subsec:cell_results}

Cell-organelle classification addresses a practical flow cytometry application where cells must be rapidly categorized based on their internal structure. This task is directly relevant to biomedical imaging, where both low illumination power (to minimize photodamage to living cells) and fast detection (to achieve high throughput in sorting applications) are essential constraints.

\begin{figure}[htp]
    \centering
    \includegraphics[width=\textwidth]{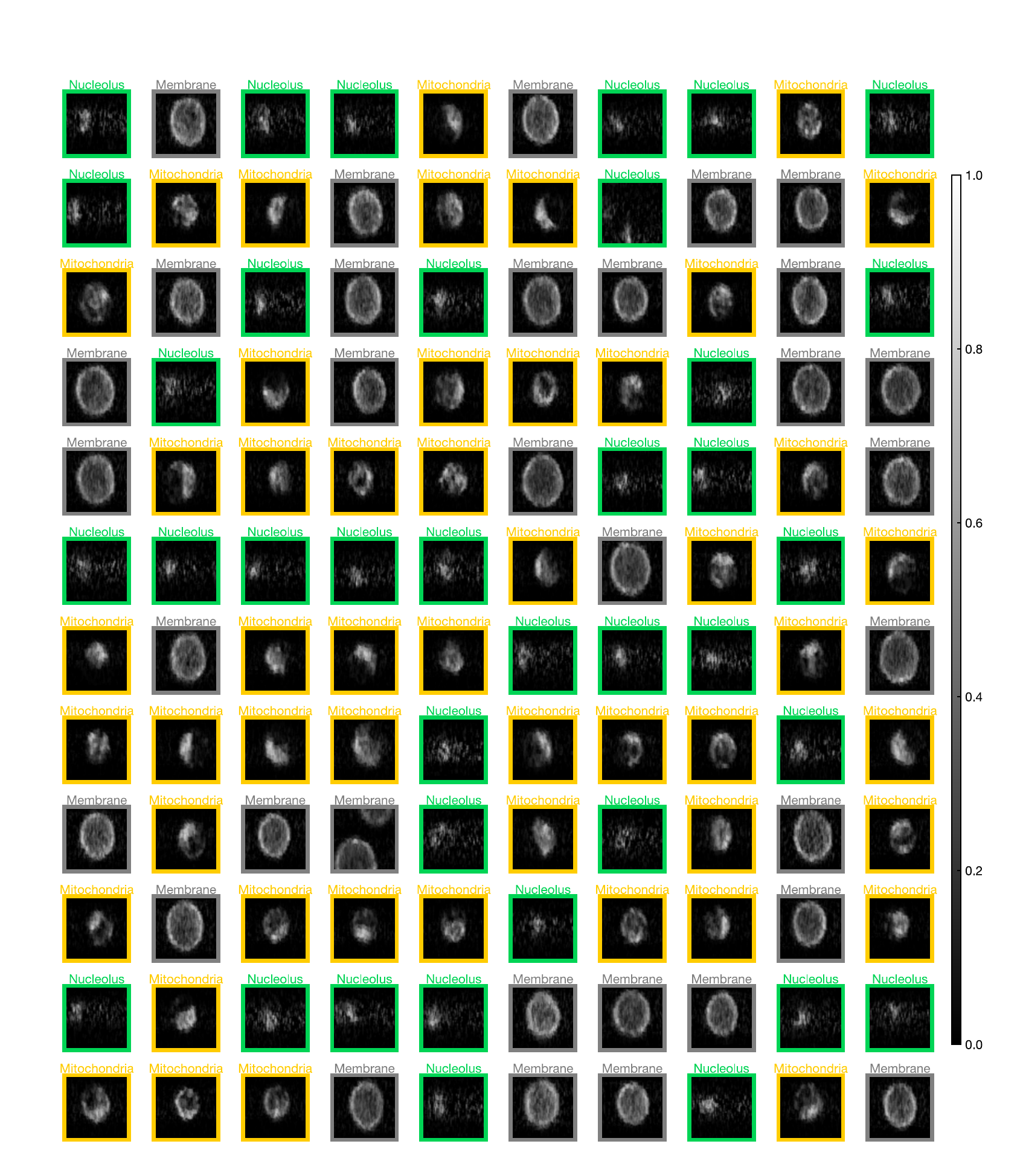}
    \caption{\textbf{Cell-organelle classification dataset.} Images randomly sampled from the test set, representing category: membrane, nucleolus, and  mitochondria. Category labels are shown above each image.}
    \label{fig:cell_examples}
\end{figure}

We use cell images derived from the dataset of Ref.~\cite{schraivogel2022high}, with four categories based on the dominant visible organelle: membrane, nucleolus, mitochondria, and a ``null'' class representing empty fields of view (no cell present). Each image is 100$\times$100 grayscale pixels, with 10,000 training images and 1,000 test images per class.
\ref{fig:cell_examples} shows representative examples randomly sampled from the test set.

The digital back-end is a three-layer MLP: $d_\text{f} \to 512 \to 512 \to 4$. Training follows the same procedure as previous tasks. We evaluated performance across different $d_\text{f}$ values and dark count rates (DCR) to assess robustness under realistic noise conditions.

Results are shown in main text Fig.~4B. Each data point reports the mean and standard deviation over 100 repetitions. Active PANS achieves $\sim90\%$ accuracy with $\sim$5 detected photons while maintaining robustness to DCR levels up to 10\%, demonstrating practical viability for photon-sensitive biological applications.

\subsection{Barcode ``1010'' identification}
\label{subsec:barcode_results}

The barcode identification task tests pattern recognition at arbitrary spatial locations---a challenge distinct from global image classification. The goal is to determine whether a 10-bin barcode segment contains the target sequence ``1010'' at any position.

\begin{figure}[h]
    \centering
    \includegraphics[width=\textwidth]{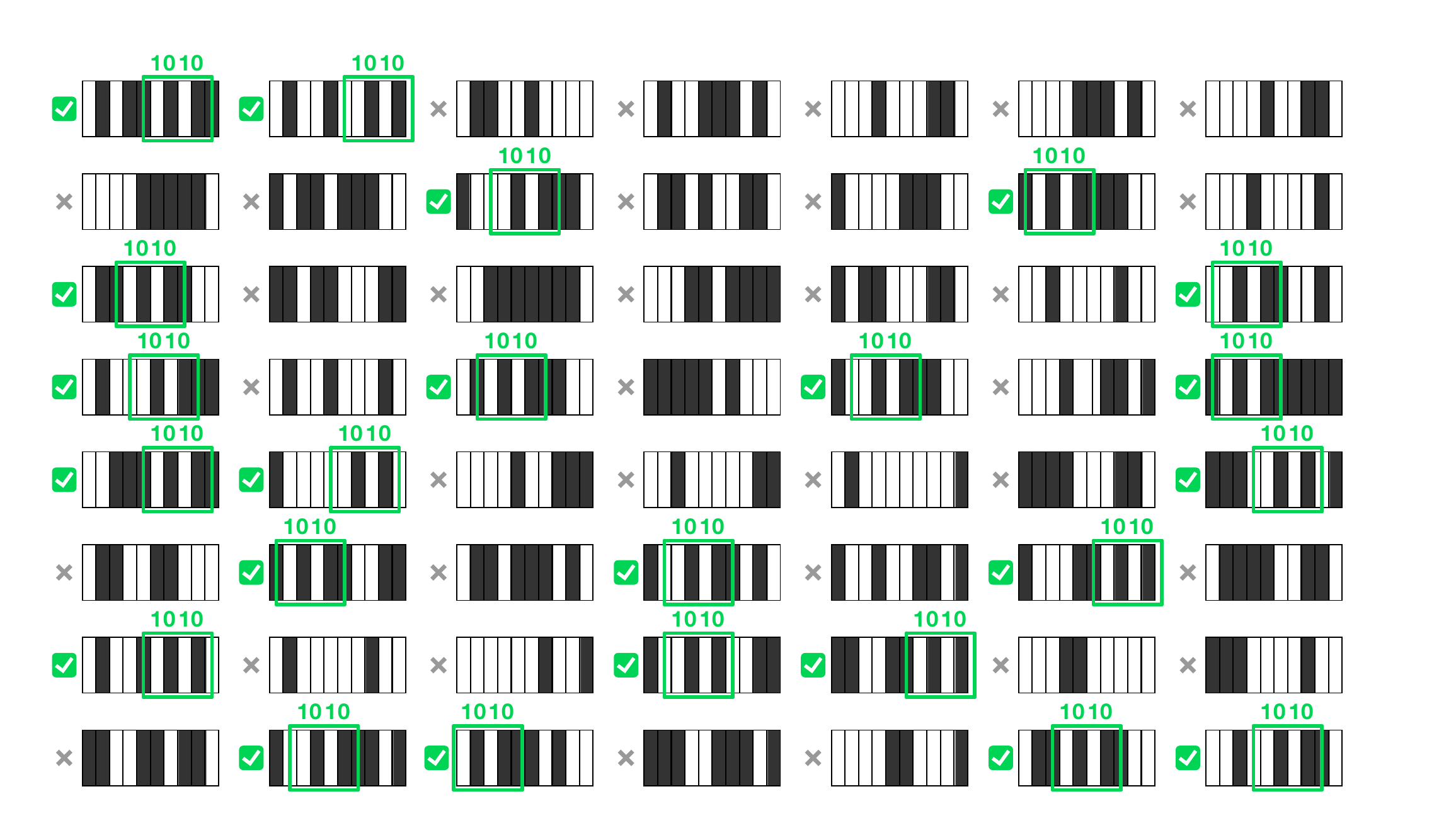}
    \caption{\textbf{Barcode ``1010'' identification dataset.} Example 10-element barcode segments randomly sampled from the test set. Positive examples (checkmark) contain the ``1010'' pattern at some position (highlighted in green box). Negative examples (cross) do not contain the target pattern. The pattern can appear at any location within the segment, requiring position-invariant feature detection.}
    \label{fig:barcode_examples}
\end{figure}

Inputs are binary barcodes of length 10 (i.e., $d_{\text{obj}} = 10$), with 4,000 training samples and 1,000 test samples. The task is binary classification: positive examples contain the ``1010'' pattern somewhere within the 10 bars, while negative examples do not. This task is nontrivial because the target pattern can appear at multiple positions and must be distinguished from similar sequences.
\ref{fig:barcode_examples} shows example inputs from both classes.

The digital back-end uses a single hidden layer: $d_\text{f} \to 512 \to 2$. Results are shown in main text Fig.~4E, with each data point reporting mean and standard deviation over 100 repetitions across different DCR values.

This task illustrates an important point: PANS's advantage does not arise solely from dimensionality reduction. Unlike image classification where $d_\text{f} \ll d_{\text{obj}}$, here the number of detectors $d_\text{f}$ can \emph{exceed} the input dimension $d_{\text{obj}} = 10$. Yet PANS still dramatically outperforms direct imaging at equivalent photon budgets. The advantage comes from learning illumination patterns that extract task-relevant features---in this case, patterns sensitive to the ``1010'' sequence regardless of position---rather than from dimension reduction. Active PANS achieves near-perfect accuracy ($>$99\%) with only $\sim$5 detected photons, a regime where direct imaging fails entirely.

\newpage
\section{Passive PANS: Diverse sensing applications}
\label{sec:passive_pans_tasks}

Active PANS controls the illumination patterns projected onto objects, optimizing how we probe the scene. Many sensing scenarios, however, involve optical signals from sources we cannot control---scattered light through turbid media, astronomical observations, monitoring reflected or transmitted light from passive objects. Passive PANS addresses these scenarios by applying learnable transformations to the \emph{incoming} optical field before detection.

\ref{sec:passive_pans} detailed the mathematical formulation and training procedure for passive PANS with linear optical processors. For the demonstrations in this section, we use coherent optical encoders with real-valued transmission matrices $W$---a choice motivated by experimental accessibility and compatibility with established optical platforms, as discussed in \ref{subsec:passive_coherent}. The detected intensity is the squared magnitude of the transformed field (Eq.~\ref{eq:passive_coherent_lambda}), and training follows the same photon-aware framework as active PANS.

This section presents numerical simulation results across five passive sensing tasks shown in main text Fig.~5. Unlike the active PANS demonstrations which focused on classification, passive PANS also demonstrates \emph{image reconstruction}---recovering spatial structure from scattered light with only a handful of detected photons. This reconstruction capability highlights the broader applicability of the framework beyond categorical inference, showing that PANS can preserve rich information through the detection bottleneck.

All results are from numerical simulation; each data point reports mean and standard deviation over 100 repetitions. For direct imaging baselines, we use an AlexNet-style CNN classifier. Dark count rates (DCR) are varied to assess robustness under realistic noise conditions.

\subsection{MNIST through multimode fiber: classification and reconstruction}
\label{subsec:mmf_results}

Multimode fibers (MMFs) scramble spatial information through mode mixing, transforming coherent input images into seemingly random speckle patterns at the output. This presents a challenging sensing scenario: the spatial structure of the original image is encoded in the complex interference pattern of the speckle, requiring either careful calibration or learned decoding to recover useful information.

We simulate MMF propagation using a random unitary scattering matrix that transforms input field amplitudes into output speckle patterns. The passive PANS encoder then applies a learnable linear transformation to this speckle before single-photon detection. Both classification (identifying which digit) and reconstruction (recovering the original image) are demonstrated.

\begin{figure}[htp]
    \centering
    \includegraphics[width=.98\textwidth]{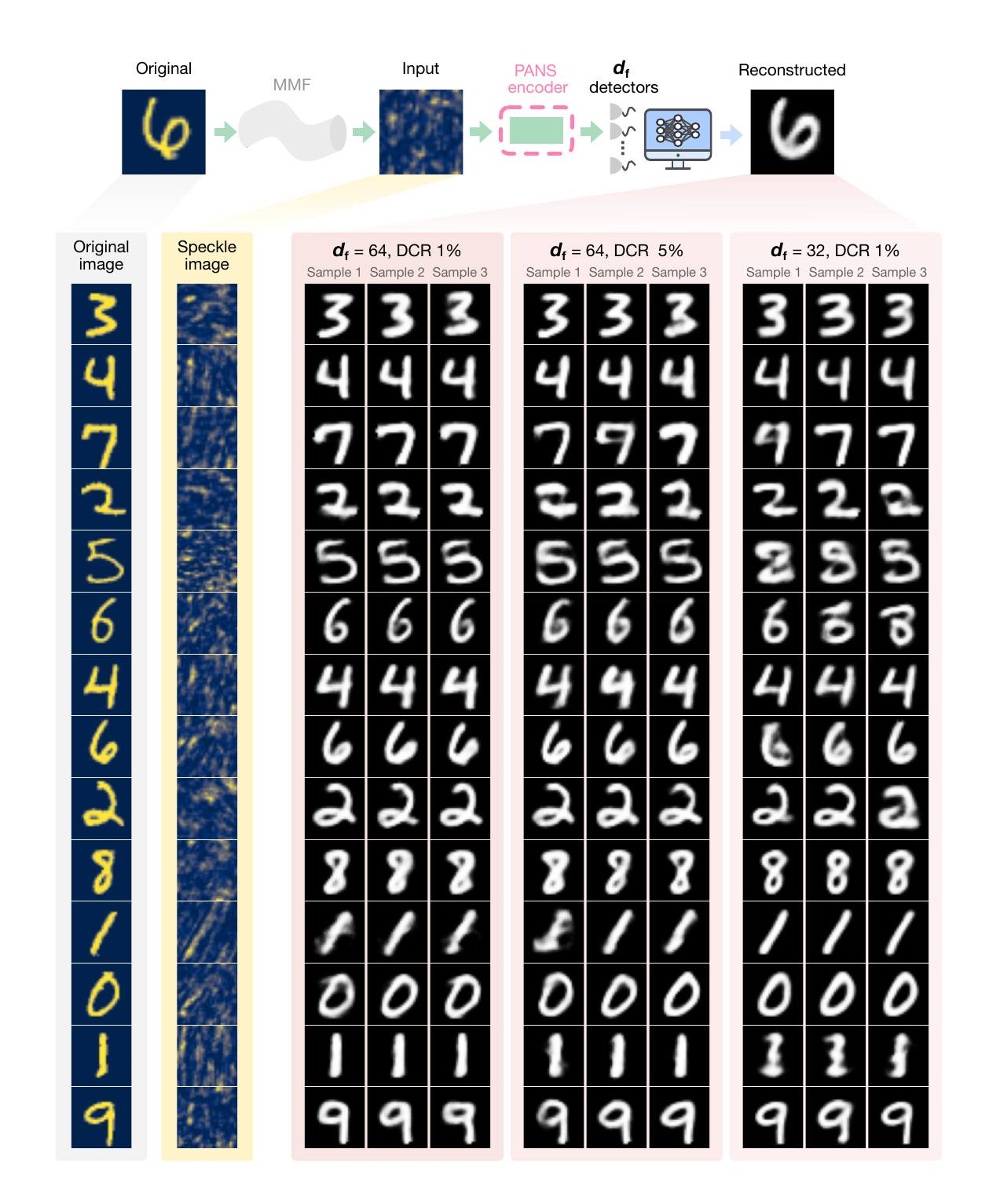}
    \caption{\textbf{Image sensing through multimode fiber.} Top: schematic showing original image $\rightarrow$ MMF propagation $\rightarrow$ speckle pattern $\rightarrow$ PANS encoder $\rightarrow$ single-photon detection (SPD) $\rightarrow$ reconstruction. Bottom: example results showing original MNIST digits (gray shade), corresponding speckle patterns after MMF propagation (yellow shade), and reconstructed images from passive PANS at different configurations ($d_\text{f} = 64$ with DCR 1\%, $d_\text{f} = 64$ with DCR 5\%, and $d_\text{f} = 32$ with DCR 1\%). Three independent samples per digit demonstrate consistency and variation across stochastic detection events. Spatial structure can be recovered even from apparently random speckle at few-photon levels.}
    \label{fig:mmf_pipeline}
\end{figure}

The input images of this task are the 28$\times$28 grayscale speckle images transformed from the original MNIST digits (see \ref{fig:mmf_pipeline} and main text Fig.~5A). For classification, the digital back-end is a two-layer MLP ($d_\text{f} \to 512 \to 10$) trained with $d_\text{f} \in \{4, 5, 6, 8, 10, 16, 24, 32\}$; training was performed with 60,000 images, and test accuracy is evaluated on 10,000 images. For reconstruction, the digital back-end is a decoder network ($d_\text{f} \to 512 \to 512 \to 784$, reshaped to 28$\times$28) trained with $d_\text{f} \in \{4, 6, 8, 10, 32, 64\}$; reconstruction quality is measured by structural similarity index (SSIM) averaged over 10,000 test images.
\ref{fig:mmf_pipeline} illustrates the sensing pipeline and shows example results across different configurations.

Classification results are shown in main text Fig.~5B. Passive PANS achieves $\sim$90\% accuracy with only $N_{\text{det}} \sim 10$ photons, while direct imaging of the speckle patterns requires hundreds of photons to exceed 50\%. This gap arises because direct speckle imaging distributes photons across the full spatial pattern, whereas the optimized passive encoder concentrates detection into task-relevant features.

Reconstruction results are shown in main text Fig.~5C--D. Passive PANS achieves SSIM $\sim$0.7 with $N_{\text{det}} \sim 10$ photons. Reconstruction is inherently more demanding than classification: the system must preserve enough information to recover spatial structure, not merely identify which of 10 categories applies. Nevertheless, meaningful image recovery is possible in regimes where direct imaging produces only noise. This demonstrates that PANS can also preserve rich, continuous information through the detection bottleneck, not just categorical distinctions.

\subsection{Transient event detection}
\label{subsec:transient_results}

Transient event detection involves identifying brief objects appearing against a fluctuating background. This scenario is relevant to diverse applications: contamination monitoring in cleanrooms, defect detection on moving conveyor belts, pest identification in food processing, and security perimeter surveillance. The challenge is that the transient signal contributes minimal energy compared to background fluctuations---the signal is ``buried'' in noise.

\begin{figure}[htp]
    \centering
    \includegraphics[width=\textwidth]{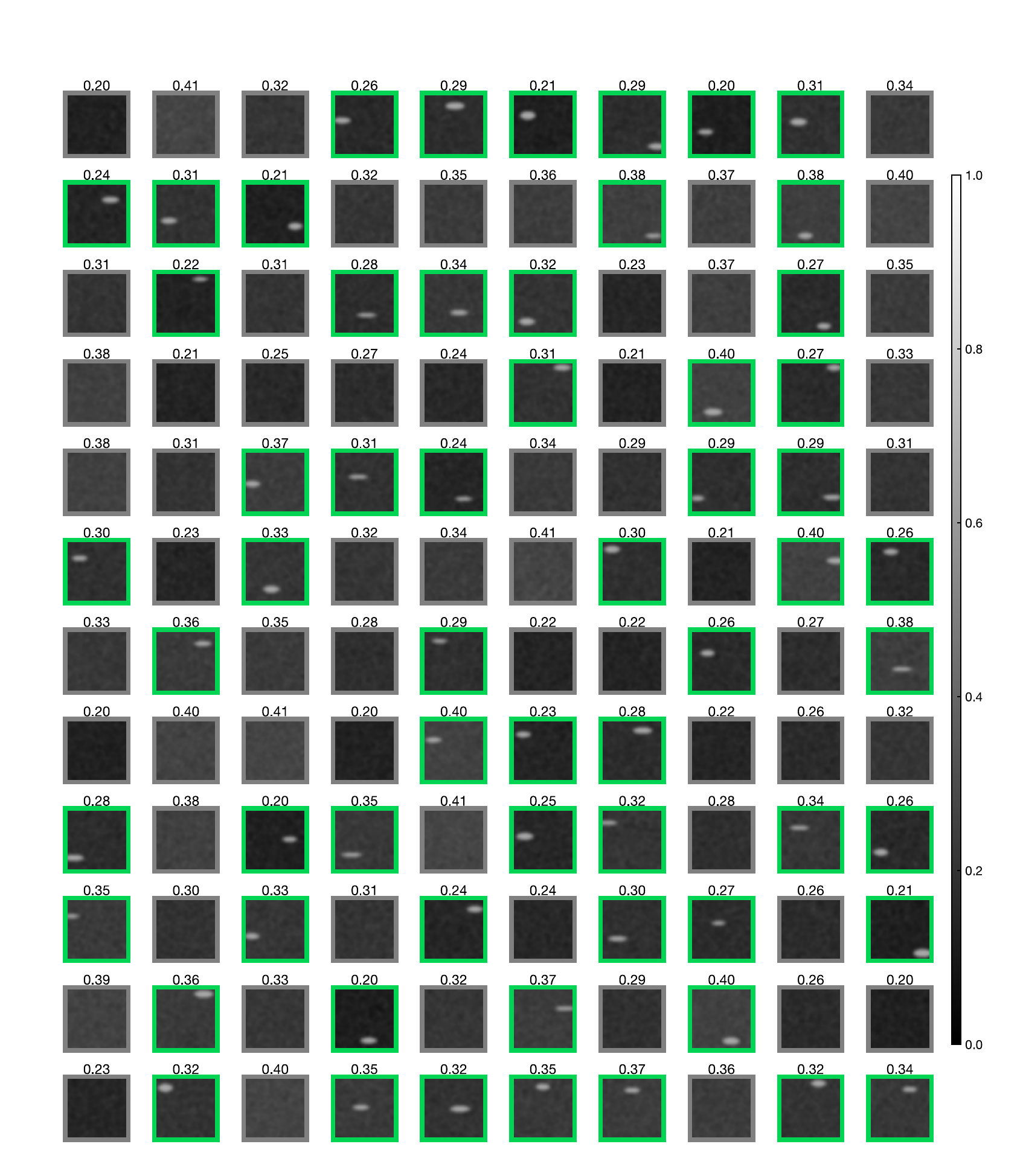}
    \caption{\textbf{Transient event detection dataset.} Example frames showing background-only (gray border) and transient-present (green border) cases. The mean pixel intensity is annotated above each frame. The similar mean values across both classes demonstrate that simple intensity thresholding cannot distinguish transient events, inferring that detection requires learning the spatial signature of the transient against background fluctuations.}
    \label{fig:transient_examples}
\end{figure}

We use synthetic data where transient objects (small patterns) appear briefly against a spatially varying, noisy background. \ref{fig:transient_examples} shows example frames with mean pixel intensity annotated.
The annotated mean values reveal why this task is nontrivial: the mean pixel intensity of transient-present frames overlaps substantially with transient-absent frames due to background variability. Simple intensity thresholding cannot distinguish the cases. Successful detection requires learning the spatial distribution characteristic of transient events---precisely what the optimized passive encoder achieves by concentrating photons onto discriminative spatial features.

The images are 64$\times$64 grayscale, with 4,000 training and 1,000 test samples. The digital back-end is a two-layer MLP ($d_\text{f} \to 512 \to 2$) trained with $d_\text{f} \in \{2, 4, 5, 6, 8, 10\}$.
Results are shown in main text Fig.~5E. Passive PANS achieves $>$95\% detection accuracy with only a few detected photons, while direct imaging fails because the transient signal is indistinguishable from background fluctuations without optimized feature extraction.

\subsection{Tissue blood flow detection}
\label{subsec:tissue_results}

Laser speckle contrast imaging detects blood flow by exploiting the temporal dynamics of scattered light: perfused tissue with flowing blood produces rapidly fluctuating speckle (appearing blurred in time-averaged images), while static ischemic tissue produces stable, high-contrast speckle. This technique has clinical applications in monitoring cerebral blood flow during neurosurgery, assessing burn depth, and evaluating tissue viability.

Conventional speckle contrast analysis requires sufficient photons to reliably estimate the speckle statistics. Passive PANS offers the potential for perfusion assessment with dramatically reduced illumination---important for minimizing tissue heating and enabling longer monitoring periods.

\begin{figure}[htp]
    \centering
    \includegraphics[width=\textwidth]{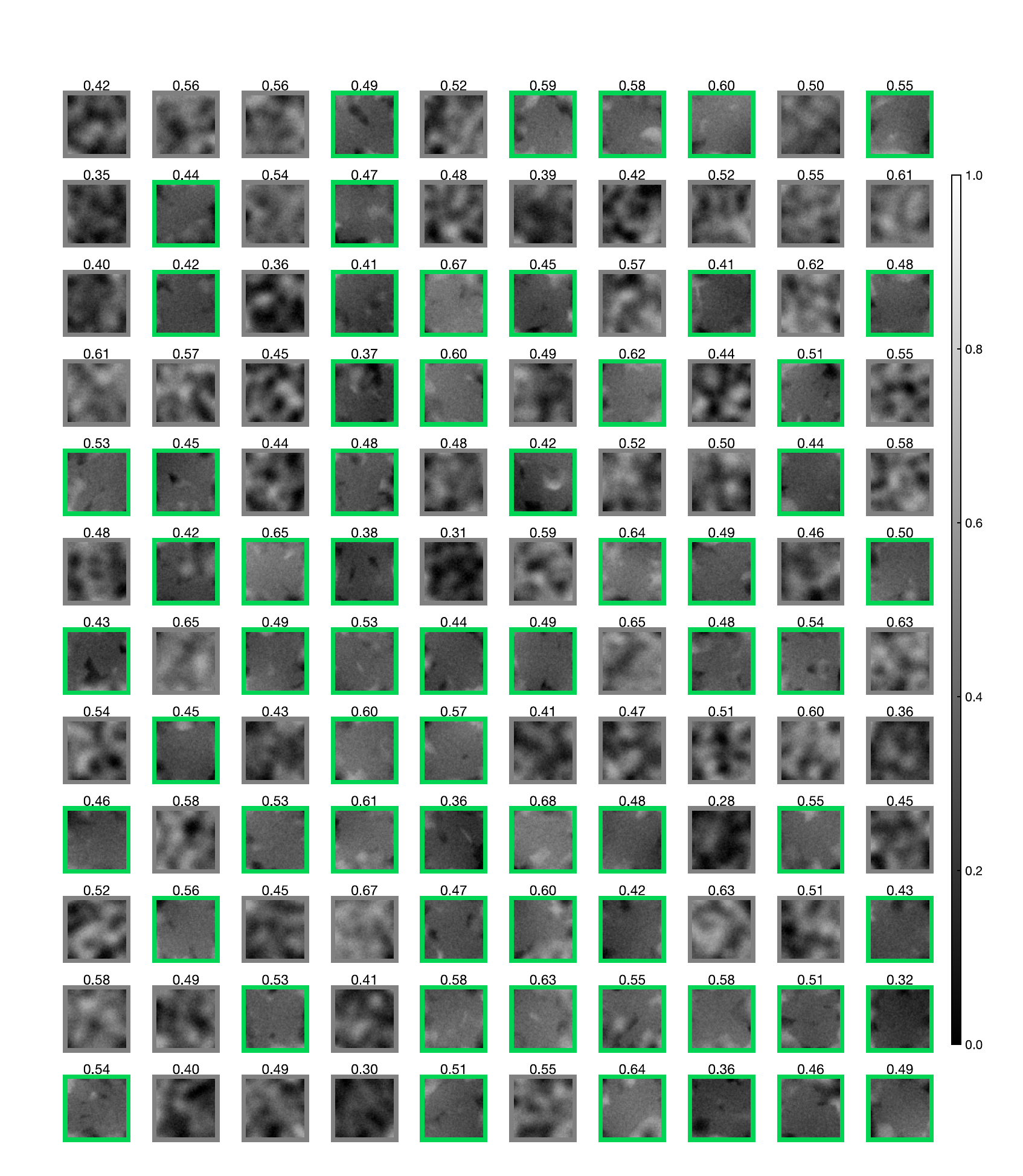}
    \caption{\textbf{Tissue speckle imaging dataset.} Images randomly sampled from the test set, showing static tissue (gray border) with high speckle contrast versus perfused tissue (green border) with reduced speckle contrast due to blood flow. The mean pixel intensity is annotated above each frame. Classification performance relies on speckle texture analysis---the contrast difference between sharp (static) and blurred (perfused) patterns---rather than intensity measurements.}
    \label{fig:tissue_examples}
\end{figure}

We use synthetic speckle images capturing the contrast difference between static and perfused tissue. More representative examples highlighting the speckle contrast difference are shown in \ref{fig:tissue_examples}. Images are 64$\times$64 grayscale, with 4,000 training and 1,000 test samples per class. The digital back-end is a two-layer MLP ($d_\text{f} \to 512 \to 2$) trained with $d_\text{f} \in \{2, 3, 6, 8, 12, 16\}$. Results are shown in main text Fig.~5F. Passive PANS enables reliable perfusion classification with photon budgets far below what conventional speckle contrast analysis would require.

\subsection{Compact nebula classification}
\label{subsec:nebula_results}

Astronomical imaging frequently operates in photon-starved conditions due to the vast distances to celestial objects and the faintness of many sources. Compact nebula classification---distinguishing planetary nebulae (shells ejected by dying stars) from emission nebulae (illuminated interstellar gas clouds)---exemplifies a task where morphological features must be extracted from extremely limited photon budgets.
These objects are typically observed through narrowband filters (H$\alpha$, [OIII], etc.) that isolate specific emission lines, providing quasi-monochromatic light suitable for coherent processing. Their compact angular sizes (arcseconds to arcminutes) produce spatially coherent wavefronts at the telescope aperture.

\begin{figure}[htp]
    \centering
    \includegraphics[width=\textwidth]{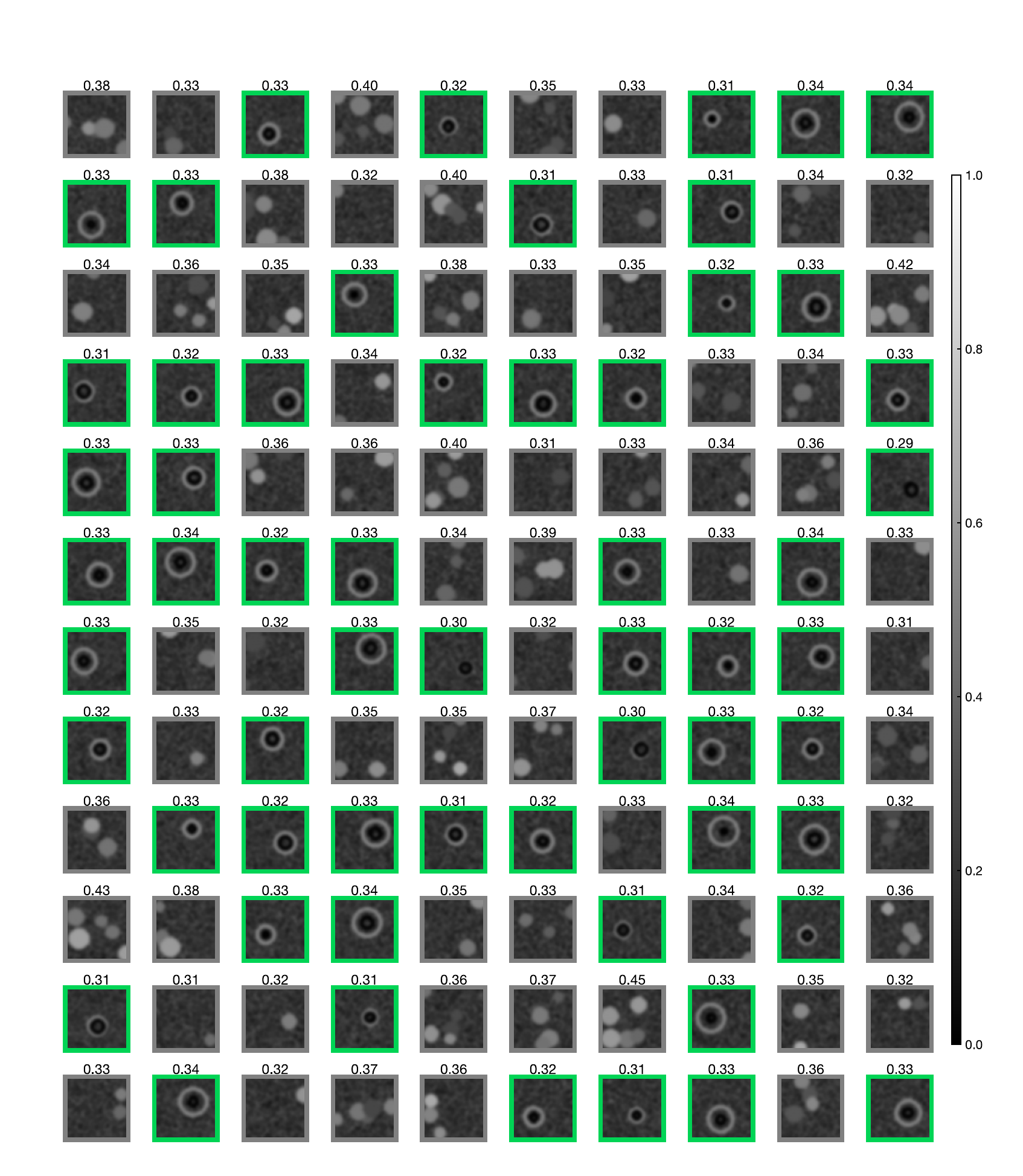}
    \caption{\textbf{Nebula classification dataset.} Images randomly sampled from the test set, showing planetary nebulae (green border), characterized by roughly circular or ring-like morphology from expanding stellar ejecta, versus emission nebulae (gray border), showing more irregular structure from illuminated interstellar medium. The mean pixel intensity is annotated above each frame. The overlapping intensity distributions demonstrate that classification requires morphological analysis, not simple photometry.}
    \label{fig:nebula_examples}
\end{figure}

We use synthetic nebula images capturing the morphological differences between classes (\ref{fig:nebula_examples}). Images are 64$\times$64 grayscale, with 4,000 training and 1,000 test samples per class. The digital back-end is a two-layer MLP ($d_\text{f} \to 1024 \to 2$, larger hidden layer due to the subtler class distinctions). We trained models with $d_\text{f} \in \{2, 3, 4, 6, 8, 10, 12\}$. Results are shown in main text Fig.~5G. Passive PANS achieves reliable classification with photon budgets representative of faint astronomical sources, demonstrating applicability to observational astronomy.

\subsection{Fiber end-face inspection}
\label{subsec:fiber_results}

\begin{figure}[htp]
    \centering
    \includegraphics[width=\textwidth]{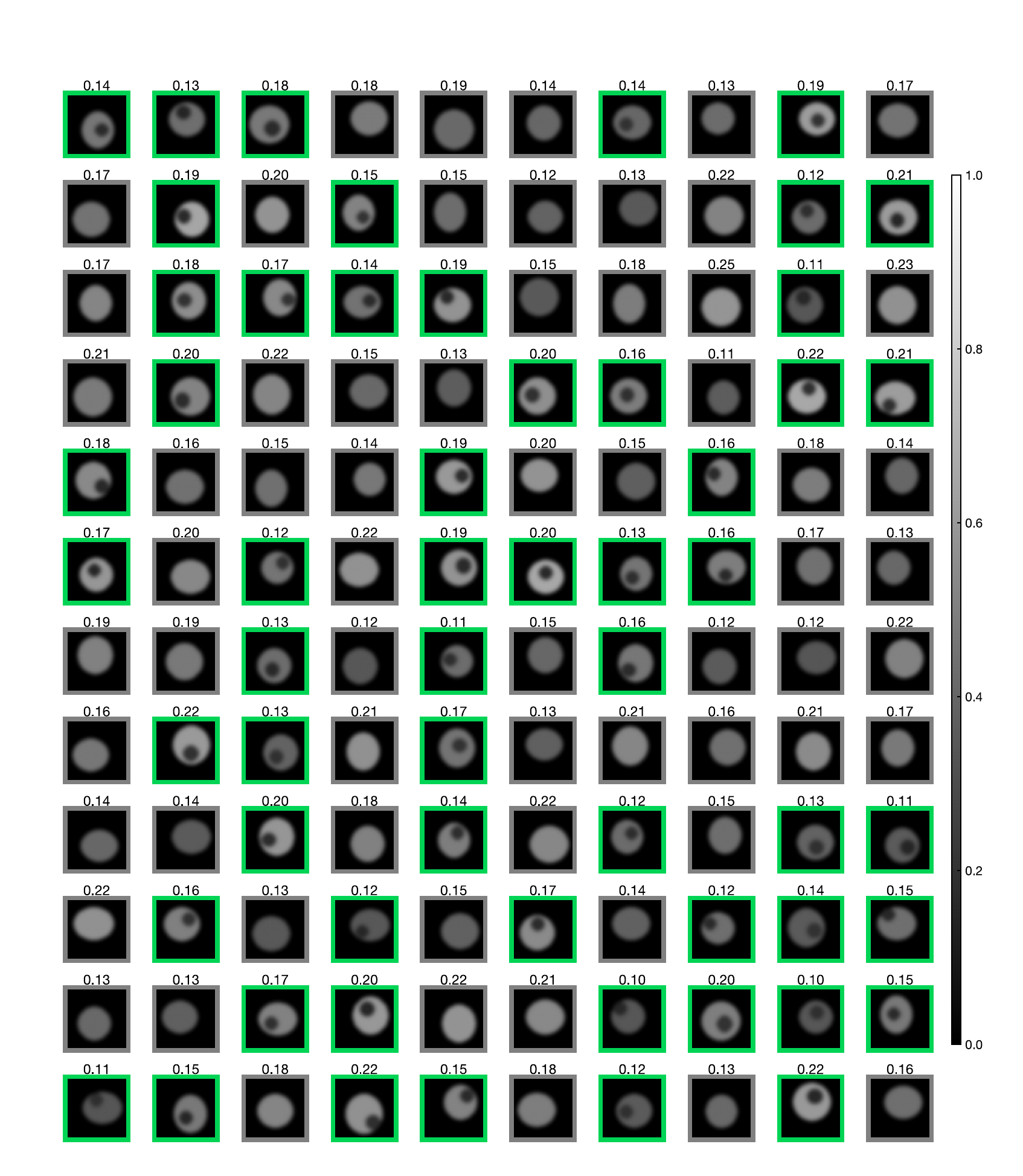}
    \caption{\textbf{Fiber end-face inspection dataset.} Images randomly sampled from the test set, showing clean fiber ends (gray border) with uniform transmission versus contaminated fiber ends (green border) with debris, scratches, or surface damage causing localized intensity variations. The mean pixel intensity is annotated above each frame. Detecting contamination requires identifying localized anomalies against the fiber's baseline profile, not intensity thresholding.}
    \label{fig:fiber_examples}
\end{figure}

The final demonstration addresses surface anomaly detection---identifying localized defects, contamination, or transient features against an otherwise expected background. We use optical fiber end-face inspection as a concrete example: detecting surface contamination (debris, scratches, dust, residue) that can degrade optical transmission quality. Clean fiber ends show uniform, predictable patterns; contaminated ends exhibit localized intensity anomalies (\ref{fig:fiber_examples}). 
The images are 64$\times$64 grayscale, with 4,000 training and 1,000 test samples (clean vs. contaminated). The digital back-end is a two-layer MLP ($d_\text{f} \to 512 \to 2$) trained with $d_\text{f} \in \{2, 3, 4, 5, 8\}$. Results are shown in main text Fig.~5H. 

This toy-model application may be relevant to telecommunications infrastructure, laboratory optical systems, and precision manufacturing quality control. A possibly useful application of passive PANS for this type of application is enabling continuous inline monitoring: by tapping only a minimal fraction of the transmitted signal, anomalies can be detected without disrupting primary operation. One might imagine similar approaches finding use in other contexts, such as semiconductor wafer inspection, detecting minute periodic dimming in stellar light curves, or general industrial surface monitoring.
In each case, the core challenge is the same: extract a weak, localized signal against background variation using minimal optical resources. Passive PANS provides a unified framework for optimizing this extraction.

\subsection{Summary and outlook}
\label{subsec:passive_summary}

Across the tasks presented here, we found that passive PANS consistently outperformed direct imaging baselines at low photon budgets, maintained reasonable performance when tested below the training photon regime, and tolerated realistic dark count rates without severe degradation. The image reconstruction results are worth noting separately, as they suggest the framework can preserve continuous spatial information through the detection bottleneck, not just categorical labels.

We emphasize that the coherent linear encoders used throughout this section represent one particular choice, selected because such transformations are already well-established in optical hardware. Other front-end implementations---complex-valued transformations, intensity-based encoders for incoherent sources, nonlinear optical operations, or even quantum optical states and novel sensing materials---could in principle be incorporated, though we did not explore these directions here.

The passive PANS demonstrations in this section complement the active PANS results of \ref{sec:active_pans_tasks}, together illustrating the flexibility of the photon-aware optimization framework. As discussed in \ref{subsec:pans_frontend}, the PANS framework is agnostic to the specific optical front end---what matters is the detection bottleneck, where physical signals are converted to digital data with inevitable information loss under photon constraints. Any programmable optical transformation before this bottleneck can be optimized using the same methodology. Active illumination and passive linear encoding are simply two concrete instantiations: they share the same optimization target (maximizing information flow through the detection bottleneck) while differing only in the physical implementation of the front end.

The consistent results we observed across our test cases are encouraging, and we hope these demonstrations provide a useful reference point for researchers interested in adapting similar ideas to their own sensing platforms.

\newpage

\part*{Part III: \\Experimental implementation}
\addcontentsline{toc}{part}{III\quad Experimental implementation}
\label{part:experiment}

\vspace{48pt}
\section{Structured illumination platform}
\label{sec:setup}

We implemented active PANS using an incoherent optical system that projects programmable illumination patterns onto an object plane and performs single-photon detection of the transmitted light. The hardware is adapted from a setup previously used for optical neural network inference~\cite{wang2022optical,ma2025quantum}; here we repurpose it as an active sensing platform. We summarize the key components and their roles; detailed characterization can be found in the earlier works.

\subsection{System overview}

\begin{figure}[h]
    \centering
    \includegraphics[width=\textwidth]{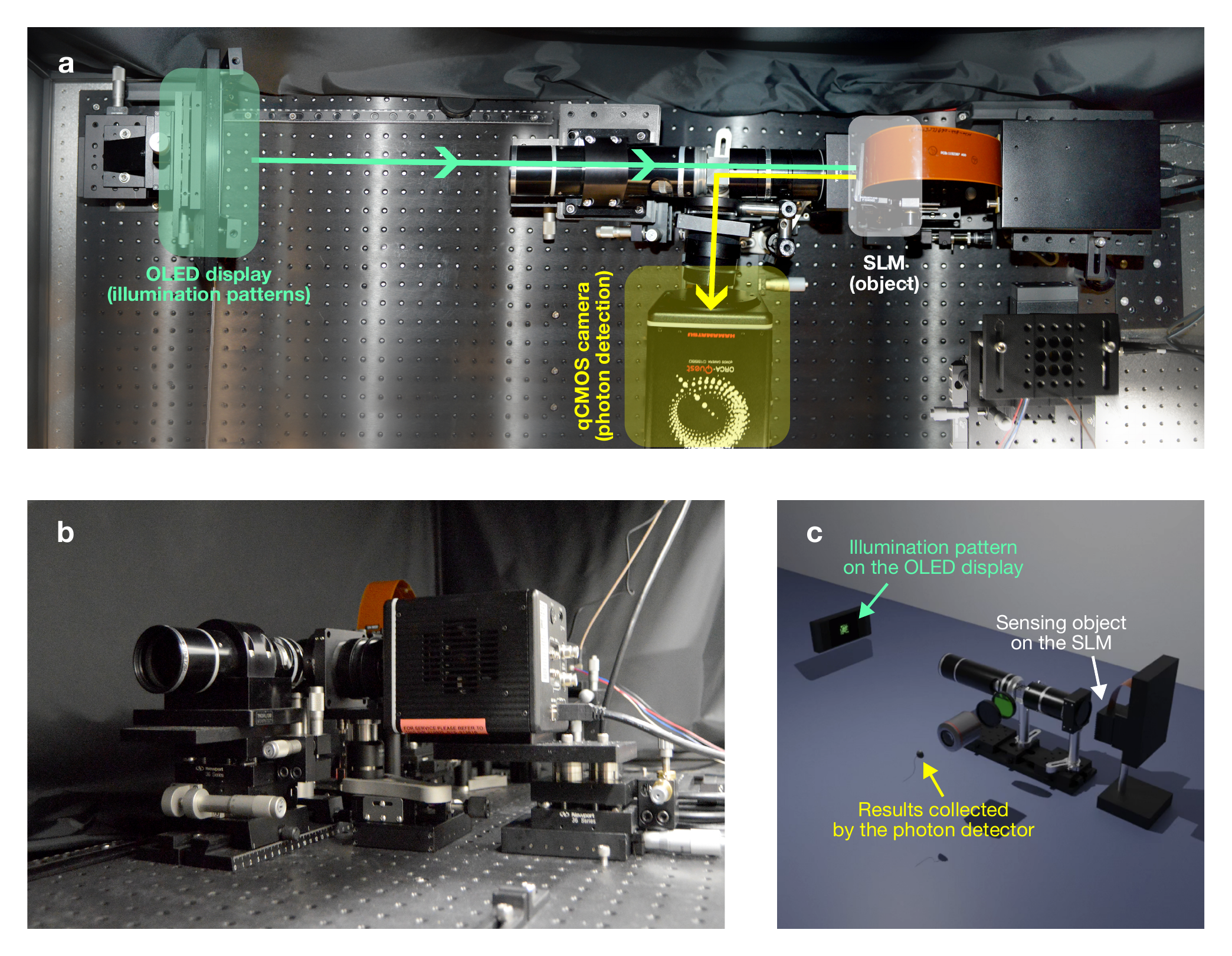}
    \caption{\textbf{Experimental apparatus for active PANS.} \textbf{a},~Top-down photograph of the optical system with the three main components highlighted: OLED display for programmable illumination patterns (green), SLM for object emulation (white), and qCMOS camera for single-photon detection (yellow). Arrows indicate the light path from illumination source through the object plane to the detector. \textbf{b},~Side-view photograph of the setup. \textbf{c},~Schematic rendering illustrating the sensing pipeline: learned illumination patterns displayed on the OLED probe the object on the SLM, and the transmitted light is measured by the photon detector to produce classification results.}
    \label{fig:setup}
\end{figure}

\ref{fig:setup} shows photographs and a schematic rendering of the apparatus. The system consists of three functional stages corresponding to the active PANS pipeline: (1)~a programmable organic light-emitting diode (OLED) display encoding the learned illumination patterns $\{\vec{w}_i\}$, (2)~a spatial light modulator (SLM) serving as the object plane where illumination interacts with the sample, and (3)~a quantitative scientific CMOS (qCMOS) camera performing single-photon detection. As illustrated by the light path in \ref{fig:setup}a, light propagates from the OLED through a zoom lens system to the SLM, then through polarization optics and spectral filtering before reaching the camera.

\subsection{Programmable illumination source}

The OLED display is taken from a commercial smartphone (Google Pixel 2016), providing $1920 \times 1080$ pixels with individually controllable intensity. We use only the green sub-pixels ($\lambda \approx 532$~nm), arranged in a square lattice with 57.5~µm pitch. The OLED's high extinction ratio and true black pixels enable high dynamic range modulation; after calibration, approximately 7-bit linear intensity control (124 levels) is achieved per pixel~\cite{wang2022optical}.

A zoom lens system (Resolv4K, Navitar) images the OLED onto the object plane, with magnification adjusted to provide one-to-one correspondence between OLED pixels and SLM pixels. Spatial uniformity across OLED pixels is characterized and corrected during calibration, ensuring consistent intensity output across the display region used in experiments.

\subsection{Object plane and intensity modulation}

For controlled experimental validation, we emulate object transmission using a reflective liquid-crystal SLM (Meadowlark P1920-500-1100-HDMI), which provides $1920 \times 1152$ pixels with 9.2~µm pitch. Combined with a half-wave plate (HWP) and polarizing beam splitter (PBS), the SLM functions as a pixel-wise intensity modulator: incident horizontally-polarized light acquires a pixel-dependent polarization rotation, and the PBS selects the vertical component, yielding intensity modulation with approximately 50:1 extinction ratio and 8-bit control resolution after calibration.

Spatial uniformity across SLM pixels is similarly characterized and corrected. This configuration allows arbitrary grayscale test images to be displayed as ``objects'' whose transmission we seek to classify.

\subsection{Single-photon detection}

The intensity-modulated light is demagnified by a telescope (Navitar rear adapter + Olympus 4$\times$ objective) and focused onto the qCMOS camera (Hamamatsu ORCA-Quest, C15550-20UP), which has $4096 \times 2304$ pixels with 4.6~µm pitch.

The qCMOS operates in ultra-quiet photon-counting mode, where readout noise is suppressed below 0.3 photoelectrons, enabling detection of individual photon events. In this mode, each camera pixel functions as an independent single-photon detector, directly outputting photon counts. At 532~nm, quantum efficiency is approximately 86\% with dark count rate $\sim$0.006 photoelectrons per pixel per second at $-35$°C. After accounting for detection imperfections, the effective detection efficiency is $\sim$68\% with effective dark count rate $\sim$0.01 photoelectrons per pixel per second.

Through optical fan-in, light from the object interaction region is focused onto a small area of the camera sensor spanning a few pixels.

\newpage
\section{Implementation of active PANS}
\label{sec:implementation}

This section describes the experimental procedures for mapping trained PANS models to the optical hardware. We implement active PANS for two classification tasks: FashionMNIST (10 clothing categories) and MNIST (10 handwritten digits), both using $28 \times 28$ pixel images. The PANS models deployed here are the same as evaluated in \ref{sec:sim_testing}.

\subsection{Normalization and parameter mapping}

Test images from FashionMNIST and MNIST are displayed on the SLM as amplitude modulation patterns. Each image $\vec{x}$ is normalized such that the maximum pixel value corresponds to maximum SLM transmission (1) and the minimum to minimum transmission (0).

The trained illumination matrix $W$ is similarly normalized so that the maximum element equals 1, corresponding to maximum OLED pixel intensity. Converting this normalized value to actual detected photon numbers requires calibration, described next.

\subsection{Photon level calibration}

The critical calibration step establishes the scaling factor $\eta$ (\ref{subsec:eta_scaling}) that converts normalized parameter values to expected detected photons. We fix the SLM to uniform maximum transmission and display random binary patterns on the OLED, where each pixel is either dark (0) or at maximum intensity (1). Since the number of illuminated pixels is known for each pattern, measuring the detected signal establishes the relationship between OLED pixel value and detected photon number. Averaging over many random patterns yields robust calibration across the field of view, accounting for any residual spatial nonuniformity.

This procedure determines the conversion factor from the normalized value ``1'' in the trained illumination patterns to expected detected photons—precisely the $\eta$ parameter that governs the operating photon budget.

\subsection{Spatial configuration}

We use one-to-one pixel mapping with no macro-pixel binning: each OLED pixel encodes one element of the illumination pattern, and each SLM pixel represents one pixel of the object. For both datasets, this corresponds to a $28 \times 28$ pixel region of interest on both devices.

This compact configuration serves two purposes. Maintaining a small region of interest reduces the effort required to ensure spatial uniformity, as correcting nonuniformities such as aberrations through additional calibration often comes at the cost of precision. Additionally, since PANS operates at single-photon levels, the available optical power from the OLED is sufficient for individual pixels without requiring macro-pixel binning to accumulate signal.

\subsection{Detector configuration}

Through optical fan-in via the objective lens, light from the entire $28 \times 28$ interaction region is focused onto a small area of the camera sensor spanning a few pixels. We select a single pixel and calibrate the system around it; this same pixel is used for all measurements, ensuring consistent detection characteristics throughout the experiment.

\subsection{Measurement protocol}

For each test image $\vec{x}$, we sequentially display the $d_\text{f}$ trained illumination patterns $\{\vec{w}_1, \vec{w}_2, \ldots, \vec{w}_{d_\text{f}}\}$ on the OLED. For each pattern, we acquire 30 independent single-shot measurements, recording the binary detection outcome (1 for photon detected, 0 for no detection). Exposure times range from approximately 0.6 to 3~ms depending on the target photon budget.

The 30 repeated measurements sample the same optical configuration multiple times, which is standard practice for single-photon detection and allows characterization of the stochastic detection process. The complete measurement for one test image yields a $d_\text{f} \times 30$ matrix of binary values, representing 30 independent single-shot feature vectors that can each be used for classification.

We evaluate 200 test images for each $d_\text{f}$ configuration, yielding a total of $200 \times d_\text{f} \times 30$ binary measurements per configuration.

\section{Evaluation of experimental results}
\label{sec:evaluation}

Following the measurement protocol described in \ref{sec:implementation}, we now present the collected experimental data and evaluate classification performance on FashionMNIST and MNIST.

\subsection{Visualization of collected SPD readouts}

\begin{figure}[htpb]
    \centering
    \includegraphics[width=.92\textwidth]{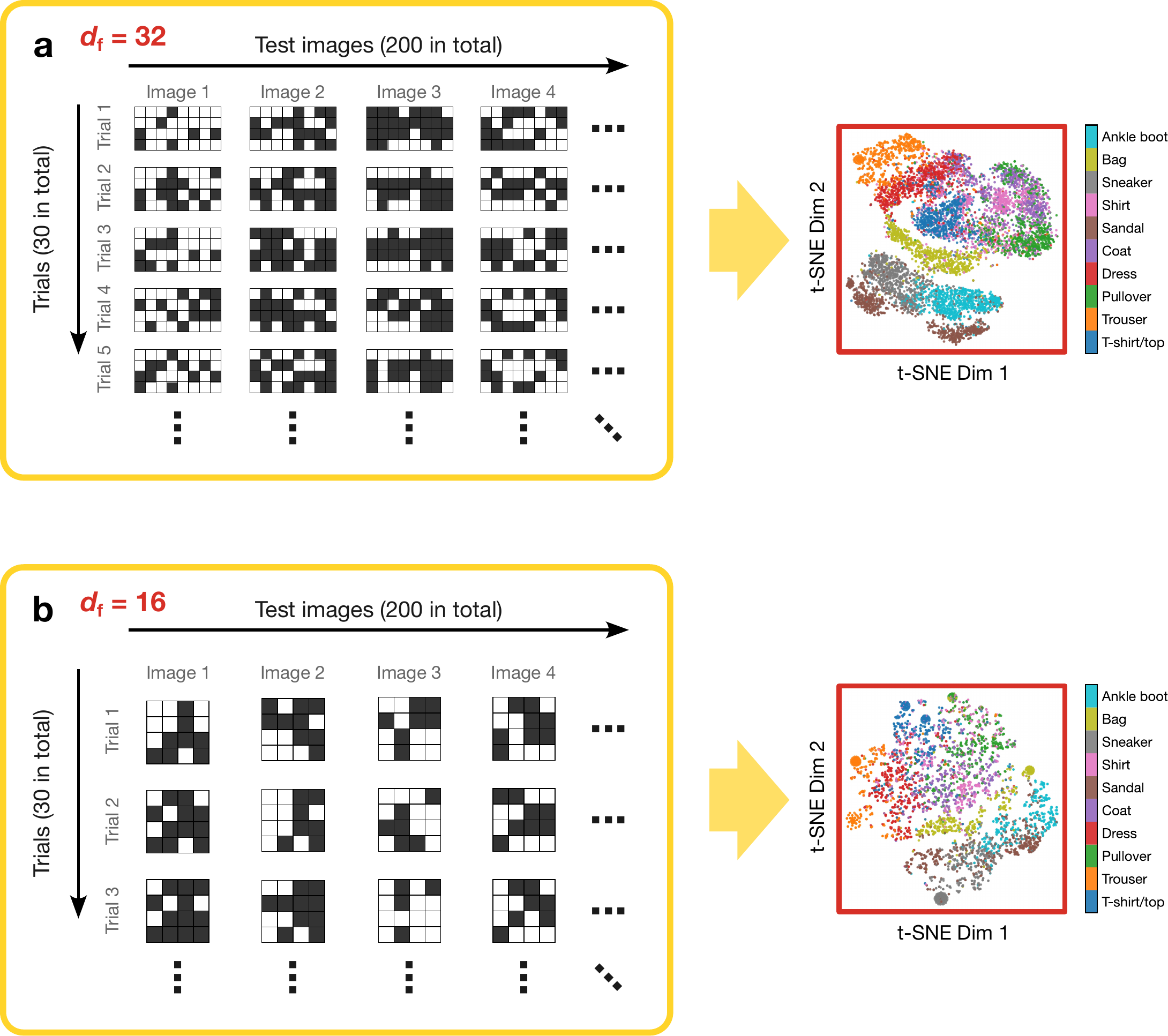}
    \caption{\textbf{Experimentally collected single-photon detection data for FashionMNIST.} \textbf{a}, Data for $d_\text{f} = 32$: each $8 \times 4$ grid displays the binary detection outcomes from one single-shot measurement, with white indicating photon detected and black indicating no detection. Columns correspond to different test images; rows correspond to repeated trials of the same image. \textbf{b}, Data for $d_\text{f} = 16$: each $4 \times 4$ grid represents one measurement. Right panels show t-SNE visualizations of the feature distributions across all test images, with colors indicating class labels.}
    \label{fig:raw_activations}
\end{figure}

\ref{fig:raw_activations} displays representative examples of the binary detection data collected from the single-photon detector. For each test image, the $d_\text{f}$ binary values from one single-shot measurement are arranged into a grid for visualization: an $8 \times 4$ grid for $d_\text{f} = 32$ or a $4 \times 4$ grid for $d_\text{f} = 16$, where each cell corresponds to one illumination pattern and shows whether a photon was detected (white) or not (black). Results from different test images are shown across columns, with different trials (repeated measurements of the same test image) shown across rows.

An interesting feature of these measurements is their high trial-to-trial variability. Even for the same input image measured under identical conditions, the binary detection patterns differ substantially across the 30 repeated trials. This is a direct consequence of the detection bottleneck in the few-photon regime: with expected detection counts below one photon per pattern on average, each single-shot measurement is dominated by shot noise. 
This stochasticity is not a limitation of our implementation but a fundamental physical constraint: with such limited photon budgets, high signal-to-noise detection is simply not possible, and PANS is designed precisely to address this regime.

This behavior differs notably from conventional machine learning inference, which is typically deterministic: the same input produces the same output. Here, the intrinsic randomness of photon detection at the detection bottleneck means that repeated measurements of the same object yield different feature vectors, and consequently may produce different classification outcomes. At first glance, such variability might seem to preclude reliable sensing.

However, the t-SNE visualizations in the right panels of \ref{fig:raw_activations} reveal that task-relevant information remains well preserved. Each point represents one single-shot feature vector, colored by the true class label of the corresponding test image. Despite the apparent randomness of individual measurements, distinct class clusters emerge: feature vectors from images of the same category cluster together, while different categories occupy separate regions of the feature space (as in Fig.~2F in the main text and \ref{fig:tsne_direct_imaging}). This demonstrates that the optimized illumination patterns successfully encode discriminative information into the statistics of photon detection. Even though any single realization appears highly noisy due to the stochastic detection process, the underlying class structure is retained and can be extracted by the digital back end. This is precisely what PANS achieves by optimizing the optical front end to maximize information flow through the detection bottleneck.

\subsection{From feature vectors to classification}

\begin{figure}[htpb]
    \centering
    \includegraphics[width=.8\textwidth]{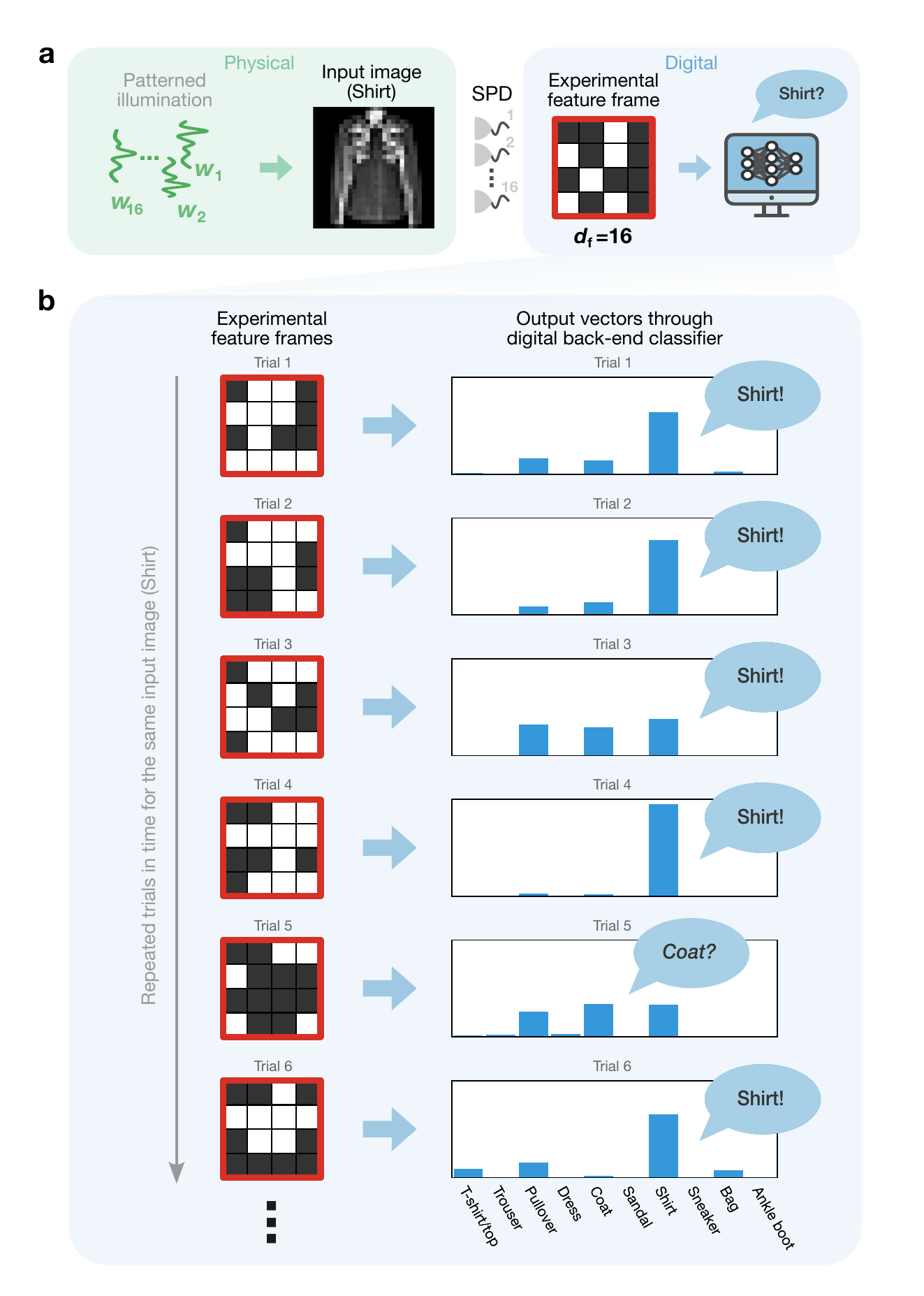}
    \caption{\textbf{Single-shot inference pipeline with experimental data.} \textbf{a}, Diagram of the inference process (compare to Fig.~2D in the main text): $d_\text{f} = 16$ trained illumination patterns probe the input image (a ``Shirt'' from FashionMNIST), single-photon detection produces a $4 \times 4$ binary feature frame, and the digital back end outputs a 10-dimensional vector for classification. \textbf{b}, Six trials randomly sampled from the 30 repeated measurements of the same input image. Left column: experimentally collected feature frames showing trial-to-trial variability. Right column: corresponding output vectors from the digital classifier, with the predicted class indicated. Five of six trials correctly predict ``Shirt''; one trial (Trial 5) incorrectly predicts ``Coat''.}
    \label{fig:inference_pipeline}
\end{figure}

To obtain classification results, each single-shot feature vector is processed by the trained digital back end (MLP). \ref{fig:inference_pipeline} illustrates this inference pipeline using $d_\text{f} = 16$ on FashionMNIST as a representative example.

\ref{fig:inference_pipeline}a shows the inference diagram, following the same structure as Fig.~2D in the main text: the trained illumination patterns $\{w_1, w_2, \ldots, w_{16}\}$ sequentially probe the input image, single-photon detection produces a binary feature vector (displayed as a $4 \times 4$ grid), and the digital classifier maps this to a 10-dimensional output vector where each dimension corresponds to one class.

\ref{fig:inference_pipeline}b displays six trials randomly sampled from the 30 repeated measurements of the same shirt image. The feature frames vary visibly across trials due to the stochastic detection process, yet the corresponding output vectors consistently peak at the correct class (Shirt) in most cases. Trial 5 produces an incorrect prediction (Coat), illustrating that classification errors do occur—expected given the model accuracy of 77.2\% for $d_\text{f}=16$ (\ref{tab:exp_fmnist_K}). This example demonstrates the characteristic behavior of PANS inference: despite substantial variability in the raw detection data, the jointly optimized system achieves correct classification in the majority of trials.

The output vectors also exhibit some variation in their numerical values across trials, though this variation is less pronounced than in the raw feature frames. The digital back end maps the binary features into a learned representation space, which partially stabilizes the outputs. Nevertheless, the variation is visible, and for classification what matters is which dimension achieves the highest value rather than the exact magnitudes.

\subsection{Output distribution across trials}

\begin{figure}[htpb]
    \centering
    \includegraphics[width=.85\textwidth]{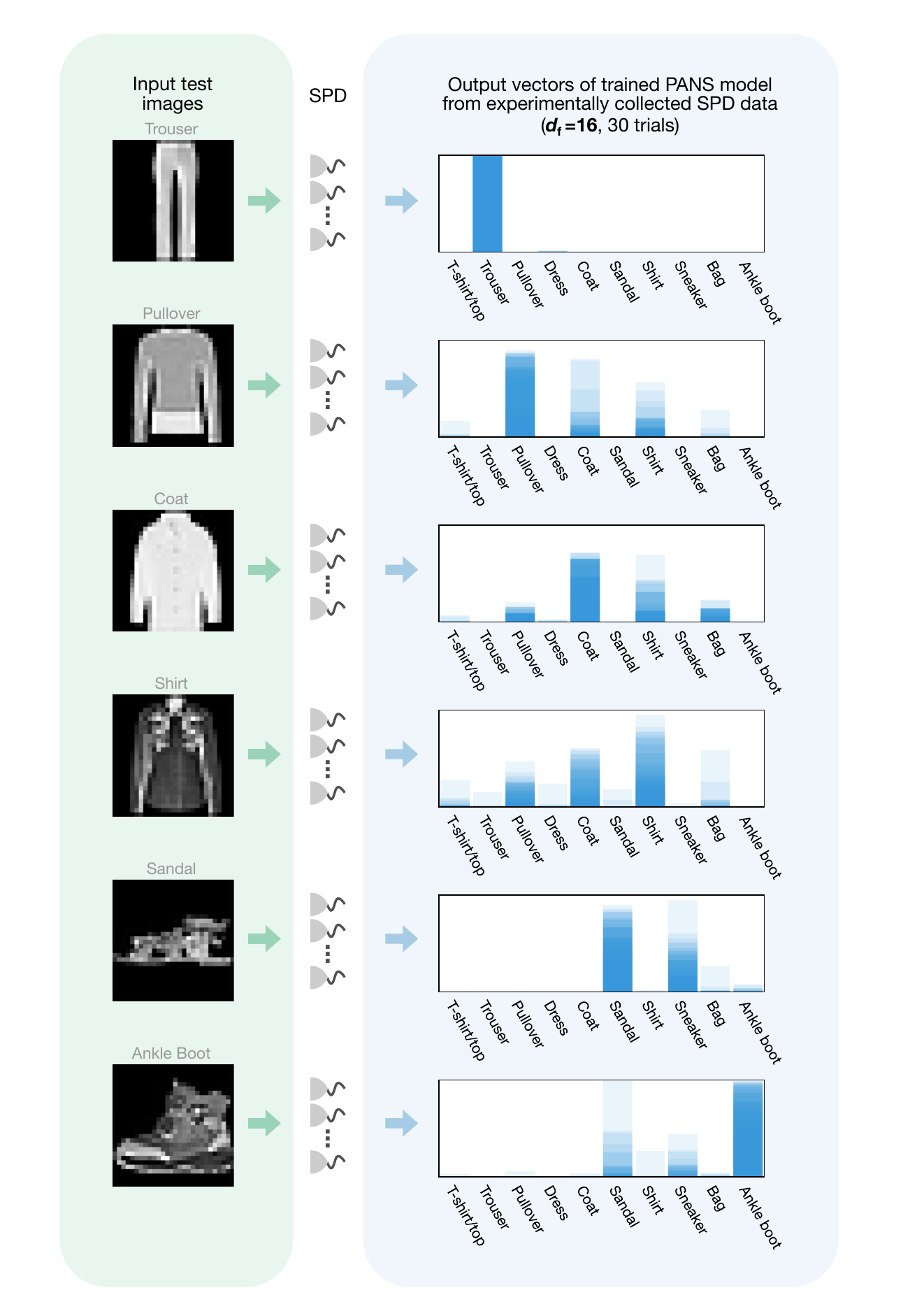}
    \caption{\textbf{Distribution of classifier outputs across 30 trials for representative test images.} Each row corresponds to one input image from FashionMNIST (with corresponding category labeled on the top). For each image, the output vectors from all 30 experimentally implemented single-shot inferences ($d_\text{f} = 16$) are overlaid with transparency to show the density of outputs across trials.}
    \label{fig:output_distributions}
\end{figure}

The six trials in \ref{fig:inference_pipeline}b provide a glimpse of the inference behavior for one input image. To better characterize the distribution of classifier outputs, \ref{fig:output_distributions} shows all 30 experimentally collected trials for several representative test images at $d_\text{f} = 16$.

For each input image, all 30 output vectors are plotted with low opacity so that overlapping regions appear darker, revealing the distribution across trials. Trouser shows a highly concentrated distribution: the correct class consistently receives a high value while all others remain near zero. Categories with similar visual appearance show greater spread. The shirt image has the correct class receiving the highest output in most trials, but with noticeable probability mass at Coat and Pullover—consistent with the incorrect Coat prediction seen in \ref{fig:inference_pipeline}b. Pullover and Coat similarly show mutual confusion, while the footwear categories (Sandal, Sneaker, Ankle boot) exhibit overlap among themselves as well.

These patterns preview the confusion structure quantified in the next subsection, where we aggregate results across all 200 test images rather than examining individual examples.

\subsection{Confusion matrices}

\begin{figure}[htp]
    \centering
    \includegraphics[width=.92\textwidth]{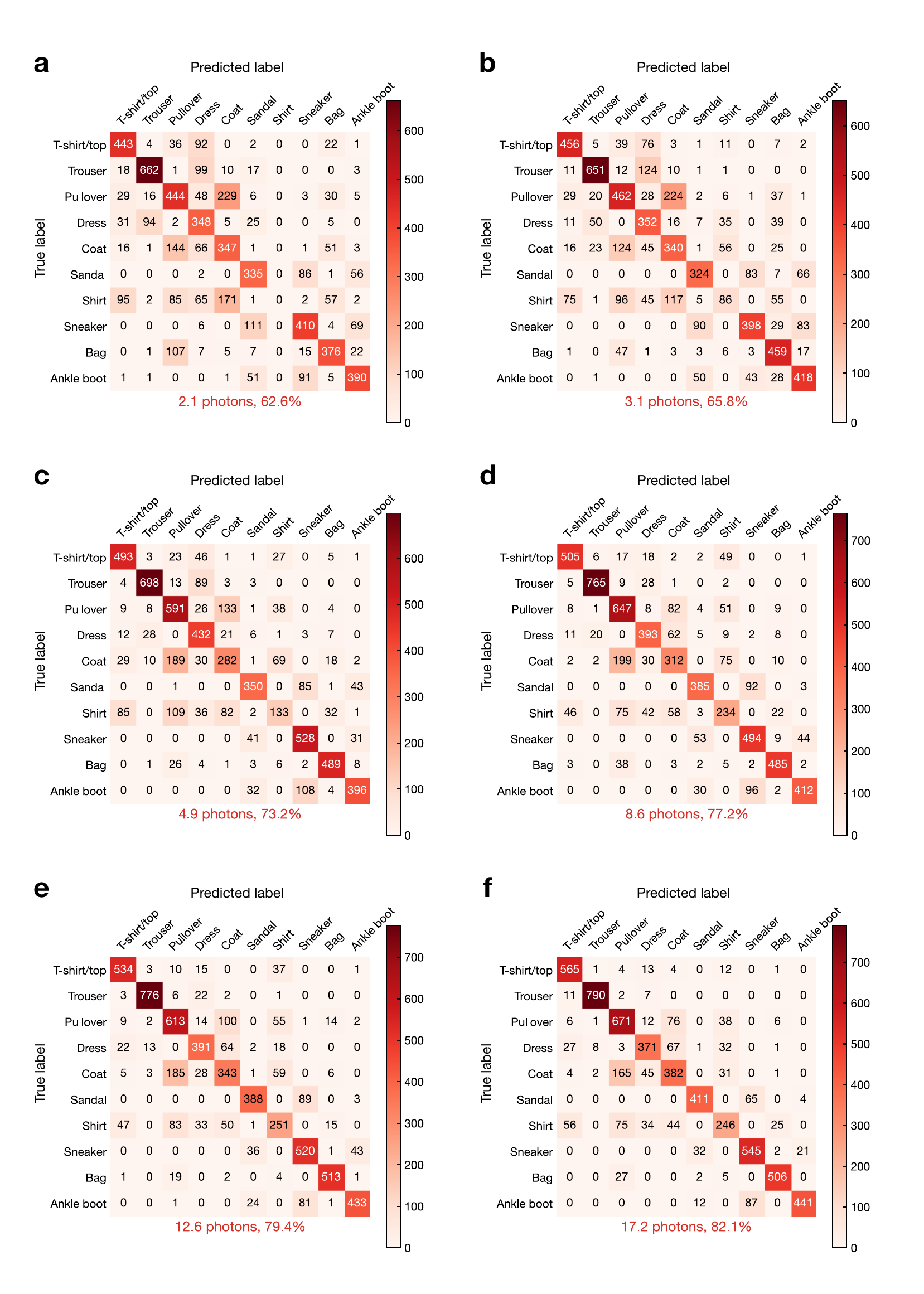}
    \caption{\textbf{Experimental confusion matrices for FashionMNIST at $K=1$.} \textbf{a}--\textbf{f} show results for $d_\text{f} = 4, 6, 10, 16, 24, 32$, with the average detected photon count and overall accuracy indicated below each matrix. Rows indicate true labels; columns indicate predicted labels. }
    \label{fig:confusion_mats_fashion}
\end{figure}

\begin{figure}[htp]
    \centering
    \includegraphics[width=.92\textwidth]{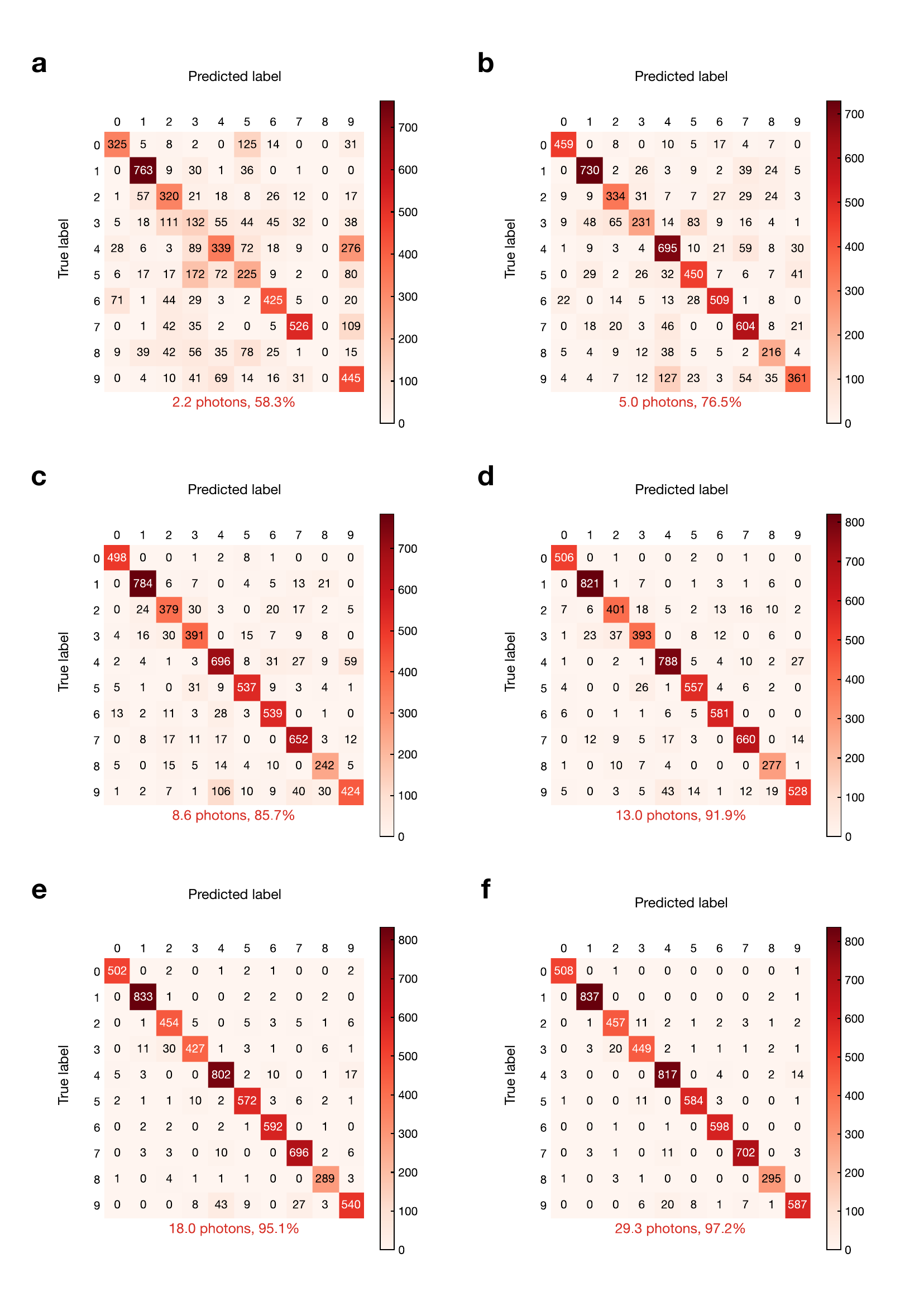}
    \caption{\textbf{Experimental confusion matrices for MNIST at $K=1$.} \textbf{a}--\textbf{f} show results for $d_\text{f} = 4, 9, 16, 25, 36, 64$, with the average detected photon count and overall accuracy indicated below each matrix.}
    \label{fig:confusion_mats_mnist}
\end{figure}

The output distributions in \ref{fig:output_distributions} characterize the behavior for individual test images. To examine classification performance across all test images, \ref{fig:confusion_mats_fashion} and \ref{fig:confusion_mats_mnist} present confusion matrices that aggregate 6000 classification outcomes (200 test images $\times$ 30 single-shot trials) at $K = 1$ for each $d_\text{f}$ configuration.

The confusion matrices confirm and extend the observations from the output distribution analysis. For FashionMNIST, Trouser achieves consistently high accuracy across all $d_\text{f}$ values. The footwear categories (Sandal, Sneaker, Ankle boot) show moderate mutual confusion. The most challenging distinctions involve upper-body garments: T-shirt/top, Pullover, Shirt, and Coat require higher $d_\text{f}$ to separate reliably. These patterns agree with the output distributions in \ref{fig:output_distributions} and the t-SNE clustering in \ref{fig:raw_activations}.

For MNIST, digits 1 and 4 achieve high accuracy even at low $d_\text{f}$, while digits sharing similar stroke structures show elevated confusion. As $d_\text{f}$ increases, the matrices become increasingly diagonal, with most off-diagonal entries approaching zero.

\subsection{$K$-shot integration}

The results presented so far use single-shot inference ($K = 1$), where each binary feature vector constitutes one independent classification attempt. In practical deployments, single-photon detectors operate with a fixed integration window and repeatedly collect counts—the natural operating mode that motivates $K$-shot measurements, where $K$ independent single-shot outcomes are summed before classification.

Importantly, $K$-shot scaling differs fundamentally from extending the integration time itself. Changing the integration duration would correspond to $\eta$ scaling (continuous adjustment of the expected photon level per measurement), whereas $K$-shot scaling keeps the integration window fixed and accumulates discrete, independent measurements. This distinction matters because the models are trained at a specific photon level ($\eta = 1$), and $K$-shot integration preserves this operating point while reducing measurement variance through averaging.

As discussed in \ref{subsec:K_scaling}, increasing $K$ improves classification accuracy but is not optimal with respect to total photon budget: for a fixed number of measurements, using more independently optimized features ($d_\text{f}$ scaling) generally outperforms repeated sampling of fewer features ($K$ scaling).

To evaluate experimental performance at different $K$ values, we group the 30 measurements per image: $K = 2$ yields 15 trials, $K = 3$ yields 10 trials, and $K = 5$ yields 6 trials, where each trial sums $K$ consecutive binary outcomes per pattern before classification. \ref{tab:exp_fmnist_K} and \ref{tab:exp_mnist_K} present the results.

\begin{table}[h]
\centering
\renewcommand{\arraystretch}{1.3}
\setlength{\tabcolsep}{10pt}
\caption{Experimental test accuracy (\%) on FashionMNIST for different $d_\text{f}$ and $K$. Each entry shows mean $\pm$ standard deviation across 200 test images.}
\label{tab:exp_fmnist_K}
\begin{tabular}{lcccc}
\hline
 & $K = 1$ & $K = 2$ & $K = 3$ & $K = 5$ \\
\hline
$d_\text{f} = 3$  & $52.35 \pm 1.43$ & $54.53 \pm 2.41$ & $56.30 \pm 1.35$ & $56.17 \pm 1.14$ \\
$d_\text{f} = 4$  & $62.58 \pm 2.03$ & $65.20 \pm 1.84$ & $66.95 \pm 1.42$ & $69.08 \pm 1.10$ \\
$d_\text{f} = 6$  & $65.77 \pm 2.35$ & $69.87 \pm 1.86$ & $71.95 \pm 1.77$ & $72.50 \pm 1.08$ \\
$d_\text{f} = 10$ & $73.20 \pm 2.22$ & $76.63 \pm 1.32$ & $77.80 \pm 1.82$ & $78.92 \pm 1.30$ \\
$d_\text{f} = 16$ & $77.20 \pm 2.20$ & $80.47 \pm 1.72$ & $82.05 \pm 1.75$ & $83.00 \pm 1.19$ \\
$d_\text{f} = 24$ & $79.37 \pm 2.06$ & $81.53 \pm 1.02$ & $83.15 \pm 1.42$ & $83.67 \pm 1.03$ \\
$d_\text{f} = 32$ & $82.13 \pm 1.27$ & $83.77 \pm 1.45$ & $84.90 \pm 0.94$ & $85.67 \pm 1.21$ \\
\hline
\end{tabular}
\end{table}

\begin{table}[h]
\centering
\renewcommand{\arraystretch}{1.3}
\setlength{\tabcolsep}{10pt}
\caption{Experimental test accuracy (\%) on MNIST for different $d_\text{f}$ and $K$. Each entry shows mean $\pm$ standard deviation across 200 test images.}
\label{tab:exp_mnist_K}
\begin{tabular}{lcccc}
\hline
 & $K = 1$ & $K = 2$ & $K = 3$ & $K = 5$ \\
\hline
$d_\text{f} = 4$  & $58.33 \pm 2.29$ & $61.83 \pm 2.53$ & $64.85 \pm 1.64$ & $68.67 \pm 1.14$ \\
$d_\text{f} = 9$  & $76.48 \pm 2.30$ & $82.57 \pm 2.12$ & $85.40 \pm 0.83$ & $87.75 \pm 0.95$ \\
$d_\text{f} = 16$ & $85.70 \pm 2.06$ & $90.87 \pm 1.37$ & $91.85 \pm 1.57$ & $93.08 \pm 0.98$ \\
$d_\text{f} = 25$ & $91.87 \pm 1.70$ & $95.50 \pm 1.10$ & $96.60 \pm 0.86$ & $96.58 \pm 1.54$ \\
$d_\text{f} = 36$ & $95.12 \pm 1.41$ & $97.73 \pm 0.96$ & $98.05 \pm 0.69$ & $98.67 \pm 0.37$ \\
$d_\text{f} = 64$ & $97.23 \pm 1.00$ & $99.00 \pm 0.66$ & $99.20 \pm 0.68$ & $99.75 \pm 0.25$ \\
\hline
\end{tabular}
\end{table}

The experimental results exhibit the same trends observed in simulation: accuracy 
increases with both $d_\text{f}$ and $K$, with diminishing returns as either 
parameter grows large. \ref{fig:exp_Ks} summarizes these results across all 
evaluated $d_\text{f}$ and $K$ configurations.

\subsection{Comparison to simulation}

Experimental accuracy can be compared directly to simulation predictions using the corresponding tables: FashionMNIST experimental results (\ref{tab:exp_fmnist_K}) versus simulation (\ref{tab:sim_fmnist_K}), and MNIST experimental results (\ref{tab:exp_mnist_K}) versus simulation (\ref{tab:sim_mnist_k}). The tables use identical formats and configurations with the same trained optical front end and digital back end; the difference is that simulation was performed on a digital computer while experimental data was collected from the physical setup.

Across all configurations, experimental accuracy closely tracks simulation predictions with a consistent gap of approximately 1--2 percentage points. 
Fig.~3B--C of the main text visualizes this comparison at $K = 1$, with a restricted y-axis range to highlight the relationship (unlike Fig.~2E in the main text, which spans the full range from zero to include the direct imaging 
baseline).

\begin{figure}[h]
    \centering
    \includegraphics[width=.98\textwidth]{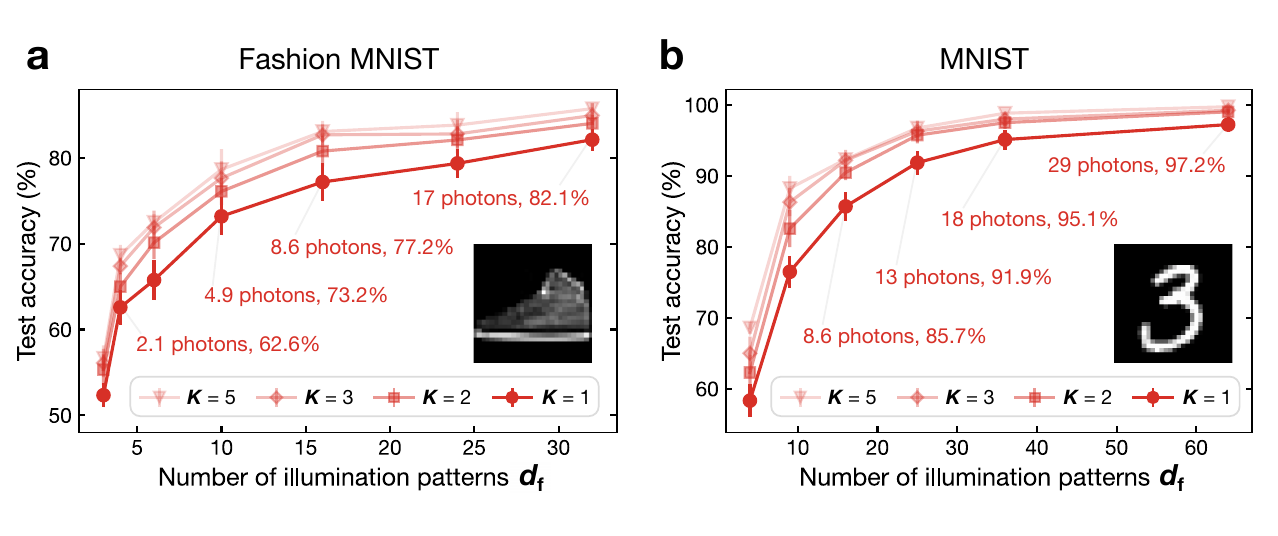}
    \caption{\textbf{Experimental classification accuracy versus number of 
    illumination patterns $d_\text{f}$ for different shot counts $K$.} 
    \textbf{a}, FashionMNIST and \textbf{b}, MNIST. Markers with error bars 
    show experimental results (mean $\pm$ standard deviation); curves connect 
    points for visual clarity. The four curves correspond to $K = 1, 2, 3, 5$ 
    accumulated single-shot measurements per pattern before classification. Red 
    annotations indicate the average total detected photon budget $N_{\text{det}}$ 
    per inference at selected $d_\text{f}$ values for $K=1$.}
    \label{fig:exp_Ks}
\end{figure}

This modest performance gap is expected given practical considerations in the experimental implementation. As shown in the robustness analysis, PANS models tolerate dark counts and ambient noise (\ref{fig:dcr_sweep}) as well as photon level fluctuations (\ref{fig:eta_sweep}), though small accuracy degradation does occur under these conditions. Additional factors include residual spatial nonuniformities in the OLED display and imaging system, and temporal drift in alignment and calibration over the course of sequential measurements. These contributions remain minor, partly due to careful calibration and more importantly because the PANS framework is inherently robust to such variations.

The qualitative scaling behavior is identical between experiment and simulation: accuracy increases monotonically with $d_\text{f}$, exhibits the same pattern of diminishing returns, and the relative ordering of all configurations matches exactly. 
The $K=1$ experimental data in Fig.~2 and Fig.~3 of the main text and the full $K$-scaling 
results in \ref{fig:exp_Ks} both correspond to the measurements reported in 
\ref{tab:exp_fmnist_K} and \ref{tab:exp_mnist_K}.

\subsection{Summary}

The experimental results validate active PANS under realistic hardware conditions. Despite the pronounced stochasticity of single-shot measurements at the detection bottleneck—where repeated trials of the same input yield visibly different detection patterns—the optimized illumination patterns successfully encode task-relevant information that the digital classifier reliably extracts. Classification accuracy closely tracks simulation predictions, confusion patterns concentrate on genuinely ambiguous categories, and the framework scales consistently with both feature dimension $d_\text{f}$ and shot count $K$.

These results demonstrate that PANS can be successfully deployed in a physical optical system, achieving high-accuracy classification with only a handful of detected photons per inference. The close agreement between simulation and experiment, despite practical imperfections inherent to real hardware, suggests that PANS provides a robust approach for photon-starved sensing applications in realistic conditions.

\bibliographystyle{mcmahonlab}
\bibliography{references}